\documentclass[11pt, leqno, a4]{article}
\usepackage{latexsym}
\usepackage{amsmath}
\usepackage{amssymb}
\usepackage{amsfonts}
\usepackage{stmaryrd}
\usepackage[T1]{fontenc}
\def\utw{\smash{\rlap{\lower5pt\hbox{$\sim$}}}}
\def\udtw{\smash{\rlap{\lower6pt\hbox{$\approx$}}}}

\def\bbbone{{\mathchoice {\rm 1\mskip-4mu l} {\rm 1\mskip-4mu l}
{\rm 1\mskip-4.5mu l} {\rm 1\mskip-5mu l}}}
\def\bbbc{{\mathchoice {\setbox0=\hbox{$\displaystyle\rm C$}\hbox{\hbox
to0pt{\kern0.4\wd0\vrule height0.9\ht0\hss}\box0}}
{\setbox0=\hbox{$\textstyle\rm C$}\hbox{\hbox
to0pt{\kern0.4\wd0\vrule height0.9\ht0\hss}\box0}}
{\setbox0=\hbox{$\scriptstyle\rm C$}\hbox{\hbox
to0pt{\kern0.4\wd0\vrule height0.9\ht0\hss}\box0}}
{\setbox0=\hbox{$\scriptscriptstyle\rm C$}\hbox{\hbox
to0pt{\kern0.4\wd0\vrule height0.9\ht0\hss}\box0}}}}
\def\bbbe{{\mathchoice {\setbox0=\hbox{\smalletextfont e}\hbox{\raise
0.1\ht0\hbox to0pt{\kern0.4\wd0\vrule width0.3pt
height0.7\ht0\hss}\box0}}
{\setbox0=\hbox{\smalletextfont e}\hbox{\raise
0.1\ht0\hbox to0pt{\kern0.4\wd0\vrule width0.3pt
height0.7\ht0\hss}\box0}}
{\setbox0=\hbox{\smallescriptfont e}\hbox{\raise
0.1\ht0\hbox to0pt{\kern0.5\wd0\vrule width0.2pt
height0.7\ht0\hss}\box0}}
{\setbox0=\hbox{\smallescriptscriptfont e}\hbox{\raise
0.1\ht0\hbox to0pt{\kern0.4\wd0\vrule width0.2pt
height0.7\ht0\hss}\box0}}}}
\def\bbbq{{\mathchoice {\setbox0=\hbox{$\displaystyle\rm Q$}\hbox{\raise
0.15\ht0\hbox to0pt{\kern0.4\wd0\vrule height0.8\ht0\hss}\box0}}
{\setbox0=\hbox{$\textstyle\rm Q$}\hbox{\raise
0.15\ht0\hbox to0pt{\kern0.4\wd0\vrule height0.8\ht0\hss}\box0}}
{\setbox0=\hbox{$\scriptstyle\rm Q$}\hbox{\raise
0.15\ht0\hbox to0pt{\kern0.4\wd0\vrule height0.7\ht0\hss}\box0}}
{\setbox0=\hbox{$\scriptscriptstyle\rm Q$}\hbox{\raise
0.15\ht0\hbox to0pt{\kern0.4\wd0\vrule height0.7\ht0\hss}\box0}}}}
\def\bbbt{{\mathchoice {\setbox0=\hbox{$\displaystyle\rm
T$}\hbox{\hbox to0pt{\kern0.3\wd0\vrule height0.9\ht0\hss}\box0}}
{\setbox0=\hbox{$\textstyle\rm T$}\hbox{\hbox
to0pt{\kern0.3\wd0\vrule height0.9\ht0\hss}\box0}}
{\setbox0=\hbox{$\scriptstyle\rm T$}\hbox{\hbox
to0pt{\kern0.3\wd0\vrule height0.9\ht0\hss}\box0}}
{\setbox0=\hbox{$\scriptscriptstyle\rm T$}\hbox{\hbox
to0pt{\kern0.3\wd0\vrule height0.9\ht0\hss}\box0}}}}
\def\bbbs{{\mathchoice
{\setbox0=\hbox{$\displaystyle     \rm S$}\hbox{\raise0.5\ht0\hbox
to0pt{\kern0.35\wd0\vrule height0.45\ht0\hss}\hbox
to0pt{\kern0.55\wd0\vrule height0.5\ht0\hss}\box0}}
{\setbox0=\hbox{$\textstyle        \rm S$}\hbox{\raise0.5\ht0\hbox
to0pt{\kern0.35\wd0\vrule height0.45\ht0\hss}\hbox
to0pt{\kern0.55\wd0\vrule height0.5\ht0\hss}\box0}}
{\setbox0=\hbox{$\scriptstyle      \rm S$}\hbox{\raise0.5\ht0\hbox
to0pt{\kern0.35\wd0\vrule height0.45\ht0\hss}\raise0.05\ht0\hbox
to0pt{\kern0.5\wd0\vrule height0.45\ht0\hss}\box0}}
{\setbox0=\hbox{$\scriptscriptstyle\rm S$}\hbox{\raise0.5\ht0\hbox
to0pt{\kern0.4\wd0\vrule height0.45\ht0\hss}\raise0.05\ht0\hbox
to0pt{\kern0.55\wd0\vrule height0.45\ht0\hss}\box0}}}}

\def\bbbz{{\mathchoice {\hbox{$\sf\textstyle Z\kern-0.4em Z$}}
{\hbox{$\sf\textstyle Z\kern-0.4em Z$}}
{\hbox{$\sf\scriptstyle Z\kern-0.3em Z$}}
{\hbox{$\sf\scriptscriptstyle Z\kern-0.2em Z$}}}}

\def\diameter{{\ifmmode\oslash\else$\oslash$\fi}}

\def\init{\setcounter{equation}{0}}

\newtheorem{theoreme}{Theorem }[section]
\newtheorem{proposition}[theoreme]{Proposition}
\newtheorem{lemma}[theoreme]{Lemma}
\newtheorem{definition}[theoreme]{Definition}

\newtheorem{remark}[theoreme]{Remark}

\def\rr{\mathbb{R}}
\def\cc{\mathbb{C}}
\def\nn{\mathbb{N}}
\def\zz{\mathbb{Z}}
\def\one{\bbbone}
\def\dd{{\bf d}}
\def\DD{{\bf D}}

\def\fin{{\rm fin}}

\def\t{{\langle t\rangle}}

\newcounter{smallarabics}

\newcommand{\ben}{\begin{enumerate}}
\newcommand{\een}{\end{enumerate}}

\def\Wick{{\rm Wick}}

\def\pp{{\rm pp}}

\def\fld{\rightarrow}

\def\e{{\rm e}}

\def\i{{\rm i}}
\def\d{{\rm d}}
\def\12{\frac{1}{2}}
\def\cinf{C^{\infty}}

\def\proof{\noindent{\bf  Proof. }}
\def\slim{\hbox{\rm s-}\lim}

\def\coinf{C_{0}^{\infty}}

\def\qed{$\Box$}

\def\Ran{{\rm Ran}}
\def\supp{{\rm supp\,}}
\def\cH{{\cal H}}
\def\cS{{\cal S}}
\def\G{\Gamma}

\def\cK{{\cal K}}
\def\cD{{\cal D}}
\def\K{{\cal K}}

\def\ch{{\mathfrak h}}
\def\cf{{\mathfrak f}}

\def\p{\partial}
\def\s{{\rm s}}
\def\ext{{\rm ext}}

\newcommand{\dega}[1]{\d\Gamma\left(#1\right)}
\def\rt{\frac{|x|}{t}}

\def\x{\langle x\rangle}
\def\xt{\frac{\x}{t}}
\def\hc{{\rm h.c.}}
\def\lPi{\mathop{\Pi}\limits}

\def\wlim{{\rm w}-\lim}

\def\invWave{W}

\def\pfi2{P(\varphi)_{2}}
\def\cO{{\check O}}

\def\ad{{\rm ad}}

\def\xr{\frac{\x}{R}}
\newcommand{\beq}{\begin{equation}}
\newcommand{\eeq}{\end{equation}}
\newcommand{\bet}{\begin{theoreme}}
\newcommand{\eet}{\end{theoreme}}
\newcommand{\bel}{\begin{lemma}}
\newcommand{\eel}{\end{lemma}}
\newcommand{\bep}{\begin{proposition}}
\newcommand{\eep}{\end{proposition}}

\newcommand{\bear}[1]{\begin{array}{#1}}
\newcommand{\ear}{\end{array}}

\begin{document}
\def\triple{\interleave}
\def\Gh{\Gamma(\ch)}
\title{Spectral and scattering theory for \\ some abstract QFT
Hamiltonians}\author{C. G\'erard, A. Panati, \\ Laboratoire de math\'ematiques, Universit\'e de Paris XI,\\
91\,405 Orsay Cedex France}\date{June 2008}
\maketitle
\begin{abstract}
We introduce an abstract class of bosonic QFT
Hamiltonians and study their spectral and scattering theories. These
Hamiltonians are of the form $H=\d\G(\omega)+ V$ acting on the bosonic
Fock space $\G(\ch)$, where $\omega$ is a massive one-particle
Hamiltonian acting on $\ch$ and $V$ is a Wick polynomial $\Wick(w)$
for a kernel $w$ satisfying some decay properties at infinity.

We describe the essential spectrum of $H$, prove a Mourre estimate
outside a set of thresholds and prove the existence of asymptotic
fields. Our main result is the {\em asymptotic completeness} of the
scattering theory, which means that the CCR representations given by
the asymptotic fields are of Fock type, with the asymptotic vacua equal
to the bound states of $H$. As a consequence $H$ is unitarily equivalent
to a collection of second quantized Hamiltonians.
\end{abstract}
\section{Introduction}\label{sec0}
\subsection{Introduction}

In recent years a lot of effort was devoted to the spectral and
scattering theory of various  models of Quantum Field
Theory like models of non-relativistic matter coupled to quantized
radiation or self-interacting relativistic models in dimension 1+1 
(see among many others the  papers \cite{ahh}, \cite{DG},
\cite{DG1}, \cite{FGSch}, \cite{FGS}, \cite{LL}, \cite{P}, \cite{Spohnbook} and
references therein). Substantial progress was made by applying to
these models methods originally developed in the study of
$N-$particle Schroedinger operators, namely the Mourre positive
commutator method and the method of propagation observables to
study the behavior of the unitary group $\e^{-\i tH}$ for large
times.

Up to now, the most  complete results (valid for example for arbitrary
coupling constants) on the spectral and scattering theory for
these models are available only for  massive models and for
localized interactions. (For results on massless models see for example
\cite{FGS} and references therein).

It turns out that for this type of models, the details of the
interaction are often irrelevant. The essential feature of the
interaction is that it can be written as a {\em Wick polynomial}, with a
symbol (see below) which decays sufficiently fast at infinity.

The conjugate operator (for the Mourre theory), or the propagation
observables (for the proof of propagation estimates), are chosen
as second quantizations of corresponding operators on the
one-particle space $\ch$.

In applications the one-particle kinetic energy is
usually  the operator $(k^{2}+ m^{2})^{\12}$ acting on $L^{2}(\rr^{d},\d k)$,
which clearly has a nice spectral and scattering theory.
Therefore the necessary one-particle operators  are easy to construct.

Our goal in this paper is to describe an abstract class of
bosonic QFT Hamiltonians to which the methods and results of
\cite{DG1}, \cite{DG} can be naturally extended.

Let us first briefly describe this class of models.
We consider
Hamiltonians  of the form:
\[
H= H_{0}+ V,\hbox{ acting on the bosonic Fock space }\G(\ch),
\]
where $H_{0}=\d\G(\omega)$ is the second quantization of the
one-particle kinetic energy $\omega$ and $V=\Wick(w)$ is a {\em Wick
polynomial}. To define $H$ without ambiguity, we assume that $H_{0}+V$
is essentially selfdjoint and bounded below on $\cD(H_{0})\cap
\cD(V)$.

The Hamiltonian $H$  is assumed to be {\em massive}, namely we
require that $\omega\geq m>0$ and moreover that  powers of the  number
operator $N^{p}$ for $p\in \nn$ are controlled by sufficiently high
powers of the resolvent $(H+b)^{-m}$. These bounds are usually called
{\em higher order estimates}.

The interaction $V$  is supposed to be a {\em Wick polynomial}. If for
example $\ch= L^{2}(\rr^{d}, \ d k)$, this means that $V$ is a finite
sum $V=\sum_{p, q\in I}\Wick(w_{p,q})$ where $\Wick(w_{p,q})$ is
formally defined as:
\[
\Wick(w_{p,q})=\int a^{*}(K)a(K')w_{p, q}(K, K')\d K \d K',
\]
for
\[
K= (k_{1}, \dots, k_{p}), \: K'=(k'_{1},\dots, k'_{q}), \:
a^{*}(K)=\Pi_{i=1}^{p}a^{*}(k_{i}), \ \
a^{*}(K')=\Pi_{i=1}^{q}a(k'_{i}),
\]
and $w_{p,q}(K , K')$ is a scalar function separately symmetric in $K$
and $K'$. To define $\Wick(w)$ as an unbounded operator on $\G(\ch)$,
the functions $w_{p,q}$ are supposed to be in $L^{2}(\rr^{(p+q)d})$.
The functions $w_{p,q}$ are then the distribution kernels of a
Hilbert-Schmidt operator $w_{p,q}$ from $\otimes_{\s}^{q}\ch$ into
$\otimes_{\s}^{p}\ch$. Putting together these operators we obtain a
Hilbert-Schmidt operator $w$ on $\G(\ch)$ which is called the {\em Wick
symbol} of the interaction $V$.

In physical situations, this corresponds to an interaction which has
both a space and an ultraviolet cutoff (in one space dimension, only a
space cutoff is required).

As said above, it is necessary to assume  that the one-particle
energy $\omega$ has a nice spectral and scattering theory.  It is
possible to formulate the necessary properties of $\omega$ in a
very abstract framework, based on the existence of only two
auxiliary Hamiltonians on $\ch$. The first one is a {\em conjugate
operator} $a$ for $\omega$, in the sense of the Mourre method. The
second one is a {\em weight operator} $\x$, which is used both to
control the 'order' of various operators on $\ch$ and as a way to
localize bosons in $\ch$.  Note that the one-particle energy
$\omega$ may have bound states.

The first basic result on spectral theory that we obtain is the {\em
HVZ theorem}, which describes the essential spectrum of $H$. If
$\sigma_{\rm ess}(\omega)=[m_{\infty}, +\infty[$ for some
$m_{\infty}\geq m>0$, then we show that
\[
 \sigma_{\rm ess}(H)= [\inf\sigma(H)+ m_{\infty}, +\infty[,
\]
in particular $H$ always has a ground state.

We then consider the Mourre theory and prove that the second quantized
Hamiltonian $A= \d\G(a)$ is a conjugate operator for $H$. In
particular this proves the local finiteness of point spectrum outside
of the set of {\em thresholds}, which is equal to
\[
\tau(H)= \sigma_{\rm pp}(H)+ \d\G^{(1)}(\tau(\omega)),
\]
where $\tau(\omega)$ is the set of thresholds of $\omega$ for $a$ and
$\d\G^{(1)}(E)$ for $E\subset \rr$ is the set of all finite sums of elements of $E$.

The scattering theory for our abstract Hamiltonians follows the
standard approach based on the {\em asymptotic Weyl operators}. These
are defined as the limits:
\[
W^{\pm}(h)=\slim_{t\to \pm \infty}\e^{\i tH}W(h_{t})\e^{-\i tH}, \ \
h\in \ch_{\rm c}(\omega),
\]
where $\ch_{\rm c}(\omega)$ is the continuous spectral subspace 
 for $\omega$ and $h_{t}= \e^{-\i t \omega}h$. The asymptotic Weyl
operators define two CCR representations over $\ch_{\rm
c}(\omega)$.  Due to the fact that the theory is massive, it is
rather easy to see that these representations are of Fock type.
The main problem of scattering theory is to describe their {\em
vacua}, i.e. the spaces of vectors annihilated by the asymptotic
annihilation operators $a^{\pm}(h)$ for $h\in \ch_{\rm
c}(\omega)$.

The main result of this paper is that the vacua coincide with the {\em
bound states} of $H$. As a consequence one sees that $H$ is unitarily
equivalent to the asymptotic Hamiltonian:
\[
H_{|\cH_{\rm pp}(H)}\otimes\one + \one\otimes\d\G(\omega), \hbox{
acting on }\cH_{\rm pp}(H)\otimes\G(\ch_{\rm c}(\omega)).
\]
This result is usually called the {\em asymptotic completeness} of wave
operators.  It implies that $H$ is unitarily equivalent to a direct
sum of $E_{i}+ \d\G(\omega_{|\ch_{\rm c}(\omega)})$, where $E_{i}$
are the eigenvalues of $H$. In more physical terms, asymptotic
completeness means that for large times any initial state
asymptotically splits into  a bound state and a finite number of free
bosons.

We conclude the introduction by describing the examples of
abstract QFT Hamiltonians to which our results apply.

The first example is the
space-cutoff $P(\varphi)_{2}$ model with a {\em variable metric},
which  corresponds to the quantization of a non-linear
Klein-Gordon equation with variable coefficients in one space
dimension.

The one-particle space is  $\ch= L^{2}(\rr, \d x)$ and the usual
relativistic kinetic energy $(D^{2}+ m^{2})^{\12}$ is replaced by the square root
$h^{\12}$ of a second order differential operator $h= D a(x)D + c(x)$,
where $a(x)\to 1$  and $c(x)\to m_{\infty}^{2}$  for $m_{\infty}>0$
when $x\to \infty$. (It is also possible to treat functions $c$ having
different limits $m^{2}_{\pm\infty}>0$ at $\pm\infty$).

The interaction is of the form:
\[
V= \int_{\rr}g(x):\!P(x, \varphi(x))\!:\d x,
\]
where $g\geq 0$  is a function on $\rr$ decaying sufficiently fast at
$\infty$, $P(x, \lambda)$ is a bounded below polynomial of even degree
with variable coefficients,  $\varphi(x)=
\phi(\omega^{-\12}\delta_{x})$ is the relativistic field operator and
$:\ \ :$ denotes the Wick ordering.

This model is  considered in details in 
\cite{GP}, applying the abstract arguments in this paper. Note that
some conditions on the eigenfunctions and generalized eigenfunctions
of $h$ are necessary in order to prove the higher order estimates.

The analogous model for constant coefficients was considered in
\cite{DG}. Even in the constant coefficient case we
improve the results in \cite{DG} by  removing an unpleasant
technical assumption on $g$, which excluded to take $g$ compactly
supported.

The second example is the generalization to higher dimensions. The
one-particle energy $\omega$ is:
\[
\omega= (\sum_{1\leq i,j\leq d}D_{i}a_{ij}(x)D_{j}+ c(x))^{\12},
\]
where $h= \sum_{1\leq i,j\leq d}D_{i}a_{ij}(x)D_{j}+ c(x)$ is an
elliptic second order differential operator converging to $D^{2}+
m_{\infty}^{2}$ when $x\to \infty$. The interaction is now
\[
\int_{\rr}g(x)P(x, \varphi_{\kappa}(x))\d x,
\]
where $P$ is as before and $\varphi_{\kappa}(x)=
\phi(\omega^{-\12}F(\omega\leq \kappa)\delta_{x})$ is now the  UV-cutoff
relativistic field. Here because of the UV cutoff, the Wick ordering
is irrelevant.
Again some conditions on eigenfunctions and generalized eigenfunctions
of $h$ are necessary.

We believe that our set of hypotheses should be sufficiently general
to consider also Klein-Gordon equations on other Riemannian manifolds, like for
example manifolds equal to the union of a compact piece and a cylinder
$\rr^{+}\times M$, where the metric on $\rr^{+}\times M$ is of product
type.

\subsection{Plan of the paper}
We now describe briefly the plan of the paper.

Section \ref{sec1} is a collection of various auxiliary results
needed in the rest of the paper. We first recall in Subsects.
\ref{sec1.3} and \ref{sec1.4} some  arguments connected with the
abstract Mourre theory and a convenient functional calculus
formula. In Subsect. \ref{sec1.5} we fix some notation 
 connected with  one-particle operators.
Standard results taken from \cite{DG}, \cite{DG1} on bosonic Fock
spaces and Wick polynomials are recalled in Subsects. \ref{sec1.1}
and \ref{sec1.2}.

The class of abstract QFT Hamiltonians that we will consider in the
paper is described in Sect. \ref{sec3}.  The results of the paper are
summarized in Sect. \ref{sec1bis}. In Sect. \ref{sec2} we give
examples of abstract QFT Hamiltonians to which all our results apply,
namely the  space-cutoff $P(\varphi)_{2}$ model with a
variable metric,  and the analogous models in higher dimensions, where now an ultraviolet cutoff is imposed on the
polynomial interaction.

Sect. \ref{sec3b} is devoted to the proof of commutator estimates
needed in  various localization arguments.
The spectral theory of abstract QFT Hamiltonians is studied in Sect.
\ref{sec4}. The essential spectrum is described in Subsect.
\ref{sec4.1}, the virial
theorem and Mourre's positive commutator estimate are proved in
Subsects. \ref{sec4.2}, \ref{sec4.4} and  \ref{sec4.5}. 
The results of
Sect. \ref{sec4} are related to those  of 
\cite{G}, where abstract bosonic and fermionic QFT Hamiltonians are
considered using a $C^{*}-$algebraic approach instead of the 
geometrical approach used in our paper. Our result on essential
spectrum can certainly be deduced from the results in \cite{G}. However
the Mourre theory in \cite{G} requires  that the one-particle
Hamiltonian $\omega$ has no eigenvalues and also that $\omega$ is
affiliated to an abelian $C^{*}-$algebra ${\cal O}$ such that $\e^{\i
t a}{\cal O}\e^{-\i t a}={\cal O}$, where $a$ is the one-particle
conjugate operator. In concrete examples, this second assumption seems
adapted to constant coefficients one-particle Hamiltonians and not
satisfied by  the examples we describe in Sect. \ref{sec2}.

In Sect. \ref{sec5} we describe the scattering theory for abstract QFT
Hamiltonians. The existence of asymptotic Weyl operators and 
asymptotic fields is shown in Subsect. \ref{sec5.1}. Other natural objects, like the wave
operators and extended wave operators are defined in Subsects.
\ref{sec5.2}, \ref{sec5.3}.

Propagation estimates are shown in Sect. \ref{sec6}. The most
important are the phase-space propagation estimates in  Subsect.
\ref{sec6.2}, \ref{sec6.3} and the minimal velocity estimate in
Subsect. \ref{sec6.4}.

Finally asymptotic completeness is proved in Sect. \ref{sec7}. The two
main steps is the proof of {\em geometric asymptotic completeness} in
Subsect. \ref{sec7.4},
identifying the vacua with the states for which no bosons escape to
infinity. The asymptotic completeness itself is shown in Subsect.
\ref{sec7.5}.

Various technical proofs are collected in the Appendix.

\section{Auxiliary results}\label{sec1}\init
In this section we collect various auxiliary results which will be
used in the sequel. 
\subsection{Commutators}\label{sec1.3}
Let $A$ be a selfadjoint operator on a Hilbert space $\cH$.
If $B\in B(\cH)$ on says that $B$ is {\em of class} $C^{1}(A)$ \cite{ABG} if the map
\[
\rr\ni t\mapsto \e^{\i t A}B\e^{-\i tA}\in B(\cH)
\]
is $C^{1}$ for the strong topology.

If $H$ is selfadjoint on $\cH$,
one says that $H$ is {\em of class }$C^{1}(A)$ \cite{ABG} if for some (and hence all) $z\in
\cc\backslash \sigma(H)$, $(H-z)^{-1}$ is of class $C^{1}(A)$.  The
classes $C^{k}(A)$ for $k\geq 2$ are defined similarly.

If $H$ is of class $C^{1}(A)$, the
commutator $[H, \i A]$ defined as a quadratic form on $\cD(A)\cap \cD(H)$
extends then uniquely as a bounded quadratic form on $\cD(H)$. The
corresponding operator in $B(\cD(H), \cD(H)^{*})$ will be denoted by $[H,
\i A]_{0}$.

If $H$ is
of class $C^{1}(A)$ then the {\em virial relation} holds (see
\cite{ABG}):
\[
\one_{\{\lambda\}}(H)[H, \i A]_{0}\one_{\{\lambda\}}(H)=0, \ \ \lambda\in
\rr.
\]
An estimate of the form
\[
\one_{I}(H)[H, \i A]_{0}\one_{I}(H)\geq c_{0} \one_{I}(H)+K,
\]
where $I\subset \rr$ is a compact interval, $c_{0}>0$ and $K$ a compact operator on $\cH$,
or:
\[
\one_{I}(H)[H, \i A]_{0}\one_{I}(H)\geq c_{0} \one_{I}(H),
\]
is called a (strict) {\em Mourre estimate} on $I$. An operator $A$
such that the Mourre estimate holds on $I$ is called a {\em
conjugate operator} for $H$ (on $I$).  Under an additional
regularity condition of $H$ w.r.t. $A$ (for example if $H$ is of
class $C^{2}(A)$), it has several important consequences like
weighted estimates on $(H-\lambda\pm\i 0)^{-1}$ for $\lambda\in I$
(see e.g. \cite{ABG}) or abstract propagation estimates (see e.g.
\cite{HSS}).

We now recall some useful machinery from \cite{ABG} related with the
best constant $c_{0}$ in the Mourre estimate. Let $H$
be a selfadjoint operator on  a Hilbert space $\cH$  and $B$
be a  quadratic form with domain $\cD(H^{M})$ for some $M\in \nn$ such
that the {\em virial relation}
\beq\label{e2.18bis}
\one_{\{\lambda\}}(H)B\one_{\{\lambda\}}(H)=0, \ \ \lambda\in \rr,
\eeq
is satisfied.
We set
\[
\rho_{H}^{B}(\lambda):=\sup\{a\in \rr| \: \exists\:\chi\in \coinf(\rr),
\: \chi(\lambda)\neq 0, \ \ \chi(H)B\chi(H)\geq a\chi^{2}(H)\},
\]
\[
\tilde{\rho}^{B}_{H}(\lambda):=\sup\{a\in \rr| \: \exists\:\chi\in \coinf(\rr),
\: \chi(\lambda)\neq 0, \exists\: K\hbox{ compact, }\ \
\chi(H)B\chi(H)\geq a\chi^{2}(H)+K\}.
\]
The functions, $\rho_{H}^{B}$, $\tilde{\rho}_{H}^{B}$ are lower
semi-continuous and it follows from the virial relation that
$\rho^{B}_{H}(\lambda)<\infty$ iff $\lambda\in \sigma(H)$,
$\tilde{\rho}^{B}_{H}(\lambda)<\infty$ iff $\lambda\in \sigma_{\rm
ess}(H)$ (see \cite[Sect. 7.2]{ABG}). One sets:
\[
\tau_{B}(H):=\{\lambda|\: \tilde{\rho}^{B}_{H}(\lambda)\leq 0\}, \ \
\kappa_{B}(H):=\{\lambda|\: \rho^{B}_{H}(\lambda)\leq 0\},
\]
which are closed subsets of $\rr$, and
\[
\mu_{B}(H):= \sigma_{\rm pp}(H)\backslash
\tau_{B}(H).
\]
The virial relation and the usual argument shows that the eigenvalues
of $H$ in $\mu_{B}(H)$ are of finite multiplicity and are not
accumulation points of eigenvalues. In the next lemma we collect
several abstract results adapted from \cite{ABG}, \cite{BG}.
\begin{lemma}\label{2.3}
{\it i)}  if $\lambda\in \mu_{B}(H)$ then $\rho^{B}_{H}(\lambda)=0$. If
$\lambda\not\in\mu_{B}(H)$ then $\rho^{B}_{H}(\lambda)=
\tilde{\rho}^{B}_{H}(\lambda)$.

{\it ii)} $\rho^{B}_{H}(\lambda)>0$ iff
$\tilde{\rho}^{B}_{H}(\lambda)>0$ and $\lambda\not\in \sigma_{\rm
pp}(H)$, which implies that
\[
\kappa_{B}(H)= \tau_{B}(H)\cup \sigma_{\rm pp}(H).
\]
{\it iii)} Let $\cH=\cH_{1}\oplus\cH_{2}$, $H=H_{1}\oplus H_{2}$,
$B=B_{1}\oplus B_{2}$, where  $B_{i}$, $H, B$ are as above and satisfy
(\ref{e2.18bis}). Then
\[
\rho_{H}^{B}(\lambda)=\min(\rho_{H_{1}}^{B_{1}}(\lambda),
\rho_{H_{2}}^{B_{2}}(\lambda)).
\]
{\it iv)} Let $\cH= \cH_{1}\otimes \cH_{2}$, $H=
H_{1}\otimes\one+\one\otimes H_{2}$, $B= B_{1}\otimes\one+ \one\otimes B_{2}$, where
$H_{i}, B_{i}$, $H, B$ are as above, satisfy (\ref{e2.18bis}) and
$H_{i}$ are bounded below. Then
\[
\rho^{B}_{H}(\lambda)=\mathop{\inf}\limits_{\lambda_{1}+
\lambda_{2}=\lambda}\left(\rho_{H_{1}}^{B_{1}}(\lambda_{1})+
\rho_{H_{2}}^{B_{2}}(\lambda_{2})\right).
\]
\end{lemma}
\proof {\it i)}, {\it ii)} can be found in \cite[Sect.
7.2]{ABG}, in the case $B=[H, \i A]$ for $A$ a selfjadjoint operator such
that $H\in C^{1}(A)$. This hypothesis is only needed to ensure the
virial relation (\ref{e2.18bis}). {\it iii)} is easy and
{\it iv)} can
be found in
\cite[Prop. Thm. 3.4]{BG} in the same framework. Again it is easy to see that
the proof extends verbatim to our situation. \qed

\medskip

Assume now that  $H$, $A$ are  two selfadjoint operators on a
Hilbert space $\cH$ such that the quadratic form $[H, \i A]$
defined on $\cD(H^{M})\cap \cD(A)$ for some $M$ uniquely extends
as a quadratic form $B$ on $\cD(H^{M})$ and the virial relation
(\ref{e2.18bis}) holds.  Abusing notation we will in the rest of
the paper denote by $\tilde{\rho}^{A}_{H}$, $\rho^{A}_{H}$,
$\tau_{A}(H)$, $\kappa_{A}(H)$ the  objects introduced above for
$B=[H, \i A]$. The set $\tau_{A}(H)$ is usually called the set of {\em
thresholds} of $H$ for $A$.

\subsection{Functional calculus}\label{sec1.4}
If $\chi\in \coinf(\rr)$, we denote by $\tilde{\chi}\in \coinf(\cc)$ an
almost analytic extension of $\chi$, satisfying
\[
\begin{array}{l}
\tilde{\chi}_{\mid \rr}=\chi,\\[3mm]
|\p_{\,\overline z}\tilde{\chi}(z) |\leq C_{n}|{\rm Imz}|^{n},\ \  \: n\in \nn.
\end{array}
\]
We use the following functional calculus
formula for $\chi\in \coinf(\rr)$ and $A$ selfadjoint:
\beq
\chi(A)=\frac{\i}{2\pi}\int_{\cc}\partial_{\,\overline z}\tilde{\chi}(z)
(z-A)^{-1}\d z\wedge \d\,\overline z.
\label{HS}
\eeq
\subsection{Abstract operator classes}\label{sec1.5}

In this subsection we introduce a poor man's version of 
pseudodifferential calculus tailored to our  abstract setup. It rests
on two positive selfadjoint operators $\omega$ and $\x$ on the one-particle
space $\ch$. Later $\omega$ will of course be the one-particle
Hamiltonian. The operator $\x$ will have two purposes: first as a weight to control
various operators, and second as an observable to localize
particles in $\ch$.

We fix selfadjoint operators $\omega$, $\x$ on $\ch$ such that:
\[
\begin{array}{l}
 \omega\geq m>0, \ \x\geq 1, \\[2mm] 
\hbox{there exists a dense subspace }\cS\subset \ch \hbox{ such that
}\omega, \x: \cS\to \cS.
\end{array}
\]
To understand the terminology
below the reader familiar with the standard pseudodifferential
calculus should think of the example
\[\ch= L^{2}(\rr^{d}), \ \ \omega=(D_{x}^{2}+ 1)^{\12}, \ \
 \x= (x^{2}+ 1)^{\12},  \hbox{ and }\cS=\cS(\rr^{d}).
\]

To control various commutators later it is convenient to introduce the
following classes of operators on $\ch$. If $a,b:\cS\to \cS$ we set
${\rm ad}_{a}b=[a, b]$ as an operator on $\cS$.
\begin{definition}\label{symbols}
For $m\in \rr$, $0\leq \delta<\12$ and $k\in \nn$ we set
\[
S^{m}_{(0)}=\{b:\cS\to \ch \:| \: \x^{s}b\x^{-s-m}\in B(\ch), \ \
s\in \rr\},
\]
and for $k\geq 1$:
\[
S^{m}_{\delta, (k)}=\{b:\cS\to \cS \: | \:
\x^{-s}\ad^{\alpha}_{\x}\ad^{\beta}_{\omega}b\x^{s-m+
(1-\delta)\beta-\delta\alpha}\in B(\ch) \ \
\alpha+ \beta\leq k, \ \ s\in \rr\},
\]
where the multicommutators  are considered as operators on $\cS$.
\end{definition}
The parameter $m$ control the "order" of the operator: roughly
speaking an operator in $S^{m}_{\delta, (k)}$ is controlled by
$\x^{m}$.  The parameter $k$ is the number of commutators of the
operator with $\x$ and $\omega$  that are controlled. The lower
index $\delta$ controls the behavior of multicommutators: one
looses  $\x^{\delta}$ for each commutator with $\x$ and gains
$\x^{1-\delta}$ for each commutator with $\omega$.

The operator norms of the (weighted) multicommutators above can be
used as a family of seminorms on $S^{m}_{\delta, (k)}$.

The spaces $S^{m}_{\delta, (k)}$ for $\delta=0$ will be denoted simply
by $S^{m}_{(k)}$.
We will use the following natural notation for operators depending on
a parameter:

if $b= b(R)$ belongs to $S^{m}_{\delta, (k)}$ for all $R\geq 1$ we
will say that
\[
b\in O(R^{\mu})S^{m}_{\delta,(k)},
\]
if the seminorms of $R^{-\mu}b(R)$ in $S^{m}_{\delta, (k)}$ are
uniformly bounded in $R$. The following lemma is easy.
\begin{lemma}\label{gototime}
{\it i)}
\[
 S^{m_{1}}_{\delta, (k)}\times
S^{m_{2}}_{\delta, (k)}\subset S^{m_{1}m_{2}}_{\delta, (k)}.
\]
{\it ii)} Let $b\in S^{(m)}_{(0)}$. Then $J(\frac{\x}{R})b\x^{s}\in O(R^{m+s})$ for
$m+s\geq 0$ if $J\in \coinf(\rr)$ and for all $s\in \rr$  if $J\in
\coinf(]0, +\infty[)$.
\end{lemma}
\proof {\it i)} follows from Leibniz rule applied to the operators $\ad_{\x}$
and $\ad_{\omega}$. {\it ii)} is immediate. \qed

\subsection{Fock spaces.}\label{sec1.1}
In this subsection we recall  various definitions on bosonic
Fock spaces.  We will also collect some
bounds needed later.

\medskip

{\bf Bosonic Fock spaces.}
\medskip

If $\ch$ is  a Hilbert space then
\[
\Gamma(\ch):=\bigoplus_{n=0}^\infty\otimes_{\rm s}^{n}\ch,
\]
is the {\em bosonic Fock space} over $\ch$. $\Omega\in \G(\ch)$ will denote the
{\em vacuum vector}.
The {\em number operator} $N$ is defined as
\[
N\Big|_{\bigotimes_{\rm s}^{n}\ch}=n\one.
\]
We define the space of {\em finite particle vectors}:
\[
\Gamma_{\rm fin}(\ch):=\{u\in \Gamma(\ch) \: |
\hbox{ for some }\ n\in\nn,\ \ \one_{[0,n]}(N)u=u\},
\]
The {\em creation-annihilation} operators on $\G(\ch)$ are denoted by
$a^{*}(h)$ and $a(h)$. We denote by
\[
\phi(h):=\frac{1}{\sqrt{2}}(a^{*}(h)+ a(h)),\ W(h):=\e^{\i \phi(h)},
\]
the {\em field} and {\em Weyl operators}.
\medskip

{\bf $\d\G$ operators.}

\medskip

If $r:\ch_{1}\to \ch_{2}$ is an operator one sets:
\[
\begin{array}{rl}
\d \Gamma(r)&:\Gamma(\ch_{1})\to\Gamma(\ch_{2}),\\[3mm]
\d\Gamma(r)\Big|_{\bigotimes_\s^n\ch_{1}}&
:=\sum\limits_{j=1}^n\one^{\otimes(j-1)}
\otimes r\otimes \one^{\otimes(n-j)},
\end{array}
\]
with domain $\G_{\rm fin}(\cD(r))$. If $r$ is closeable, so is
$\d\G(r)$.
\medskip

{\bf $\G$ operators.}
\medskip

If  $q:\ch_{1}\mapsto \ch_{2}$
is  bounded  one sets:
\[
\begin{array}{l}\G(q):
\G(\ch_{1})\mapsto \G(\ch_{2})\\[3mm]
\G(q)\Big|_{\bigotimes_{\rm s}^{n}\ch_{1}}= q\otimes\cdots \otimes q.
\end{array}
\]
$\G(q)$ is bounded iff $\|q\|\leq 1$ and then $\|\G(q)\|=1$.

\medskip

{\bf $\d\G(r, q)$ operators.}

If $r, q$ are as above one sets:
\[\begin{array}{rl}
\d\Gamma(q,r)&:\Gamma(\ch_1)\to\Gamma(\ch_2),\\[3mm]
\d\Gamma(q,r)\Big|_{\bigotimes_\s^n\ch_1}&:=
\sum\limits_{j=1}^nq^{\otimes(j-1)}\otimes r \otimes
q^{\otimes(n-j)},
\end{array}
\]
with domain $\G_{\rm fin}(\cD(r))$.
We refer the reader to \cite[Subsects 3.5, 3.6, 3.7]{DG} for more
details.
\medskip

{\bf Tensor products of Fock spaces.}
\medskip

If $\ch_{1}$, $\ch_{2}$ are two Hilbert spaces, one denote by $U:
\G(\ch_{1})\otimes\G(\ch_{2})\to \G(\ch_{1}\oplus\ch_{2})$ the
canonical unitary map (see e.g. \cite[Subsect. 3.8]{DG} for
details).

If $\cH=\G(\ch)$, we set
\[
\cH^{\ext}:= \cH\otimes\cH\simeq \G(\ch\oplus \ch).
\]
The second copy of $\cH$ will be the state space for bosons {\em living near
infinity} in the spectral theory of a Hamiltonian $H$ acting
on $\cH$.

Let  $H=\d\G(\omega)+ V$ be an abstract QFT Hamiltonian defined in
Subsect. \ref{sec3.1}
Then we set:
\[
\cH^{\rm scatt}:=\cH\otimes\G(\ch_{\rm c}(\omega)).
\]
 The Hilbert space $\G(\ch_{\rm c}(\omega))$ will be the state space
for
{\em free bosons} in the scattering   theory of a Hamiltonian $H$ acting
on $\cH$. We will need also:
\[
H^{\ext}:= H\otimes\one+ \one\otimes\d\G(\omega),\hbox{ acting on
}\cH^{\ext}.
\]
Clearly $\cH^{\rm scatt}\subset\cH^{\rm ext}$ and $H^{\ext}$ preserves
$\cH^{\rm scatt}$. We will use the notation
\[
N_{0}:= N\otimes \one,\ \ \ \
N_{\infty}:= \one\otimes N,\hbox{ as operators on }\cH^{\rm
ext}\hbox{ or }\cH^{\rm scatt}.
\]

\medskip

{\bf Identification operators.}

\medskip

The {\em  identification operator} is defined as
\[
\begin{array}{l}
I: \cH^{\rm ext}\to \cH, \\[3mm]
I:= \G(i)U,
\end{array}
\]
where $U$ is defined as above for $\ch_{1}=\ch_{2}=\ch$ and:
\[
\begin{array}{l}
i: \ch\oplus \ch\fld \ch ,\\[3mm]
(h_{0}, h_{\infty})\mapsto h_{0}+ h_{\infty}.
\end{array}
\]
We have:
\[
I\mathop{\Pi}\limits_{i=1}^{n}a^{*}(h_{i})\Omega \otimes
\lPi_{i=1}^{p}a^{*}(g_{i})\Omega:=
 \lPi_{i=1}^{n}a^{*}(h_{i})\lPi_{i=1}^{p}a^{*}(g_{i})\Omega,
\ \ \ \  h_{i}\in \ch, \ \ g_{i}\in \ch.
\]

If $\omega$ is a selfadjoint operator as above, we denote by $I^{\rm
scatt}$ the restriction of $I$ to $\cH^{\rm scatt}$.

Note that $\|i\|=\sqrt{2}$ so $\G(i)$ and hence $I$, $I^{\rm scatt}$
are  unbounded. As
domain for $I$ (resp. $I^{\rm scatt}$) we can choose for example $\cD(N^{\infty})\otimes\G_{\rm
fin}(\ch)$ (resp. $\cD(N^{\infty})\otimes \G_{\rm fin
}(\ch_{\rm c}(\omega))$).
We refer to \cite[Subsect. 3.9]{DG} for details.
\medskip

{\bf Operators $I(j)$ and $\d I(j,k)$.}

\medskip

Let $j_{0}, j_{\infty}\in B(\ch)$ and set $j=(j_0,j_\infty)$.
We define
\[I(j):\Gamma_\fin(\ch)\otimes\Gamma_\fin(\ch)\to\Gamma_\fin(\ch)\]
\[I(j):=I\Gamma(j_0)\otimes\Gamma(j_\infty).\]
If we identify $j$ with the operator
\beq
\begin{array}{l}
j:\ch\oplus \ch \to \ch,\\[3mm]
j(h_0\oplus h_\infty) := j_{0}h_0+j_{\infty}h_\infty,
\end{array}
\label{mapp}\eeq
then we have
\[
I(j)=\Gamma(j)U.
\]
We deduce from this identity that if $j_{0}j_{0}^{*}+
j_{\infty}j_{\infty}^{*}=\one$ (resp. $j_{0}j_{0}^{*}+
j_{\infty}j_{\infty}^{*}\leq \one$) then $I^{*}(j)$ is isometric
(resp.
is a contraction).

Let $j=(j_0,j_\infty)$, $k=(k_0,k_\infty)$ be pairs of maps from $\ch$ to
$\ch$.
We define
\[\d I(j,k):\Gamma_\fin(\ch)\otimes\Gamma_\fin(\ch)
\to\Gamma_\fin(\ch)\]
as follows:
\[\d I(j,k):=I(\d\Gamma(j_0,k_0)\otimes\Gamma(j_\infty)
+\Gamma(j_0)\otimes\d\Gamma(j_\infty,k_\infty)).\]

Equivalently, treating $j$ and $k$ as maps from $\ch\oplus\ch$ to $\ch$
as in (\ref{mapp}),
we can write
\[
\begin{array}{l}
\d I(j,k):=\d \Gamma(j,k)U.
\end{array}
\]
We refer to \cite[Subsects. 3.10, 3.11]{DG} for details.

\medskip

{\bf Various bounds.}

\medskip

\begin{proposition}\label{stup1}
{\it i)} let $a, b$ two selfadjoint operators on $\ch$ with $b\geq 0$
and $a^{2}\leq b^{2}$. Then
\[
\d\G(a)^{2}\leq \d \G(b)^{2}.
\]
{\it ii)} let $b\geq 0$, $1\leq \alpha$. Then:
\[
\d\G(b)^{\alpha}\leq N^{\alpha-1}\d\G(b^{\alpha}).
\]
{\it iii)} let $0\leq r$ and $0\leq q\leq 1$. Then:
\[
\d\G(q,r)\leq \d\G(r).
\]
{\it iv)} Let $r, r_{1}, r_{2}\in B(\ch)$  and $\|q\|\leq 1$. Then:
\[
|(u_2|\d\Gamma(q,r_2r_1)u_1)|
\leq \|\d\G(r_2r_2^*)^\12u_2\|\|\d\G(r_1^*r_1)^\12 u_1\|,
\]
\[
\|N^{-\12}\d\Gamma(q,r)u\|\leq\|\d\Gamma(r^*r)^{\12}u\|.
\]
{\it v)} Let $j_{0}j_{0}^{*}+ j_{\infty}j_{\infty}^{*}\leq 1$,
$k_{0}$, $k_{\infty}$ selfadjoint. Then:
\[
\begin{array}{rl}
|(u_{2}|\d I^*(j,k)u_{1})|&\leq \|\d\G(|k_{0}|)^{\12} \otimes
\one u_{2}\|\|\d \Gamma(|k_{0}|)^{\12} u_{1}\|\\[3mm]
&+\|\one\otimes\d\G(|k_{\infty}|)^{\12}
 u_{2}\|\|\d \Gamma(|k_{\infty}|)^{\12} u_{1}\|, \: u_{1}\in \G(\ch),
\: u_{2}\in \G(\ch)\otimes \G(\ch).
\end{array}
\]
\[
\begin{array}{l}
\|(N_0+N_\infty)^{-\12}\d I^*(j,k)u\|\leq
\|\d\Gamma(k_0k_0^*+k_\infty k_\infty^*)^{\12}u\|, \: u\in \G(\ch).
\end{array}
\]
\end{proposition}
\proof {\it i)} is proved in  \cite[Prop. 3.4]{GGM}. The other
statements can be found in \cite[Sect. 3]{DG}.

\subsection{Heisenberg derivatives}\label{sec1.3bis}
Let $H$ be a selfadjoint operator on $\G(\ch)$ such that $H=
\d\G(\omega)+V$
on $\cD(H^{m})$ for some $m\in \nn$ where $\omega$ is selfadjoint and $V$
symmetric.
We will use the following notations for various Heisenberg
derivatives:
\[
\begin{array}{l}
\dd_{0}= \frac{\p}{\p t} + [\omega, \i\cdot\: ]\hbox{ acting on }B(\ch), \\[3mm]
\DD_{0}= \frac{\p}{\p t} + [H_{0}, \i\cdot \:],\ \ \DD= \frac{\p}{\p t}
+ [H, \i\cdot\:], \hbox{ acting on } B(\G(\ch)),
\end{array}
\]
where the commutators on the right hand sides are quadratic forms.

If  $\rr\ni t\mapsto M(t)\in B(\cD(H), \cH)$ is of class $C^{1}$
then: \beq\label{e6.0}
\DD\chi(H)M(t)\chi(H)=\chi(H)\DD_{0}M(t)\chi(H)+ \chi(H)[V, \i
M(t)]\chi(H), \eeq for $\chi\in \coinf(\rr)$.

If $\rr\ni m(t)\in B(\ch)$ is of class $C^{1}$ and $H_{0}=
\d\G(\omega)$ then:
\[
\DD_{0}\d\G(m(t))= \d\G(\dd_{0}m(t)).
\]
\subsection{Wick polynomials}\label{sec1.2}
In this subsection we recall some results from \cite[Subsect.
3.12]{DG}.

 We set
\[
B_{\rm fin}(\Gamma(\ch)):=\{B\in B(\Gh)\: | \hbox{ for some }n\in \nn\
\ \one_{[0, n]}(N)B\one_{[0, n]}(N)=B\}.
\]
Let $w\in B(\otimes_\s^p\ch,\otimes_\s^q\ch)$. We define the operator
\[
\Wick(w):\Gamma_\fin(\ch)\to \Gamma_\fin(\ch)
\]
as follows:
\beq
\Wick(w)\Big|_{\bigotimes_\s^n\ch}:
=\frac{\sqrt{n!(n+q-p)!}}{(n-p)!}w\otimes_\s\one^{\otimes(n-p)}.
\label{sec.wick.e1}\eeq
The operator $\Wick(w)$ is called a {\em Wick monomial of order} $(p,
q)$.
This definition extends to $w\in B_\fin(\Gamma(\ch))$ by linearity. 
The operator $\Wick (w)$ is called a {\em Wick polynomial} and
the operator $w$ is called the {\em symbol} of the Wick polynomial
$\Wick (w)$. If $w=\sum_{(p,q)\in I}w_{p,q}$ for $w_{p,q}$ of order
$(p,q)$ and $I\subset\nn$ finite, then
\[
{\rm deg}(w):=\sup_{(p,q)\in I}p+q
\]
is called the {\em degree} of $\Wick(w)$. If $h_{1}, \dots, h_{p},
g_{1},\dots, g_{q}\in \ch$ then:
\[
\Wick\left(
|\ g_1\otimes_\s\cdots\otimes_\s g_q)
(h_p\otimes_\s\cdots\otimes_\s h_1)|\right)
=a^*(q_1)\cdots a^*(g_q)
a( h_p)\cdots a( h_1).
\]

We recall some basic properties of Wick polynomials.
\begin{lemma}\label{1.1}
\[
i)\ \ \Wick(w)^{*}= \Wick(w^{*})\hbox{ as a identity on }\Gamma_{\fin}(\ch).
\]
{\it ii)} If $\slim w_s=w$, for $w_{s}, w$ of order $(p,q)$ then for
$k+m\geq (p+q)/2$:
\[
\slim_{s}(N+1)^{-k}\Wick(w_s)(N+1)^{-m}
=(N+1)^{-k}\Wick(w)(N+1)^{-m}.
\]
\[
{\it iii)}\ \ \|(N+1)^{-k}\Wick(w)(N+1)^{-m}\|\leq C\|w\|_{B(\Gh)},
\]
uniformly for $w$ of degree less than $p$ and $k+m\geq p/2$.
\end{lemma}
Most of the time the symbols of Wick polynomials will be {\em
Hilbert-Schmidt} operators. Let us introduce some more notation in
this context:
we set
\[
B_{\rm fin}^{2}(\Gh):= B^{2}(\Gh)\cap B_{\rm fin}(\Gh),
\]
where $B^{2}(\cH)$ is the set of Hilbert-Schmidt operators on the
Hilbert space $\cH$. Recall that  extending the map:
\[
B^{2}(\cH)\ni |u)(v|\mapsto u\otimes \overline{v}\in
\cH\otimes\overline{\cH}
\]
by linearity and density allows to unitarily identify $B^{2}(\cH)$  with
$\cH\otimes\overline{\cH}$, where $\overline{\cH}$ is the 
Hilbert space conjugate to $\cH$.
Using this identification, $B^{2}_{\rm fin}(\Gh)$ is identified with
$\Gamma_{\fin}(\ch)\otimes \Gamma_{\fin}(\overline{\ch})$ or
equivalently to $\Gamma_{\fin}(\ch\oplus\overline{\ch})$. We will
often use
this identification in the sequel.

If $u\in\otimes_\s^m\ch$, $v\in\otimes _s^n\ch$, $
w\in B(\otimes_\s^p\ch,\otimes_\s^q\ch)$ with $m\leq p$, $n\leq q$,
then one defines the contracted symbols:
\[
\begin{array}{l}
(v|w:= \left((v|\otimes_\s\one^{\otimes(q-n)}\right)w\:
\in B(\otimes_\s^{p}\ch,\otimes_\s^{q-n}\ch),\\[3mm]
w|u):= w\left(|u)\otimes_s\one^{\otimes(p-m)}\right)
\:\in B(\otimes_\s^{p-m}\ch,\otimes_\s^{q}\ch),\\[3mm]
(v|w|u):
=\left((v|\otimes_\s\one^{\otimes(q-n)}\right)w
\left(|u)\otimes_s\one^{\otimes(p-m)}\right)\:\in B(\otimes_\s^{p-m}\ch,
\otimes_\s^{q-n}\ch).
\end{array}
\]
If $a$ is selfadjoint on $\ch$ and $w\in B^{2}_{\fin}(\Gh)$, we set
\[
\triple\d\G(a)w\triple= \sum_{1\leq
i<\infty}\|(a)_{i}\otimes \one_{\G(\overline{\ch})}w\|_{B^{2}_{\fin}(\Gh)} +\sum_{1\leq
i<\infty}\|\one_{\G(\ch)}\otimes(\overline{a})_{i}w\|_{B^{2}_{\fin}(\Gh)},
\]
where the sums are finite since $w\in B^{2}_{\fin}(\Gh)\simeq
\G_{\fin}(\ch)\otimes\G_{\fin}(\overline{\ch})$ and one uses the
convention $\|au\|=+\infty$ if $u\not\in \cD(a)$.

We collect now some bounds on various commutators with Wick
polynomials.

\begin{proposition}\label{wick-recap-bis}
{\it i)} Let $b$ a selfadjoint operator on $\ch$ and $w\in
B_{\fin}(\Gh)$. Then:
\[
[\d \G(b), \Wick(w)]= \Wick([\d \G(b), w]),
\]
as quadratic form on $\cD(\d\G(b))\cap \cD(N^{{\rm deg}(w)/2})$.

{\it ii)} Let $q$ a unitary operator on $\ch$ and $w\in B_{\rm
fin}(\Gh)$. Then
\[
\G(q)\Wick(w)\G(q)^{-1}= \Wick(\G(q)w\G(q)^{-1}).
\]
{\it iii)} Let $w\in B_{\rm fin}(\Gh)$ of order $(p,q)$ and $h\in
\ch$. Then:
\beq
[\Wick(w),a^*(h)]=p\Wick\big(w|h)\big), \: \ \ \
[\Wick(w),a(h)]=q\Wick\big((h|w\big),\label{sec.wick.e8}
\eeq
\beq W(h)\Wick(w)W(-h)
=\sum_{s=0}^{p}\sum_{r=0}^{q}\frac{p!}{s!}\frac{q!}{r!}
(\frac{i}{\sqrt{2}})^{p+q-r-s}\Wick(w_{s,r}),
\label{sec.wick.e9}
\eeq
where
\beq
w_{s,r}= (h^{\otimes(q-r)}|w|h^{\otimes(p-s)}).
\label{sec.wick.e00}
\eeq
\end{proposition}

\begin{proposition}
\label{wick-recap}
{\it i)} Let $q\in B(\ch), \: \|q\|\leq 1$ and $w\in B^{2}_{\rm
fin}(\ch)$. Then for $m+k\geq {\rm deg}(w)/2$:
\beq
\begin{array}{rl}
&\|(N+1)^{-m}[\G(q), \Wick(w)](N+1)^{-k}\|\\[3mm]
\leq &C\triple\d\G(\one -q)w\triple.
\end{array}
\label{sec.wick.e10}
\eeq
{\it ii)} Let $j=(j_{0}, j_{\infty})$ with $j_{0},j_{\infty}\in
B(\ch)$, $\|j_{0}^{*}j_{0}+ j_{\infty}^{*}j_{\infty}\|\leq 1$. Then
for $m+k\geq {\rm deg}(w)/2$:
\beq
\begin{array}{rl}
&\|(N_{0}+N_{\infty}+1)^{-m}\Big(I^{*}(j)\Wick(w)-(\Wick(w)\otimes\one)
I^{*}(j) \Big)(N+1)^{-k}\|\\[3mm]
\leq &C \triple\d\G(\one -j_{0})w\triple+ C\triple \d\G(j_{\infty})w\triple.
\end{array}
\label{sec.wick.e11}
\eeq
\end{proposition}

\section{Abstract QFT Hamiltonians}\label{sec3}\init
In this section we define the class of abstract QFT Hamiltonians that
we will consider in this paper.
\subsection{Hamiltonians}\label{sec3.1}
Let $\omega$ be a selfadjoint operator on $\ch$ and $w\in
B^{2}_{\fin}(\Gh)$ such that $w=w^{*}$. We set
\[
H_{0}:=\d\Gamma(\omega),\ \ V:=\Wick(w).
\]
Clearly $H_{0}$ is selfadjoint   and $V$ symmetric
on $\cD(N^{n})$ for $n\geq {\rm deg}(w)/2$ by Lemma \ref{1.1}.

We assume:
\[
(H1)\ \ {\rm inf}\sigma(\omega)= m>0,
\]
\[
(H2)\ \ H_{0}+ V\hbox{ is essentially selfadjoint and bounded below on
}\cD(H_{0})\cap \cD(V).
\]
We set
\[
H:=\overline{H_{0}+ V}.
\]
 In the sequel we fix
$b>0$ such that $H+b\geq 1$. We assume:
\[
(H3)\ \ \begin{array}{l}\forall n\in \nn, \exists\: p\in \nn\hbox{ such that
}\|N^{n}H_{0}(H+b)^{-p}\|<\infty,\\[3mm]
\forall P\in \nn, \ \ \exists\: P<M\in \nn\hbox{ such that
}\|N^{M}(H+b)^{-1}(N+1)^{-P}\|<\infty.
\end{array}
\]
The bounds in (H3) are often called {\em higher order estimates}.
\begin{definition}
A Hamiltonian $H$ on $\G(\ch)$ satisfying {\it (Hi)} for $1\leq i\leq
3$ will be
called an {\em abstract QFT Hamiltonian}.
\end{definition}

\subsection{Hypotheses on the one-particle Hamiltonian}
The study of the spectral and scattering theory of abstract QFT
Hamiltonians relies heavily on corresponding statements for the
one-particle Hamiltonian $\omega$. The now standard approach to such
results is through the proof of a Mourre estimate and suitable
propagation estimates on the unitary group $\e^{-\i t\omega}$.

Many of these results can be formulated in a completely abstract way.
A convenient setup is based on the introduction of only three
selfadjoint operators on the one-particle space $\ch$, the Hamiltonian
$\omega$, a conjugate operator $a$ for $\omega$ and a {\em weight
operator} $\x$. In this subsection we describe the necessary abstract
hypotheses and collect various technical results used in the sequel.
We will use the abstract operator classes introduced in Subsect.
\ref{sec1.5}.

\medskip

{\bf Commutator estimates.}

\medskip

We assume that there exists a selfadjoint operator  $\x\geq 1$ for $\omega$ such
that:

\medskip

{\it (G1 i)} {\it there exists a subspace} $\cS\subset \ch$ {\it
such that} $\cS$ {\it is a core for} $\omega$, $\omega^{2}$ {\it
and the operators }  $\omega$, $\x$ {\it for} $z\in \cc\backslash
\sigma(\x)$, $(\x-z)^{-1}$, $F(\x)$ for $F\in \coinf(\rr)$ {\it
preserve }$\cS$.

\medskip

{\it (G1 ii)} $[\x, \omega]$ {\it belongs to} $S^{0}_{(3)}$.
\medskip

\begin{definition}
An operator $\x$ satisfying {\it (G1)} will be called a {\em
weight operator} for $\omega$.
\end{definition}

\medskip

{\bf Dynamical estimates.}

\medskip

Particles living at time $t$ in $\x\geq
ct$ for some $c>0$ are interpreted as  {\em  free particles}.  The
following assumption  says  that states in $\ch_{\rm
c}(\omega)$ describe free particles:

\medskip

{\it (S)} {\it there exists a  subspace }$\ch_{0}\:${\it dense in }$\ch_{\rm
c}(\omega)$ {\it such that for all }$h\in \ch_{0}\:${\it   there exists }$\epsilon>0$
{\it such that}
\[
\|\one_{[0, \epsilon]}(\frac{\x}{|t|})\e^{-\i t \omega}h\|\in
O(t^{-\mu}), \qquad \mu>1.
\]
(We recall that $\ch_{\rm c}(\omega)$ is the continuous spectral subspace
for $\omega$).

Note that {\it (S)} can  be deduced from {\it (G1)}, {\it (M1)}
and {\it (G4)}, assuming that $\omega\in C^{3}(a)$. The standard
way to see this is  to prove first a {\em strong propagation
estimate} (see e.g. \cite{HSS}):
\[
F(\frac{|a|}{|t|}\leq \epsilon)\chi(\omega)\e^{-\i t \omega}(a+
i)^{-2}\in O(t^{-2}),
\]
in norm if $\chi\in \coinf(\rr)$ is supported away from
$\kappa_{a}(\omega)$, and then to obtain a corresponding estimate
with $a$  replaced by $\x$ using {\it (G4)} and arguments similar to
those in \cite[Lemma A.3]{GN}.

The operators $[\omega, \i \x]$ and $[\omega, \i [\omega, \i \x]]$
are respectively the instantaneous {\em velocity} and {\em
acceleration} for the weight $\x$. The following condition
 means  roughly that the acceleration is positive:

\medskip

{\it (G2)} there exists  $0<\epsilon<\12$ such that
\[
[\omega, \i [\omega, \i \x]]= \gamma^{2}+ r_{-1-\epsilon},
\]
{\it where} $\gamma=\gamma^{*}\in S^{-\12}_{\epsilon, (2)}$ and
$r_{-1-\epsilon}\in S^{-1-\epsilon}_{(0)}$.
\medskip

\medskip

{\bf Mourre theory and local compactness.}

\medskip

We now state hypotheses about the conjugate operator $a$:

\medskip

{\it (M1 i)} $\omega\in C^{1}(a)$,  $[\omega, \i a]_{0}\in
B(\ch)$.

{\it (M1 ii)} $\rho^{a}_{\omega}\geq 0$,  $\tau^{a}(\omega)$ {\it is a
closed countable set.}

\medskip

We will also need the following condition which allows to localize
the operator $[\omega, \i a]_{0}$ using the weight operator $\x$.

\medskip

{\it (G3)} $a$ {\it preserves }$\cS$ and  $[\x, [\omega, \i a]_{0}]$
belongs to $S^{0}_{(0)}$.

\medskip

Note that  if $a$ preserves $\cS$ then $[\omega,  a]_{0}=  \omega a- a \omega$
on $\cS$. Therefore $[\x, [\omega,  a]_{0}]$ in {\it (G3)} is well defined as an
operator on $\cS$.

We will also need some conditions which roughly say that $a$
is controlled by $\x$. This allows to translate propagation estimates
for  $a$ into propagation estimates for $\x$.

\medskip

{\it (G4)} $a$  {\it belongs to }$S^{1}_{(0)}$.

\medskip

Note that by Lemma \ref{gototime} {\it i)}, $a^{2}\in S^{2}_{(0)}$ hence $a\x^{-1}$
and $a^{2}\x^{-2}$ are bounded.

We state also an hypothesis on local compactness:

\medskip

{\it (G5)}  $\x^{-\epsilon}(\omega+ 1)^{-\epsilon}$ {\it is
compact on }
$\ch$ {\it  for some }$0<\epsilon\leq \12$.

\medskip

{\bf Comparison operator.}

\medskip

To get a sharp Mourre estimate for abstract QFT Hamiltonians, it is
convenient to assume the existence of a {\em comparison operator}
$\omega_{\infty}$ such that:

{\it (C i)} $C^{-1}\omega_{\infty}^{2}\leq \omega^{2}\leq C
\omega_{\infty}^{2}$, {\it for some }$C>0$,

\medskip

{\it (C ii)} $\omega_{\infty}$ satisfies {\it (G1)}, {\it (M1)}, {\it
(G3)} for the same $\x$ and $a$ and
$\kappa^{a}_{\omega_{\infty}}\subset \tau^{a}_{\omega_{\infty}}$.

\medskip

Note that the last condition in {\it (C ii)} is satisfied if
$\omega_{\infty}$ has no eigenvalues.

\medskip

{\it (C iii)}
$\omega^{-\12}(\omega-\omega_{\infty})\omega^{-\12}\x^{\epsilon}$
and $[\omega-\omega_{\infty}, \i a]_{0}\x^{\epsilon}$ are  bounded
for some $\epsilon>0$.

\medskip

\medskip

{\bf Some consequences.}

\medskip

We now state some standard consequences of {\it (G1)}.
\begin{lemma}\label{gauge}
Assume {\it (H1)}, {\it (G1)}. Then for $F\in \coinf(\rr)$:
\[
i) \ \ [F(\frac{\x}{R}), \ad^{k}_{\x}\omega]=
R^{-1}F'(\frac{\x}{R})[\x, \ad^{k}_{\x}\omega]+M(R), \ \ k=0, 1,
\]
where $M(R)\in O(R^{-2})S^{0}_{(0)}\cap O(R^{-1})S^{-1}_{(0)}$.
\[
\begin{array}{l}
ii)\ \ F(\frac{\x}{R}): \cD(\omega)\to \cD(\omega)\hbox{ and
}\omega
F(\frac{\x}{R})\omega^{-1}\in O(1), \\[3mm]
iii) \ \ [F(\frac{\x}{R}), [\omega, \x]]\in O(R^{-1}), \\[3mm]
iv)\ \ F(\frac{\x}{R})[\omega, \i \x](1-F_{1})(\frac{\x}{R})\in
O(R^{-2}), \\[3mm]
\end{array}
\]
if $F_{1}\in \coinf(\rr)$ and $FF_{1}= F$.

Assume {\it (H1)}, {\it (M1 i)}, {\it (G3)}. Then
for $F\in \coinf(\rr)$:
\[
v) \ \ [F(\frac{\x}{R}), [\omega, \i a]_{0}]\in O(R^{-1})
\]
Assume {\it (H1)}, {\it (G1)}, {\it (G2)}. Then
for $F\in \coinf(\rr)$:
\[
vi) \ \  F(\frac{\x}{R}): \cD(\omega^{2})\mapsto
\cD(\omega^{2})\hbox{ and }[\omega^{2},
F(\frac{\x}{R})]\omega^{-1}\in O(R^{-1}).
\]
Let $b\in S^{-\mu}_{\delta, (1)}$ for $\mu\geq 0$ and $F\in \coinf(\rr\backslash \{0\})$.
Then:
\[
vii)\ \ [F(\frac{\x}{R}), b]\in O(R^{-\mu-1+ \delta}).
\]
 In {\it i)} for $k=0$  the commutator
on the l.h.s. is considered as a quadratic form on $\cD(\omega)$.
\end{lemma}

\begin{lemma}\label{compare}
Let $\omega_{\infty}$ be a comparison operator satisfying {\it (C)}.
Then for  $F\in \cinf(\rr)$ with $F\equiv 0$ near $0$, $F\equiv 1$
near $+\infty$ we have:
\[ \omega^{-\12}
(\omega-\omega_{\infty})F(\xr)\omega^{-\12}, \ \
[\omega-\omega_{\infty}, \i a]F(\xr)\in o(R^{0}).
\]
\end{lemma}

The proof of Lemmas \ref{gauge}, \ref{compare}  will be given in
the Appendix.

\subsection{Hypotheses on the interaction}

We now formulate the hypotheses on the interaction $V$. If $j\in \cinf(\rr)$, we set for $R\geq 1$ $j^{R}=j(\frac{\x}{R})$.

For the scattering theory of abstract QFT Hamiltonians, we will need  the following decay hypothesis on the
symbol of $V$:
\[
(Is)\ \ \triple\d\G(j^{R})w\triple\in O(R^{-s}),\ \ s>0 \ \ \hbox{
if }j\equiv 0\hbox{ near }0, \ \ j\equiv 1\hbox{ near }\pm\infty.
\]
Note that if  $w\in B^{2}_{\fin}(\Gh)$ and $j$ is as above then
\beq\label{stupidix} \triple\d\G(j^{R})w\triple\in o(R^{0}),
\hbox{ when }R\to \infty. 
\eeq
 Another type of hypothesis
concerns the Mourre theory. We fix a conjugate operator $a$ for
$\omega$ such that {\it (M1)} holds and set
\[
A:=\d\G(a).
\]
For the Mourre theory, we will impose:
\[
(M2)\ \ w\in \cD(A\otimes\one-\one\otimes\overline{A}).
\]
If hypothesis {\it (G4)} holds then $a\x^{-1}$ is bounded. It follows
that the condition
\[
(D)\ \ \triple\d\G(\x^{s})w\triple <\infty, \hbox{ for some }s>1
\]
implies both {\it (Is)}  for $s>1$  and {\it (M2)}.

\section{Results}\init\label{sec1bis}
For the reader convenience, we summarize in this section the
results of the paper. To simplify the situation we will assume
that all the various hypotheses hold, i.e. we assume conditions
{\it (Hi)}, $1\leq i\leq 3$, {\it (Gi)}, $1\leq i\leq  5$, {\it
(S)}, {\it (M1)}, {\it (C)} and {\it (D)}. However various parts
of Thm.  \ref{mainmain} hold under  smaller sets of hypotheses, we
refer the reader to  later sections for precise statements.

The notation $\d\G^{(1)}(E)$ for a set $E\subset
\rr$ is defined in Subsect. \ref{sec4.3}.

\begin{theoreme}\label{mainmain}
Let $H$ be an abstract QFT Hamiltonian. Then:
\ben
\item if $\sigma_{\rm ess}(\omega)=[m_{\infty}, +\infty[$ then
\[
\sigma_{\rm ess}(H)= [\inf \sigma(H)+ m_{\infty}, +\infty[.
\]
\item The Mourre estimate holds for $A=\d\G(a)$ on $\rr\backslash
\tau$, where
\[
\tau= \sigma_{\rm pp}(H)+ \d\G^{(1)}(\tau_{a}(\omega)),
\]
where $\tau_{a}(\omega)$ is the set of thresholds of $\omega$ for
$a$ and $\d\G^{(1)}(E)$ for $E\subset \rr$ is defined in (\ref{e2.19}).
\item The {\em asymptotic Weyl operators}:
\[
W^{\pm}(h):=\slim_{t\pm\infty}\e^{\i tH }W(\e^{-\i t \omega}h)\e^{-\i
tH}\hbox{ exist for all }h\in \ch_{\rm c}(\omega),
\]
and define two regular CCR representations over $\ch_{\rm c}(\omega)$.

\item There exist unitary operators $\Omega^{\pm}$, called the {\em
wave operators}:
\[
\Omega^{\pm}: \cH_{\rm pp}(H)\otimes \G(\ch_{\rm c}(\omega))\to \G(\ch)
\]
such that
\[
\begin{array}{l}
W^{\pm}(h)=\Omega^{\pm} \one\otimes W(h)\Omega^{\pm*}, \ \ h\in
\ch_{\rm c}(\omega),\\[3mm]
H= \Omega^{\pm}(H_{|\cH_{\rm pp}(H)}\otimes \one + \one\otimes
\d\G(\omega))\Omega^{\pm*}.
\end{array}
\]
\een
\end{theoreme}
Parts (1), (2), (3), (4) are proved respectively in Thms. \ref{B29},
\ref{mourrebis}, \ref{4.1bis}  and \ref{asympt-comp}.

Statement (1) is the familiar HVZ theorem, describing the essential
spectrum of $H$.

Statement (2) is the well-known Mourre estimate. Under additional
conditions, it is possible to deduce from it resolvent estimates which
imply  in particular that the singular continuous spectrum of $H$ is
empty. In our case this result  follows from (4),
provided we know that $\omega$ has no singular continuous spectrum.

Statement (3) is rather easy. Statement (4) is the most important
result of this paper, namely the {\em asymptotic completeness} of wave
operators.

\begin{remark}
Assume that there exist another operator $\omega_{\infty}$ on $\ch$
such that $\omega_{|\ch_{\rm c}(\omega)}$ is unitarily equivalent to
$\omega_{\infty}$. Typically this follows  from the construction
of a nice scattering theory for the pair $(\omega, \omega_{\infty})$.
Then since $\d\G(\omega)$ restricted to $\G(\ch_{\rm c}(\omega))$ is
unitarily equivalent to $\d\G(\omega_{\infty})$, we can replace
$\omega$ by $\omega_{\infty}$ in statement (4) of Thm. \ref{mainmain}.
\end{remark}

\section{Examples}\init\label{sec2}
In this section we give examples of QFT Hamiltonians to which we can
apply Thm. \ref{mainmain}. Our two examples are space-cutoff
$P(\varphi)_{2}$ Hamiltonians for a variable metric, and similar $P(\varphi)_{d+1}$ models for $d\geq 2$ if
the interaction term has also an ultraviolet cutoff. For $\mu\in \rr$
we denote by $S^{\mu}(\rr^{d})$ the space of $C^{\infty}$ functions on
$\rr^{d}$ such that:
\[
\p^{\alpha}_{x}f(x)\in O(\langle x\rangle^{-\mu-\alpha}) \ \
\alpha\in \nn^{d},\hbox{ where }\x= (1+ x^{2})^{\12}.
\]

\subsection{Space-cutoff $P(\varphi)_{2}$ models with variable
metric}\label{sec2.3}

We fix a second order differential operator on $\ch=L^{2}(\rr)$:
\[
h:= Da(x)D + c(x),\ \ D=-\i \p_{x},
\]
where $a(x)\geq c_{0}$, $c(x)\geq c_{0}$ for some $c_{0}>0$ and
$a(x)-1, c(x)-m_{\infty}^{2}\in S^{-\mu}(\rr)$ for some $m_{\infty},
\mu>0$.
We set:
\[
\omega:= h^{\12}
\]
and consider the free Hamiltonian
\[
H_{0}= \d\G(\omega), \hbox{ acting on }\G(\ch).
\]
To define the interaction, we fix a real polynomial with $x-$dependent
coefficients:
\beq
\label{defdeP}
P(x,\lambda)=\sum_{p=0}^{2n}a_{p}(x)\lambda^{p}, \ \ a_{2n}(x)\equiv
a_{2n}>0,
\eeq
and a function $g\in L^{1}(\rr)$ with $g\geq 0$. For $x\in \rr$, one
sets
\[
\varphi(x):= \phi(\omega^{-\12}\delta_{x}),
\]
where $\delta_{x}$ is the Dirac distribution at $x$. The associated
$P(\varphi)_{2}$ interaction is formally defined as:
\[
V:=\int_{\rr}g(x):\!P(x, \varphi(x))\!:\d x,
\]
where $:\ \ :$ denotes the Wick ordering.

In \cite {GP} we  prove the following theorem. Condition {\it (B3)}
below is formulated in terms of a (generalized) basis of
eigenfunctions of $h$. To be precise we say that the families
$\{\psi_{l}(x)\}_{l\in I}$ and $\{\psi(x, k)\}_{k\in \rr}$ form a
generalized basis of eigenfunctions of $h$ if:
\[
\begin{array}{l}
\psi_{l}(\cdot)\in L^{2}(\rr), \ \ \psi(\cdot, k)\in \cS'(\rr), \\[3mm]
h \psi_{l}= \epsilon_{l}\psi_{l}, \ \ \epsilon_{l}\leq m^{2}_{\infty},
 \ \ l\in I,
\\[3mm]
h \psi(\cdot, k)= (k^{2}+ m_{\infty}^{2})\psi(\cdot, k), \ \ k\in \rr,
\\[3mm]
\sum_{l\in I}|\psi_{l})(\psi_{l}|+
\frac{1}{2\pi}\int_{\rr}|\psi(\cdot, k))(\psi(\cdot, k)|\d k=\one.
\end{array}
\]
\begin{theoreme}\label{exemple2}
Assume that:
\[
(B1)\: ga_{p}\in L^{2}(\rr), \: 0\leq p\leq 2n, \ \ g\in L^{1}(\rr), \
\ g\geq 0, \ \   g(a_{p})^{2n/(2n-p)}\in L^{1}(\rr), \:0\leq p\leq 2n-1,
\]
\[
(B2)\:\x^{s}ga_{p}\in L^{2}(\rr) \ \ \forall \: 0\leq p\leq 2n, \hbox{
for some }s>1.
\]
Assume moreover that for   a measurable function
$M:\rr\to \rr^{+}$ with   $M(x)\geq 1$
there exists a generalized basis of eigenfunctions of $h$
such that:
\[
(B3)\:\left\{
\begin{array}{l}
\sum_{l\in I}\|M^{-1}(\cdot)\psi_{l}(\cdot)\|_{\infty}^{2}<\infty, \\[3mm]
\|M^{-1}(\cdot)\psi(\cdot, k)\|_{\infty}\leq C, \ \ k\in \rr.
\end{array}
\right.
\]
\[
(B4)\: ga_{p}M^{s}\in L^{2}(\rr), \ \  g (a_{p}M^{s})^{2n/(2n-p+s)}\in
L^{1}(\rr), \ \ \forall \: 0\leq s\leq p\leq 2n-1.
\]
Then the Hamiltonian
\[
H= \d\G(\omega)+ \int_{\rr}g(x):\!P(x, \varphi(x))\!:\d x
\]
satisfies all the hypotheses of Thm. \ref{mainmain} for the weight
operator $\x= (1+ x^{2})^{\12}$ and conjugate operator $a=\12(
x\langle
D_{x}\rangle^{-1}D_{x}+ {\rm hc})$.
\end{theoreme}
\begin{remark}\label{turlututu}
If $g$ is compactly supported we can take $M(x)=+\infty$ outside
$\supp g$, and the meaning of {\it (B3)} is that the sup norms $\|\:
\|_{\infty}$ are taken only on $\supp g$.
\end{remark}
\begin{remark}
Condition (B3) is discussed in details in \cite{GP}, where many
sufficient conditions for its validity are given. As an example let us
simply mention that if  $a(x)-1$, $c(x)- m_{\infty}^{2}$ and the
coefficients $a_{p}$ are  in the Schwartz class $\cS(\rr)$, then
all conditions in Thm. \ref{exemple2} are satisfied.
\end{remark}

\subsection{Higher dimensional examples}\label{sec2.4}
We work now on $L^{2}(\rr^{d})$ for $d\geq 2$ and consider
\[
\omega= (\sum_{1\leq i,j\leq d}D_{i}a_{ij}(x)D_{j}+ c(x))^{\12}
\]
where $a_{ij}, c$ are real, $[a_{ij}](x)\geq c_{0}\one$, $c(x)\geq
c_{0}$ for some $c_{0}>0$ and $[a_{ij}]-\one \in S^{-\mu}(\rr^{d})$,
$c(x)-m_{\infty}^{2}\in S^{-\mu}(\rr^{d})$ for some $m_{\infty}, \mu>0$. 

The free Hamiltonian is
as above
\[
H_{0}= \d\G(\omega),
\]
acting on the Fock space $\G(L^{2}(\rr^{d}))$.

Since  $d\geq 2$ it is necessary to add an
ultraviolet cutoff to make sense out of the formal expression
\[
\int_{\rr^{d}}g(x)P(x, \varphi(x))\d x.
\]
We set
\[
\varphi_{\kappa}(x):=
\phi(\omega^{-\12}\chi(\frac{\omega}{\kappa})\delta_{x}),
\]
where $\chi\in \coinf([-1, 1])$ is a cutoff function equal to $1$
on $[-\12, \12]$ and $\kappa\gg 1$ is an ultraviolet cutoff
parameter.  Since
$\omega^{-\12}\chi(\frac{\omega}{\kappa})\delta_{x}\in
L^{2}(\rr^{d})$, $\varphi_{\kappa}(x)$ is a well defined
selfadjoint operator on $\G(L^{2}(\rr^{d}))$.

If $P(x, \lambda)$
is as in (\ref{defdeP}) and $g\in L^{1}(\rr^{d})$, then
\[
V:=\int_{\rr^{d}}g(x)P(x, \varphi_{\kappa}(x))\d x,
\]
is a well defined selfadjoint operator on $\G(L^{2}(\rr^{d}))$.
We have then the following theorem. As before we consider a
generalized basis $\{\psi_{l}(x)\}_{l\in I}$ and $\{\psi(x, k)\}_{k\in
\rr^{d}}$ of eigenfunctions of $h$.
\begin{theoreme}\label{exemple3}
Assume that:
\[
(B1)\; ga_{p}\in L^{2}(\rr^{d}), \: 0\leq p\leq 2n, \ \ g\in
L^{1}(\rr^{d}), \ \ g\geq 0, \ \   g(a_{p})^{2n/(2n-p)}\in
L^{1}(\rr^{d}), \: 0\leq p\leq 2n-1,
\]
\[
(B2)\:\x^{s}ga_{p}\in L^{2}(\rr^{d}) \ \ \forall \: 0\leq p\leq 2n,
\hbox{ for some }s>1.
\]
Assume moreover that for   a measurable function
$M:\rr^{d}\to \rr^{+}$ with  $M(x)\geq 1$
there exists a generalized basis of eigenfunctions of $h$
such that:
\[
(B3)\:\left\{
\begin{array}{l}
\sum_{l\in I}\|M^{-1}(\cdot)\psi_{l}(\cdot)\|_{\infty}^{2}<\infty, \\[3mm]
\|M^{-1}(\cdot)\psi(\cdot, k)\|_{\infty}\leq C, \ \ k\in \rr.
\end{array}
\right.
\]
\[
(B4)\: ga_{p}M^{s}\in L^{2}(\rr^{d}), \ \  g (a_{p}M^{s})^{2n/(2n-p+s)}\in
L^{1}(\rr^{d}), \ \ \forall \: 0\leq s\leq p\leq 2n-1.
\]
Then the Hamiltonian
\[
H= \d\G(\omega)+ \int_{\rr^{d}}g(x)P(x, \varphi_{\kappa}(x))\d x
\]
satisfies all the hypotheses of Thm. \ref{mainmain} for the weight
operator $\x= (1+ x^{2})^{\12}$ and conjugate operator $a=\12(
x\cdot\langle
D_{x}\rangle^{-1}D_{x}+ {\rm hc})$.
\end{theoreme}
\begin{remark}
Sufficient conditions for {\it (B3)} to hold with $M(x)\equiv 1$ are
given in \cite{GP}.
\end{remark}

\section{Commutator estimates}\label{sec3b}\init
In this section we collect various commutator estimates, needed in
Sect. \ref{sec4}.
\subsection{Number energy estimates}\label{sec3.4}

We  recall first some  notation from \cite{DG}:
let an operator
$B(t)$  depending on some parameter $t$
 map $\cap_n\cD(N^n)\subset\cH$ into itself.
We will write
\beq
B(t)\in (N+1)^{m}O_N(t^{p})\hbox{ for }m\in \rr\hbox{ if }
\label{om1}\eeq
\[
\|(N+1)^{-m-k}B(t)(N+1)^{k}\|\leq C_k\t^{p},\ \ \ \ k\in\zz.
\]

If (\ref{om1}) holds for any $m\in \rr$, then we will write
\[B(t)\in (N+1)^{-\infty}O_N(t^{p}).\]

Likewise, for an operator $C(t)$ that maps  $\cap_n\cD(N^n)\subset\cH$
into  $\cap_n\cD((N_0+N_\infty)^n)\subset\cH^\ext$ we will write
\beq
C(t)\in (N+1)^{m}\check O_N(t^{p})\hbox{ for }m\in \rr\hbox{ if }
\label{om2}\eeq
\[
\|(N_0+N_{\infty})^{-m-k}C(t)(N+1)^{k}\|\leq
C_k\t^{p},\ \ \ \ k\in\zz.
\]
If (\ref{om2}) holds for any $m\in \rr$, then we will write
\[
B(t)\in (N+1)^{-\infty}\check O_N(t^{p}).
\]
The notation $(N+1)o_{N}(t^{p})$, $(N+1)^{m}\check{o}_{N}(t^{p})$ are
defined similarly.

\begin{lemma}\label{1.2}Let $H$ be an abstract QFT Hamiltonian. Then:

{\it i)} for all $P\in \nn$ there exists $\alpha>0$ such that for all $0\leq
s\leq P$
\[
N^{s+\alpha}(H-z)^{-1}N^{-s}\in O(|{\rm Im}z|^{-1}), \hbox{ uniformly
for }z\in \cc\backslash \rr\cap\{|z|\leq R\}.
\]
{\it ii)} for $\chi\in \coinf(\rr)$ we have
\[
\|N^{m}\chi(H)N^{p}\|<\infty, \ \ m,p\in \nn.
\]
\end{lemma}
\proof {\it ii)} follows directly from  {\it (H3)}. It
remains to prove {\it i)}. Let us fix $P\in \nn$ and $M>P$ such that

\beq\label{ito}
N^{M}(H+b)^{-1}(N+1)^{-P}\in B(\cH).
\eeq We deduce also from {\it (H3)}
and interpolation that there exists $\alpha>0$ such that
\beq\label{itoo}
N^{\alpha}(H+b)^{-1}\in B(\cH).
\eeq
 We can choose $\alpha>0$ small
enough such that $\delta=(M-\alpha)/P>1$. Interpolating between
(\ref{ito}) and (\ref{itoo}) we obtain first
that $N^{\alpha+ \delta x}(H+b)^{-1}(N+1)^{-x}$ is bounded for all
$x\in [0, P]$. Since $\delta>1$, we get that
\beq\label{e1.3}
\|N^{\alpha(s+1)}(H+b)^{-1}(N+1)^{-s\alpha}\|<\infty, \ \ s\in
[0, P\alpha^{-1}].
\eeq
 Without loss of generality we can assume that
$\alpha^{-1}\in \nn$, and we will prove by induction on $s\in\nn$ that
\beq\label{e1.2}
N^{(s+1)\alpha}(H-z)^{-1}(N+1)^{-s\alpha}\in O(|{\rm Im}z|^{-1}),
\eeq
uniformly for $z\in \cc\backslash \rr\cap\{|z|\leq R\}$ and $0\leq s\leq P\alpha^{-1}$.

For $s=0$ (\ref{e1.2}) follows from the fact that
$N^{\alpha}(H+b)^{-1}$ is bounded. Let us assume that (\ref{e1.2})
holds for $s-1$. Then we write:
\[
\begin{array}{rl}
&N^{(s+1)\alpha}(H-z)^{-1}(N+1)^{-s\alpha}\\[3mm]
=&N^{(s+1)\alpha}(H+b)^{-1}N^{-s\alpha}N^{s\alpha}(H+b)(H-z)^{-1}(N+1)^{-s\alpha}\\[3mm]
=&N^{(s+1)\alpha}(H+b)^{-1}N^{-s\alpha}N^{s\alpha}(\one +
(b+z)(H-z)^{-1})(N+1)^{-s\alpha},
\end{array}
\]
so (\ref{e1.2}) for $s$ follows from (\ref{e1.3}) and the induction
hypothesis. We extend then (\ref{e1.2}) from integer $s\in [0,
P\alpha^{-1}]$ to all $s\in [0, P\alpha^{-1}]$ by interpolation.
Denoting $s\alpha$ by $s$ we obtain {\it i)}. \qed

\medskip

\subsection{Commutator estimates}\label{sec3.5}
\begin{lemma}\label{commut-1}
Let $H$ be an abstract QFT Hamiltonian and $\x$ a weight operator
for $\omega$. Let $q\in \coinf(\rr)$, $0\leq q\leq 1$, $q\equiv 1$
near $0$. Set for $R\geq 1$ $q^{R}= q(\frac{\x}{R})$. Then for
$\chi\in \coinf(\rr)$:
\[
[\Gamma(q^{R}), \chi(H)]\in \left\{
\begin{array}{l}(N+1)^{-\infty}O_N(R^{-\inf(s,1)})
\hbox{ under hypothesis }(Is),\\
(N+1)^{-\infty}o_{N}(R^{0})\hbox{ otherwise}.
\end{array}\right.
\]
\end{lemma}
\proof In all the proof $M$ and $P$ will denote integers chosen
sufficiently large. We prove the lemma under hypothesis {\it (Is)}
$s>0$, the general case being handled replacing hypothesis {\it
(Is)} by  the estimate (\ref{stupidix}). Clearly $\G(q^{R})$
preserves $\cD(N^{n})$. We have \beq [H_{0},
\Gamma(q^R)]=\d\Gamma(q^{R}, [\omega, q^{R}]), \label{toto} \eeq
By Lemma \ref{gauge} {\it i)},  $[\omega,q^R]\in O(R^{-1})$  and
hence $[H_0,\Gamma(q^R)](H_0+1)^{-1}$ is bounded. Therefore,
 $\Gamma(q^R)$ preserves  $\cD(H_0)$. As in \cite[Lemma 7.11]{DG} the
following identity is valid as a operator identity on $\cD(H_{0})\cap \cD(N^{P})$:
\[
[H,\Gamma(q^R)]=[H_0,\Gamma(q^R)]+[V,\Gamma(q^R)]=:T.
\]
From (\ref{toto}) and Prop. \ref{stup1} {\it iv)} we get that
\[
[\Gamma(q^R), H_{0}]\in (N+1)O_N(R^{-1}).
\]
Using Prop. \ref{wick-recap} {\it i)} and hypothesis {\it (Is)}, we
get that
\[
[\Gamma(q^R), V]\in (N+1)^{n}O_N(R^{-s}),\ \ n\geq {\rm deg}(w)/2
\]
which gives
\beq
T\in (N+1)^{n}O(R^{-\inf(s,1)}).
\label{toto2}
\eeq
Let now \[
\begin{array}{rl}
T(z)&:=[\Gamma(q^R),(z-H)^{-1}]\\[3mm]
&=-(z-H)^{-1}[\G(q^{R}), H](z-H)^{-1}.
\end{array}
\]
By {\it (H3)} $\cD(H^{M})\subset \cD(H_{0})\cap
\cD(N^{P})$, so the following identity holds on $\cD(H^{M})$:
\[
T(z)= (z-H)^{-1}T(z-H)^{-1}.
\]
Let now $\chi_{1}\in \coinf(\rr)$ with  $\chi_{1}\chi=\chi$ and
$\tilde{\chi}_{1}$, $\tilde{\chi}$ be almost analytic extensions of
$\chi_{1}$, $\chi$. We write:
\[
\begin{array}{rl}
&N^{m}[\chi(H), \Gamma(q^R)]N^p\\[3mm]
=& N^{m}\chi_{1}(H)[\chi(H), \Gamma(q^R)]N^p+
N^{m}[\chi_{1}(H),  \Gamma(q^R)]\chi(H)N^p\\[3mm]
=&\frac{\i}{2\pi}\int_{\cc}\partial_{\,\overline z}\tilde{\chi}(z)
N^{m}\chi_{1}(H)T(z)N^p\d z\wedge \d\,\overline z\\[3mm]
&+ \frac{\i}{2\pi}\int_{\cc}\partial_{\,\overline z}\tilde{\chi_{1}}(z)
N^{m}T(z)\chi(H)N^p\d z\wedge \d\,\overline z.
\end{array}
\]
Using Lemma \ref{1.2} {\it i)} and (\ref{toto2}), we obtain that for
all $n_{1}\in \nn$ there exists $n_{2}\in \nn$ such that
\[
N^{n_{1}}T(z)(N+1)^{-n_{2}}, \ \ (N+1)^{-n_{2}}T(z)N^{n_{1}}\in O(|{\rm Im}z|^{-2}),
\hbox{ uniformly for } z\in \cc\backslash \rr\cap\{|z|\leq R\}.
\]
Using also Lemma \ref{1.2} {\it ii)}, we obtain that
\[
N^{m}[\chi(H), \Gamma(q^R)]N^p\in O(R^{-\inf(s,1)}),
\]
which completes the proof of the Lemma. \qed

\medskip

Let $j_0\in C_0^\infty(\rr)$,
$j_\infty\in C^\infty(\rr)$, $0\leq j_0$, $0\leq
j_\infty$, $j_0^2+j_\infty^2\leq 1$, $j_0=1$ near $0$
(and hence $j_\infty=0$
near $0$). Set  for $R\geq 1$
$j^R=(j_0(\frac{\x}{R}), j_{\infty}(\frac{\x}{R}))$.
\begin{lemma}
Let $H$ be an abstract QFT Hamiltonian and $\x$ a weight operator
for $\omega$. Then for $\chi\in \coinf(\rr)$:
\[
\chi(H^\ext)I^*(j^R)
-I^*(j^R)\chi(H)\in \left\{
\begin{array}{l}(N+1)^{-\infty}\check O(R^{-\inf(s,1)})\hbox{ under
hypothesis }(Is),\\
(N+1)^{-\infty}\check o(R^{0})\hbox{otherwise}.\end{array}\right.
\]
\label{Fw200}
\end{lemma}
\proof  Again we will only prove the lemma under hypothesis {\it (Is)}.
As in \cite[Lemma 7.12]{DG}, we have:
\[
H_{0}^{\ext}I^*(j^R) -I^*(j^R) H_{0}\in (N+1)O(\| [\omega,
j_{0}^{R}]\|+ \|[\omega, j_{\infty}^{R}] \|).
\]
Writing $[\omega, j_{\infty}^{R}]= [(1-j_{\infty})^{R},
\omega]$, we obtain that $\| [\omega,
j_{0}^{R}]\|+ \|[\omega, j_{\infty}^{R}] \|\in O(R^{-1})$, hence:
\beq
H_{0}^{\ext}I^*(j^R) -I^*(j^R) H_{0}\in (N+1)\cO_N(R^{-1}).
\label{ez.1}
\eeq
This implies that $I^*(j^R)$ sends $\cD(H_{0})$ into $\cD(H_{0}^{\ext})$,
and since $I^*(j^R) N= (N_{0}+ N_{\infty})I^*(j^R)$, $I^*(j^R)$  sends
also $\cD(N^{n})$ into $\cD((N_{0}+N_{\infty})^{n})$.

Next by Prop. \ref{wick-recap} {\it ii)} and condition {\it (Is)}
we have \beq (V\otimes \one) I^*(j^R) -I^*(j^R) V\in
(N+1)^{n}\cO_N(R^{-s}), \ \ n\geq{\rm deg}(w)/2. \label{ez.2} \eeq
This and (\ref{ez.1}) show that as an operator identity on
$\cD(H_{0})\cap \cD(N^{n})$ we have \beq H^{\ext}I^{*}(j^{R})-
I^{*}(j^{R})H \in (N+1)^{n}\cO_{N}(R^{-\min(1,s)}). \label{ez.3}
\eeq Using then  {\it (H3)} and the fact that $I^*(j^R)$ sends
$\cD(H_{0})$ into $\cD(H_{0}^{\rm ext})$ and $\cD(N^{n})$ into
$\cD((N_{0}+N_{\infty})^{n})$, we obtain the following operator
identity on $\cD(H^{M})$ for $M$ large enough:
\[
\begin{array}{rl}
T(z)&:=(z-H^{\ext})^{-1}I^*(j^R) -I^*(j^R) (z-H)^{-1}\\[3mm]
&=(z-H^{\ext})^{-1}\Big(I^*(j^R)  H-H^{\rm ext}I^*(j^R)\Big)
(z-H)^{-1},
\end{array}
\]
uniformly for $z\in \cc\backslash \rr\cap\{|z|\leq R\}$.

Using then  Lemma \ref{1.2} {\it i)} (and its obvious extension for
$H^{\ext}$), we obtain that for
all $n_{1}\in \nn$ there exists $n_{2}\in \nn$ such that
\beq
(N_0+N_\infty)^{n_{1}}T(z)(N+1)^{-n_{2}}, \ \ (N_{0}+
N_{\infty}+1)^{-n_{2}}T(z)N^{n_{1}}\in
O(|{\rm Im}z|^{-2})R^{-\inf(s,1)}.
\label{ez.4}
\eeq
Let us again pick  $\chi_{1}\in \coinf(\rr)$ with $\chi_{1}\chi=\chi$.
We have:
\[
\begin{array}{rl}
&(N_{0}+N_{\infty})^{m}\left(\chi(H^{\ext})I^*(j^R) -I^*(j^R)
\chi(H)\right)N^m\\[3mm]
= &(N_{0}+N_{\infty})^{m}
\chi_{1}(H^{\ext})\Big(\chi(H^{\ext})I^*(j^R) -I^*(j^R) \chi(H)\Big)N^m\\[3mm]
&+ (N_{0}+N_{\infty})^{m}
\Big(\chi_{1}(H^{\ext})I^*(j^R) -I^*(j^R)
\chi_{1}(H)\Big)\chi(H)N^m\\[3mm]
=& \frac{\i}{2\pi}\int_{\cc}\partial_{\,\overline z}\tilde{\chi}(z)
(N_{0}+N_{\infty})^{m}
\chi_{1}(H^{\ext})T(z)N^m\d z\wedge \d\,\overline z\\[3mm]
&+ \frac{\i}{2\pi}\int_{\cc}\partial_{\,\overline z}\tilde{\chi_{1}}(z)
(N_{0}+N_{\infty})^{m}T(z)\chi(H)N^m\d z\wedge \d\,\overline z.
\end{array}
\]
Using Lemma \ref{1.2} {\it i)},  (\ref{ez.4}), the above operator is $O(R^{-\inf(s,1)})$ as
claimed. \qed
\section{Spectral analysis of abstract QFT
Hamiltonians}\label{sec4}\init
In this section we study the spectral theory of our abstract QFT
Hamiltonians. The essential spectrum is described in Subsect.
\ref{sec4.1}. The Mourre estimate is proved in Subsect. \ref{sec4.4}.
An improved version with a smaller threshold set is proved in Subsect.
\ref{sec4.5}.
\subsection{HVZ theorem and existence of a ground state}\label{sec4.1}

\bet Let $H$ be an abstract QFT Hamiltonian and let $\x$ be a
weight operator for $\omega$. Assume hypotheses {\it (G1)}, {\it
(G5)}.  Then

{\it i)} if $\sigma_{\rm ess}(\omega)\subset [m_{\infty}, +\infty[$ then
\[
\sigma_{\rm ess}(H)\subset  [\inf \sigma(H)+m_{\infty}, +\infty[.
\]
{\it ii)} if $\sigma_{\rm
ess}(\omega)=[m_{\infty}, +\infty[$ then
\[
\sigma_{\rm ess}(H)= [\inf \sigma(H)+m_{\infty}, +\infty[.
\]
\label{B29}
\eet
\proof
Let us pick functions $j_{0}, j_{\infty}\in \cinf(\rr)$ with $0\leq
j_{0}\leq 1$, $j_{0}\in\coinf(\rr)$, $j_{0}\equiv 1$ near $0$ and
$j_{0}^{2}+ j_{\infty}^{2}=1$. For $R\geq 1$, $j^{R}$ is defined as in
Subsect. \ref{sec3.5} and we set $q^{R}=
(j_{0}^{R})^{2}$. From Subsect. \ref{sec1.1} we know that
\[
I(j^{R})I^{*}(j^{R})=\one.
\]

We first prove {\it i)}. Let $\chi\in
\coinf(]-\infty, {\rm inf}\sigma(H)+m_{\infty}[)$.  Using Lemma
\ref{Fw200} we get:
\beq\label{e2.4}
\begin{array}{rl}
\chi(H)=&\chi(H)I(j^{R})I^{*}(j^{R})\\[3mm]
=& I(j^{R})\chi(H^{\rm ext}) I^{*}(j^{R})+ o(R^{0})\\[3mm]
=&\sum_{k=0}^{M}I(j^{R})\one_{\{k\}}(N_{\infty})\chi(H^{\rm
ext})I^{*}(j^{R})+ o(R^{0}),
\end{array}
\eeq
for some $M$, using the fact that $H$ is bounded below and $\omega\geq
m>0$.  Using again Lemma \ref{Fw200}, we have:
\beq\label{e2.5}
\begin{array}{rl}
&I(j^{R})\one_{\{0\}}(N_{\infty})\chi(H^{\rm ext})I^{*}(j^{R})\\[3mm]
=&I(j^{R})\one_{\{0\}}(N_{\infty})I^{*}(j^{R})\chi(H)+ o(R^{0})\\[3mm]
=&\G(q^{R})\chi(H)+ o(R^{0}).
\end{array}
\eeq
It remains to treat the other terms in (\ref{e2.4}). Because of the
support of $\chi$ and using again Lemma \ref{Fw200}, we have:
\[
\begin{array}{rl}
&I(j^{R})\one_{\{k\}}(N_{\infty})\chi(H^{\rm
ext})I^{*}(j^{R})\\[3mm]
=&I(j^{R})\one_{\{k\}}(N_{\infty})\one\otimes F(\d\G(\omega)<m_{\infty})\chi(H^{\rm
ext})I^{*}(j^{R})\\[3mm]
=&I(j^{R})\one_{\{k\}}(N_{\infty})\one\otimes
F(\d\G(\omega)<m_{\infty})I^{*}(j^{R})\chi(H)+o(R^{0}),
\end{array}
\]
where $F(\lambda<m_{\infty})$ is a cutoff function supported in
$]-\infty, m_{\infty}[$.

From hypothesis {\it (H3)}, it follows that $\one_{[P,
+\infty[}(N)\chi(H)$ tends to $0$ in norm when $P\to +\infty$.
Since $I^{*}(j^{R})$ is isometric, we obtain:
\[
\begin{array}{rl}
&I(j^{R})\one_{\{k\}}(N_{\infty})\one\otimes
F(\d\G(\omega)<m_{\infty})I^{*}(j^{R})\chi(H)\\[3mm]
=& I(j^{R})\one_{\{k\}}(N_{\infty})\one\otimes
F(\d\G(\omega)<m_{\infty})I^{*}(j^{R})\one_{[0, P]}(N)\chi(H) + o(R^{0})+ o(P^{0}),
\end{array}
\]
where the error term $o(P^{0})$ is uniform in $R$. Next we use the
following identity from \cite[Subsect. 2.13]{DG1}:
\[
\one_{\{k\}}(N_{\infty})I^{*}(j^{R})\one_{\{n\}}(N)=
I_{k}(\frac{n!}{(n-k)!k!})^{\12}\underbrace{j^{R}_{0}\otimes
\cdots \otimes
j^{R}_{0}}_{n-k}\otimes\underbrace{j^{R}_{\infty}\otimes\cdots \otimes
j^{R}_{\infty}}_{k},
\]
where $I_{k}$ is the natural isometry between $\bigotimes^{n}\ch$ and
$\bigotimes^{n-k}\ch\otimes\bigotimes^{k}\ch$.

We note next that if $F\in \coinf(\rr)$ is supported in $]-\infty,
m_{\infty}[$, $F(\omega)$ is compact on $\ch$, so
$F(\omega)j_{\infty}^{R}$ tends to $0$ {\em in norm} when $R\to\infty$
since $\slim_{R\to\infty}j_{\infty}^{R}=0$. It follows from this
remark that for each $k\geq 1$ and $n\leq P$:
\[
I(j^{R})\one_{\{k\}}(N_{\infty})\one\otimes
F(\d\G(\omega)<m_{\infty})I^{*}(j^{R})\one_{\{n\}}(N)=o_{P}(R^{0}),
\]
and hence
\beq\label{e2.6}
I^{*}(j^{R})\one_{\{k\}}(N_{\infty})\chi(H^{\rm
ext})I(j^{R})= o(P^{0})+ o(R^{0})+ o_{P}(R^{0})= o(R^{0}),
\eeq
if we choose first $P$ large enough and then $R$ large enough.
Collecting (\ref{e2.4}), (\ref{e2.5}) and (\ref{e2.6}) we finally get
that
\[
\chi(H)= \G(q^{R})\chi(H)+  o(R^{0}).
\]
We use now that for each $R$ $\G(q^{R})(H_{0}+ 1)^{-\12}$ is
compact on $\Gh$, which follows easily from  {\it (H1)} and {\it (G5)} (see e.g.
\cite[Lemma 4.2]{DG1}). We obtain that $\chi(H)$ is compact as a
norm limit of compact operators. Therefore $\sigma_{\rm
ess}(H)\subset [{\rm inf}\sigma(H)+ m_{\infty}, +\infty[$.

Let us now prove {\it ii)}.  Note that it follows from {\it i)}
that $H$ admits a ground state. Let $\lambda={\rm inf}\sigma(H)+
\varepsilon$ for $\varepsilon>m_{\infty}$. Since $\varepsilon\in
\sigma_{\rm ess}(\omega)$, there exists unit vectors $h_{n}\in
\cD(\omega)$ such that $\lim_{n\to \infty}(\omega-\varepsilon)
h_{n}= 0$ and $\wlim_{n\to \infty} h_{n}= 0$. Let $u\in \Gh$ a
normalized ground state of $H$ and set
\[
u_{n}= a^{*}(h_{n})u.
\]
Since $u\in \cD(N)$ by {\it (H3)} $u_{n}$ is well defined. Moreover
since $\wlim h_{n}=0$, we obtain that $\lim \|u_{n}\|=1$ and $\wlim
u_{n}=0$.
 Since $u\in \cD(H^{\infty})$, we know from {\it
(H3)} that $u, Hu\in \cD(N^{\infty})$ and hence the following identity
is valid:
\[
H_{0}a^{*}(h_{n})u= a^{*}(h_{n})H_{0}u + a^{*}(\omega h_{n})u=
a^{*}(h_{n})Hu-a^{*}(h_{n})Vu + a^{*}(\omega h_{n})u,
\]
which shows that $u_{n}= a^{*}(h_{n})u\in \cD(H_{0})$. Clearly $u_{n}\in
\cD(N^{\infty})$, so $u_{n}\in \cD(H)$ and
\[
\begin{array}{rl}
(H-\lambda)u_{n}=& (H_{0}+ V-\lambda)u_{n}\\[3mm]
=& a^{*}(h_{n})(H-\lambda)u+ a^{*}(\omega h_{n})u+ [V,a^{*}(h_{n})]u\\[3mm]
=&a^{*}((\omega-\varepsilon)h_{n})u+ [V,a^{*}(h_{n})]u.
\end{array}
\]
We can compute the Wick symbol of $[V, a^{*}(h_{n})]$ using Prop.
\ref{wick-recap-bis}. Using the fact that $h_{n}$ tends weakly to
$0$ and Lemma \ref{1.1} {\it iii)} we obtain that $[V,
a^{*}(h_{n})]u$ tends to $0$ in norm. Similarly the term
$a^{*}((\omega-\varepsilon)h_{n})u$ tends to $0$ in norm.
Therefore $(u_{n})$ is a Weyl sequence for $\lambda$. \qed

\subsection{Virial theorem}\label{sec4.2}
Let $H$ be an abstract QFT Hamiltonian.
We fix a selfadjoint operator $a$ on $\ch$ such that hypothesis {\it
(M1 i)} holds and set
\[
A:=\d\G(a).
\]
On the interaction $V$ we impose hypothesis {\it (M2)}.
\begin{lemma}\label{2.1}
Assume {\it (M1 i)} and set $\omega_{t}=\e^{\i t a}\omega\e^{-\i ta}$. Then:

{\it i)} $\e^{\i ta}$ induces a strongly continuous group on
$\cD(\omega)$ and
\[
\sup_{|t|\leq 1}\|\omega_{t}(\omega+ 1)^{-1}\|<\infty, \ \
\sup_{|t|\leq 1} \|\omega(\omega_{t}+ 1)^{-1}\|<\infty.
\]
\[
ii)\ \ \sup_{0<|t|\leq
1}|t|^{-1}\|(\omega-\omega_{t})\|<\infty, \ \ \slim_{t\to 0}t^{-1}(\omega-\omega_{t})= -[\omega, \i
a]_{0}.
\]
\end{lemma}
\proof The first statement of {\it i)}
follows from \cite[Appendix]{GG}. This fact clearly implies the first
bound in {\it i)}. The second follows from
 $\omega(\omega_{t}+ 1)^{-1}= \e^{-\i
ta}\omega_{t}(\omega+1)^{-1}\e^{\i ta}$. We deduce then from {\it i)}
that
\beq
\sup_{|t|, |s|\leq 1}\|\omega_{s}(\omega_{t}+ 1)^{-1}\|<\infty.
\label{e2.10}
\eeq
Since $\omega\in C^{1}(a)$ we have:
\[
(\omega_{t}+ 1)^{-1}-(\omega+1)^{-1}=\int_{0}^{t}\e^{\i sa}(\omega+
1)^{-1}[\omega, \i a]_{0}(\omega+ 1)^{-1}\e^{-\i sa}\d s,
\]
as a strong integral, and hence:
\[
\begin{array}{rl}
(\omega-\omega_{t})=&(\omega_{t}+1)\left((\omega_{t}+
1)^{-1}-(\omega+1)^{-1}\right)(\omega+1)\\[3mm]
=&\int_{0}^{t}(\omega_{t}+1)(\omega_{s}+ 1)^{-1}\e^{\i s
a}[\omega, \i a]_{0}\e^{-\i sa}(\omega_{s}+1)^{-1}(\omega+1)\d s.
\end{array}
\]
Using (\ref{e2.10}) we obtain {\it ii)}. \qed

\medskip

We set now
\[
A:=\d\G(a),\ \ H_{s}= \e^{\i sA}H\e^{-\i sA}, \ \
H_{0,s}= \e^{\i sA}H_{0}\e^{-\i sA}, \ \ V_{s}= \e^{\i sA}V\e^{-\i
sA},
\]
and introduce the quadratic forms $[H_{0}, \i A]$,  $[V, \i A]$,
$[H, \i A]$ with domains  $\cD(H_{0})\cap
\cD(A)$, $\cD(N^{n})\cap \cD(A)$ and $\cD(H^{m})\cap \cD(A)$ for
$n\geq {\rm deg} w/2$ and $m$ large enough.

\begin{proposition}\label{2.2}
Let $H$ be an abstract QFT Hamiltonian such that {\it (M1 i)}, {\it
(M2)} hold. Then:

{\it i)} $[H_{0}, \i A]$ extends uniquely as a bounded operator from
$\cD(N)$ to $\cH$,  denoted by $[H_{0}, \ i A]_{0}$,

\medskip

{\it ii)} $[V, \i A]$ extends uniquely as a bounded operator from
$\cD(N^{M})$ to $\cH$ for $M$ large enough,  denoted by $[V, \i A]_{0}$,

\medskip

{\it iii)} $[H, \i A]$ extends uniquely as a bounded operator from
$\cD(H^{P})$ to $\cH$ for $P$ large enough,   denoted by $[H, \i A]_{0}$ and equal to
$[H_{0}, \i A]_{0}+ [V, \i A]_{0}$,

\medskip

{\it iv)} for $r$ large enough $(H+b)^{-r}$ is in $C^{1}(A)$ and
the following identity is valid as a bounded operators identity
from $\cD(A)$ to $\cH$: \beq A(H+b)^{-r}= (H+b)^{-r}A +\i
\frac{\d}{\d s}(H_s+b)^{-r}_{|s=0}, \label{e2.06} \eeq where \beq
\frac{\d}{\d s}(H_s+b)^{-r}_{|s=0}
=\sum_{j=0}^{r-1}(H+b)^{-r+j}([H_{0}, \i A]_{0}+ [V, \i A]_{0})
(H+b)^{-j-1} \label{piu} \eeq is a bounded operator on $\cH$.
\end{proposition}

\proof
We have $[H_{0}, \i A]=\d\G([\omega, \i a])$, which
using hypothesis {\it (M1 i)} and Prop. \ref{stup1} {\it i)} implies that
$[H_{0}, \i A](N+ 1)^{-1}$ is bounded. The fact that the
extension is unique follows from the fact that $\cD(a)\cap \cD(\omega)$ is
dense in $\ch$ since $\omega\in C^{1}(a)$.

Let us now check
{\it ii)}. Through the identification of $B^{2}_{\fin}(\ch)$ with
$\G_{\rm fin}(\ch)\otimes\G_{\rm fin}(\overline{\ch})$, we get from Prop.
\ref{wick-recap-bis} that
\[
[V, \ i A]= [\Wick(w), \ i A]= \Wick(w^{(1)})
\]
where $w^{(1)}= \left(\d\G(a)\otimes\one -
\one\otimes\d\G(\overline{a})\right)w$. By {\it (M2)} $w^{(1)}\in
B^{2}_{\rm fin}(\ch)$ which implies that $[V, \ i A](N+1)^{-n}$ is
bounded for $n\geq {\rm deg}w/2$ using  Lemma
\ref{1.1}. The fact that the extension is unique is obvious.

By the higher order estimates we have $[H, \i A]= [H_{0}, \i A]+ [V, \i A]$
on $\cD(A)\cap \cD(H^{M})$ for $M$ large enough, so  $[H, \i A]_{0}(H+b)^{-M}$
is bounded, again  by the higher order estimates. To prove that the extension
is unique we need to show that $\cD(A)\cap \cD(H^{M})$ is dense in
$\cD(H^{M})$ for $M$ large enough. Let  $u=(H+b)^{-M}v\in \cD(H^{M})$ and
$u_{\epsilon}= (H+b)^{-M}(1+\i \epsilon A)^{-1}v$. Clearly
$u_{\epsilon}\to u$ in $\cD(H^{M})$ when $\epsilon\to 0$. Next
$u_{\epsilon}$ belongs to
$\cD(H^{M})$ and to $\cD(A)$ since $(H+b)^{-M}$ is in $C^{1}(A)$ by {\it
iv)}. This completes the proof of {\it iii)}.

It remains to prove {\it iv)}. We start by proving some auxiliary
properties of  $H_{s}$. Since $H_{0, s}= \d\G(\omega_{s})$, we
obtain using Lemma \ref{2.1} {\it i)} and Prop.  \ref{stup1} {\it
i)} that 
\beq
\label{e2.11} 
\sup_{|s|\leq
1}\|H_{0}(H_{0,s}+1)^{-1}\|<\infty. 
\eeq 
The same arguments show also that $\cD(H_{0})= \cD(H_{0, s})$ ie
$\e^{\i s A}$ preserves $\cD(H_{0})$. Since $\e^{\i s A}$ preserves
$\cD(N^{n})$ we obtain from the higher order estimates that
\beq
H_{s}= H_{0 ,s}+ V_{s}\hbox{ on }\cD(H^{P}).
\label{irlit}
\eeq
Let us fix $n\geq {\rm
deg}w/2$.  Conjugating the bounds in {\it (H3)} by $\e^{\i sA}$,
we obtain that  there exists $p\in \nn$ such
that
\[
N^{2n}H_{0,s}^{2}\leq C(H_{s}+ b)^{2p}, \hbox{ uniformly in }|s|\leq 1.
\]
Using also (\ref{e2.11}) we obtain
\beq
N^{2n}H_{0}^{2}\leq C(H_{s}+ b)^{2p}, \hbox{ uniformly in }|s|\leq 1.
\label{e2.12}
\eeq
Let us show that for $r$ large enough:
\beq
\|(H_{s}+b)^{-r}-(H+b)^{-r}\|\leq C|s|, \ \ |s|\leq 1.
\label{e2.13}
\eeq
Using (\ref{irlit}),  we can write for $P$ large enough:
\beq
\begin{array}{rl}
&\left((H_{s}+b)^{-r}-(H+b)^{-r}\right)(H+b)^{-P}\\[3mm]
=&\sum_{j=0}^{r-1}(H_{s}+b)^{-r+j}(H-H_{s})(H+b)^{-j-1}(H+b)^{-P}\\[3mm]
=&\sum_{j=0}^{r-1}(H_{s}+b)^{-r+j}(H_{0}-H_{0,s}+V-V_{s})(H+b)^{-j-1}(H+b)^{-P}.
\end{array}
\label{e2.17}
\eeq
Using that $H_{0,s}-H_{0}= \d\G(\omega_{s}-\omega)$,  Lemma \ref{2.1}
{\it ii)} and Prop.  \ref{stup1} {\it i)} we obtain
that
\beq\label{e2.14}
\|(H_{0, s}-H_{0})(N+1)^{-1}\|\leq C|s|, \ \ |s|\leq 1.
\eeq
If $r\geq 2p$ then for $0\leq j\leq r-1$ then either $j+1\geq p$ or
$r-j\geq p$. Using (\ref{e2.12}) and (\ref{e2.14}) we deduce that
\beq
\|(H_{s}+b)^{-r+j}(H_{0, s}-H_{0})(H+b)^{-j-1}\|\leq C|s|, \ \ |s|\leq 1.
\label{e2.15}
\eeq

Next from Prop. \ref{wick-recap-bis}, we have:
\[
V_{s}= \Wick(\e^{\i sA}w\e^{-\i sA}).
\]
Through the identification of $B^{2}_{\fin}(\ch)$ with
$\G_{\rm fin}(\ch)\otimes\G_{\rm fin}(\overline{\ch})$, the symbol $\e^{\i sA}w\e^{-\i sA}$ is identified
with $\e^{\i sA}\otimes\e^{-\i s\overline{A}}w$.
From hypothesis {\it (M2)} and Prop. \ref{1.1}, we obtain that for
$M\geq {\rm deg}(w)/2$:
\beq
\|(V_{s}-V)(N+1)^{-M}\|\leq C|s|, \ \ |s|\leq 1.
\label{e2.16}
\eeq
By the same argument as above we obtain:
\beq
\|(H_{s}+b)^{-r+j}(V-V_{s})(H+b)^{-j-1}\|\leq C|s|, \ \ |s|\leq 1.
\label{e2.18}
\eeq
Combining (\ref{e2.17}), (\ref{e2.18}) and  (\ref{e2.15}), we obtain (\ref{e2.13}).

Next from (\ref{e2.14}) we obtain by considering first  finite particle
vectors that
\[
\slim_{s\to 0}s^{-1}(H_{0, s}-H_{0})(N+1)^{-1}\hbox{ exists}.
\]
We note next that by hypothesis {\it (M2)} we know that $s^{-1}(\e^{\i
sA}w\e^{-\i sA}-w)$ converges in $B^{2}_{\rm fin}(\Gh)$ when $s\to 0$.
Using then Lemma \ref{1.1} {\it ii)}, we obtain also that
\[
\slim_{s\to 0}s^{-1}(V_{s}-V)(N+1)^{-n}\hbox{ exists}.
\]
From (\ref{e2.17}) we obtain that for $r\geq p$ and $P$
large enough:
\[
\slim_{s\to 0}s^{-1}\left((H_{s}+b)^{-r}-(H+b)^{-r}\right)\hbox{ exists
on }\cD(H^{P})
\]
By (\ref{e2.13}) the strong limit exists on $\Gh$, which shows that
$(H+b)^{-r}$ is in $C^{1}(A)$. \qed

\begin{remark}
The same proof as in Prop. \ref{2.2} {\it iv)} shows that for $r$
large enough and $z_{i}\in \cc\backslash \rr$, the operator
$\prod_{i=1}^{r}(z_{i}-H)^{-1}$ is in $C^{1}(A)$. Using the functional
calculus formula (\ref{HS}), it is easy to deduce from this fact that
$\chi(H)$ is in $C^{1}(A)$ for all $\chi\in \coinf(\rr)$.
\end{remark}
\medskip

The following proposition is the main consequence of Prop. \ref{2.2}.
\begin{proposition}\label{viriel}
Let $H$ be an abstract QFT Hamiltonian such that {\it (M1 i)}, {\it
(M2)} hold. Then the {\em virial relation} holds:
\beq\label{virial-bis}
\one_{\{\lambda\}}(H)[H, \i A]_{0}\one_{\{\lambda\}}(H)=0, \
\lambda\in \rr.
\eeq
\end{proposition}
\proof Let us fix $r$ large enough such that $(H+ b)^{-r}\in
C^{1}(A)$ so that $(H+b)^{-r}: \cD(A)\to \cD(A)$ and $[(H+b)^{-r}, \i A]$
extends as a bounded operator on $\cH$ denoted by $[(H+b)^{-r}, \i
A]_{0}$. Moreover from Prop. \ref{2.2} {\it iv)} we have:
\[
[(H+b)^{-r}, \i A]_{0}= -\sum_{j=0}^{r-1} (H+b)^{-r+ j}[H,\i
A]_{0}(H+b)^{-j-1}.
\]
Let now $u_{1}, u_{2}\in \cH$ such that $Hu_{i}= \lambda u_{i}$. Since $(H+b)^{-r}\in
C^{1}(A)$ and $u_{i}$ is an eigenvector of $(H+b)^{-r}$, we have the virial
relation:
\[
\begin{array}{rl}
0=&(u_{1}, [(H+b)^{-r}, \i A]_{0}u_{2})\\[3mm]
=&-\sum_{j=0}^{r-1}(u_{1}, (H+b)^{-r+ j}[H,\i
A]_{0}(H+b)^{-j-1}u_{2})\\[3mm]
=&-\sum_{j=0}^{r-1}(\lambda+b)^{-r-1}(u_{1}, [H,\i
A]_{0}u_{2})\\[3mm]
=&-r(\lambda+ b)^{-r-1}(u_{1}, [H, \i A]_{0}u_{2}),
\end{array}
\]
which proves the lemma. \qed
\subsection{Mourre estimate for second quantized Hamiltonians}\label{sec4.3}
In this subsection we will apply the
abstract results in Subsect. \ref{sec1.3} to second quantized
Hamiltonians.

Let  $\omega, a$ be two selfadjoint operators on $\ch$ such that
{\it (H1)}, {\it (M1)} hold.
Note that it follows from Lemma \ref{2.3} and the results recalled
above it  that {\it (M1)} imply also that
\beq\label{e2.19bis}
\kappa_{a}(\omega)\hbox{ is a
closed countable set}.
\eeq
Clearly $\d\G(\omega)\in C^{1}(\d\G(a))$ and $[\d\G(\omega),
\i\d\G(a)]_{0}=\d\G([\omega, \i a]_{0})$. Since $\d\G(\omega)$ and
$\d\G([\omega, \i a]_{0})$ commute with $N$, we can restrict them to each
$n-$particle sector $\otimes_{\rm s}^{n}\ch$. We denote by
\[
\rho^{\d\G(A)\:(1)}_{\d\G(\omega)}
\]
the corresponding restriction of $\rho^{\d\G(A)}_{\d\G(\omega)}$ to
the range of $\one_{[1, +\infty[}(N)$.

Finally we introduce the
following natural notation for $E\subset \rr$:
\beq
\d\G^{(1)}(E)= \bigcup_{n=0}^{+\infty} \underbrace{E+\cdots + E}_{n}\ ,
\ \d\Gamma (E)= \{0\}\cup \d\G^{(1)}(E).\label{e2.19}
\eeq
\begin{remark}\label{tirli}
As an example of  use of this notation, note that if $b$ is a selfadjoint operator
on $\ch$, then:
\[
\sigma(\d\G(b))=\{0\}\cup \d\G(\sigma(b)).
\]
Note also that if  $E$ is a closed countable
set included in $[m, +\infty[$ for some $m>0$, $\d\G^{(1)}(E)$ is a closed countable set.
\end{remark}

\begin{lemma}\label{2.4}
Let $\omega, a$ be two selfadjoint operators on $\ch$ such that
{\it (M1)} holds.
Then:
\[
\begin{array}{l}
i)\ \ \rho^{\d\G(a)}_{\d\G(\omega)}\geq 0,\\[3mm]
ii) \ \ \rho^{\d\G(a)\:(1)}_{\d\G(\omega)}(\lambda)=0 \:\Rightarrow
\lambda\in
\d\G^{(1)}(\kappa_{a}(\omega)).
\end{array}
\]
\end{lemma}
\proof We have $[\d\G(\omega), \i \d\G(a)]= \d\G([\omega, \i a])$.
Since $\d\G(\omega)\in C^{1}(\d\G(a))$  the virial relation is
satisfied. Denote by $\rho_{n}$
the  restriction of $\rho^{\d\G(a)}_{\d\G(\omega)}$ to $\otimes_{\rm s}^{n}\ch$. Applying Lemma
\ref{2.3} {\it iv)} we obtain
\[
\rho_{0}(\lambda)=\left\{
\begin{array}{l}
0, \ \ \lambda=0, \\
+\infty, \ \ \lambda\neq 0
\end{array}
\right.,
\]
\[
\rho_{n}(\lambda)= \mathop{\inf}\limits_{\lambda_{1}+\cdots
\lambda_{n}=\lambda}\left(\rho^{a}_{\omega}(\lambda_{1})+\cdots+
\rho^{a}_{\omega}(\lambda_{n})\right)
\]
for $n\geq 1$.  We note next that since $\omega\geq m>0$,
$\chi(\d\Gamma(\omega))\one_{[n, +\infty[}(N)=0$ if $n$ is large enough,
where $\chi\in \coinf(\rr)$. Therefore only a finite number of
$n-$particle sectors contribute to the computation of
$\rho^{\d\G(a)}_{\d\G(\omega)}$ near an energy level $\lambda$. We
can hence apply Lemma \ref{2.3} {\it iii)} and obtain that
$\rho^{\d\G(a)}_{\d\G(\omega)}\geq 0$.

Let us now prove the  second statement of the lemma.
Since $\rho^{a}_{\omega}(\lambda)= +\infty$ if
$\lambda\not\in \sigma(\omega)$, we have
$\rho^{a}_{\omega}(\lambda)=+\infty$  for $\lambda<0$. Therefore
\[
\rho_{n}(\lambda)=\mathop{\inf}\limits_{I_{n}(\lambda)}\left(\rho^{a}_{\omega}(\lambda_{1})+\cdots+
\rho^{a}_{\omega}(\lambda_{n})\right),
\]
for $I_{n}(\lambda)=\{(\lambda_{1}, \dots, \lambda_{n})|\ \
\lambda_{1}+\cdots \lambda_{n}=\lambda, \ \ \lambda_{i}\geq 0\}$.
The
function $\rho^{a}_{\omega}(\lambda_{1})+\cdots+
\rho^{a}_{\omega}(\lambda_{n})$ is lower semicontinuous on $\rr^{n}$,
hence attains
its minimum on the compact set $I_{n}(\lambda)$.  Therefore using also
that $\rho^{a}_{\omega}\geq 0$, we see that
$\rho_{n}(\lambda)=0$ iff $\lambda\in
\kappa_{a}(\omega)+\cdots +\kappa_{a}(\omega)$ ($n$ factors).
Using Lemma \ref{2.3} {\it iii)} as above, we obtain that
$\rho^{\d\G(A)\:(1)}_{\d\G(\omega)}(\lambda)=0$ implies that  $\lambda\in
\d\G^{(1)}(\kappa_{a}(\omega))$, which proves {\it ii)}.
\qed
\subsection{Mourre estimate for abstract QFT
Hamiltonians}\label{sec4.4} In this subsection we prove the Mourre
estimate for abstract QFT Hamiltonians.  Let $H$ be an abstract
QFT Hamiltonian and $a$ a selfadjoint operator on $\ch$ such that
{\it (M1)} holds. Let also $\x$ be a weight operator for $\omega$.
\begin{theoreme}\label{mourre}
Let $H$ be an abstract QFT Hamiltonian and $a$ a selfadjoint
operator on $\ch$ such that {\it (M1)} and {\it (M2)} hold. Let
$\x$ be a weight operator for $\omega$ such that conditions {\it
(G1)}, {\it (G3)}, {\it (G5)} hold. Set
\[
\tau:=\sigma_{\rm pp}(H)+\d\G^{(1)}(\kappa_{a}(\omega))
\]
and $A=\d\G(a)$. Then:

{\it i)} Let $\lambda\in \rr\backslash\tau$. Then there exists
$\epsilon>0$, $c_{0}>0$ and a compact operator $K$ such that
\[
\one_{[\lambda-\epsilon, \lambda+ \epsilon]}(H)[H, \i
A]_{0}\one_{[\lambda-\epsilon, \lambda+ \epsilon]}(H)\geq
c_{0}\one_{[\lambda-\epsilon, \lambda+ \epsilon]}(H)+ K.
\]
{\it ii)} for all $\lambda_{1}\leq \lambda_{2}$ such that
$[\lambda_{1}, \lambda_{2}]\cap \tau=\emptyset$ one has:
\[
{\rm dim}\one_{[\lambda_{1}, \lambda_{2}]}(H)<\infty.
\]
Consequently $\sigma_{\rm pp}(H)$ can accumulate only at $\tau$, which
is a closed countable set.

{\it iii)} Let $\lambda\in \rr\backslash(\tau\cup\sigma_{\rm pp}(H))$. Then there exists
$\epsilon>0$ and  $c_{0}>0$  such that
\[
\one_{[\lambda-\epsilon, \lambda+ \epsilon]}(H)[H, \i
A]_{0}\one_{[\lambda-\epsilon, \lambda+ \epsilon]}(H)\geq
c_{0}\one_{[\lambda-\epsilon, \lambda+ \epsilon]}(H).
\]
\end{theoreme}
\proof
We note first that $[H, \i A]_{0}$ satisfies the virial relation by
Prop. \ref{viriel}.
Therefore we will be able to apply the abstract results in Lemma
\ref{2.3} in our situation.
Recall that $H^{\ext}= H\otimes\one+ \one\otimes \d\G(\omega)$ and set
\[
A^{\ext}= A\otimes\one+ \one\otimes A.
\]
By Prop. \ref{2.2} $[H, \i A]_{0}$ considered as an operator on $\cH$
with domain $\cD(H^{M})$ is equal to $H_{1}+ V_{1}$, where $H_{1}=
\d\G([\omega, \i a]_{0})$, $V_{1}= [V, \i A]_{0}$. Note that by {\it (M2)}
 $V_{1}$ is a Wick polynomial with a symbol in $B_{\rm fin}^{2}(\ch)$,
and by {\it (G3)}, $[\x, [\omega, \i a]]$ is bounded on $\ch$.
Therefore using Lemma \ref{gauge} {\it v)} we see that the analog
of  (\ref{ez.3}) holds for $[H, \i A]_{0}$. We obtain:
\[
I^{*}(j^{R})[H,\i A]_{0}= [H^{\ext}, \i A^{\ext}]_{0}I^{*}(j^{R})+
(N+1)^{n}\cO_{N}(R^{0}),
\]
for some $n$.
We recall (\ref{e2.5}):
\beq\label{e2.21bis}
\chi(H)= \G(q^{R})\chi(H)+ I(j^{R})\chi(H^{\ext})\one_{[1,
+\infty[}(N_{\infty})I^{*}(j^{R})+ o(R^{0}),
\eeq
for $q^{R}= (j_{0}^{R})^{2}$.

Using then Lemma \ref{Fw200} and the higher order estimates (which
hold also for $H^{\ext}$ with the obvious modifications), we obtain
that:
\beq
\begin{array}{rl}
\chi(H)[H, \i A]_{0}\chi(H)=& \G(q^{R})\chi(H)[H, \i A]_{0}\chi(H)\\[3mm]
&+ I(j^{R})\chi(H^{\ext})[H^{\ext}, \i A^{\ext}]_{0}\chi(H^{\rm ext})\one_{[1,
+\infty[}(N_{\infty})I^{*}(j^{R}) + o(R^{0}).
\end{array}
\label{e2.20}
\eeq

We will now prove by induction on $n\in \nn$ the following
statement:
\[
H(n)\ \ \left\{ \begin{array}{l}
i) \ \ \rho^{A}_{H}(\lambda)\geq 0, \hbox{ for }\lambda\in ]-\infty,
\inf\sigma(H)+ nm[, \\[3mm]
ii) \ \ \tau^{A}(H)\cap]-\infty,\inf\sigma(H)+ nm[ \subset \sigma_{\rm pp}(H)+
\d\G(\kappa_{a}(\omega)).
\end{array}\right.
\]
 Statement {\it  H(0)} is clearly true
since $\rho^{A}_{H}(\lambda)=+\infty$ for $\lambda<\inf\sigma(H)$.

Let us assume that {\it H(n-1)} holds. Let us denote by $\rho^{\ext\:(1)}$ the restriction of
 $\rho^{A^{\ext}}_{H^{\ext}}$  to the range of $\one_{[1,
+\infty[}(N_{\infty})$. This function is well defined since $H^{\ext}$ and
$[H^{\ext}, \i A^{\ext}]_{0}$ commute with $N_{\infty}$.

Let  $\lambda\in ]-\infty, \inf\sigma(H)+ nm[$. Using Lemma
\ref{2.3} {\it iv)} and the fact that $\omega\geq m$ we obtain:
\[
\rho^{\ext\:(1)}(\lambda)=\mathop{\inf}\limits_{(\lambda_{1},
\lambda_{2})\in I^{(n)}(\lambda)}\left(\rho^{A}_{H}(\lambda_{1})+
\rho^{A\:(1)}_{H_{0}}(\lambda_{2})\right),
\]
where
\[
I^{(n)}(\lambda)=\{(\lambda_{1}, \lambda_{2})|\: \lambda_{1}+
\lambda_{2}=\lambda, \: \inf\sigma(H)\leq \lambda_{1}\leq \inf\sigma(H)+
(n-1)m, \ \ 0\leq \lambda_{2}\leq -\inf\sigma(H)\},
\]
and the function $\rho^{A\:(1)}_{H_{0}}$ is defined in Subsect.
\ref{sec4.3}.
Note that by {\it H(n-1)} {\it i)} and Lemma \ref{2.4} {\it i)}    the
two functions  $\rho^{A}_{H}(\lambda_{1})$ and
$\rho^{A\:(1)}_{H_{0}}(\lambda_{2})$ are positive for $(\lambda_{1},
\lambda_{2})\in I^{(n)}(\lambda)$. We deduce first from this fact that:
\beq
\label{e2.21ter}
\rho^{\ext\:(1)}(\lambda)\geq 0\hbox{ for }\lambda\in ]-\infty,
\inf\sigma(H)+ nm[.
\eeq
 Moreover  using that the lower semicontinuous
function $\rho^{A}_{H}(\lambda_{1})+
\rho^{A\:(1)}_{H_{0}}(\lambda_{2})$ attains its minimum on the compact set
$I^{(n)}(\lambda)\subset \rr^{2}$, we obtain that
\beq
\label{e2.22bis}
\begin{array}{l}
\rho^{\ext\:(1)}(\lambda)=0, \ \ \lambda\in
]-\infty, \inf\sigma(H)+ nm[  \:\Rightarrow\\[3mm]
\lambda=\lambda_{1}+ \lambda_{2},
\hbox{ where }(\lambda_{1}, \lambda_{2})\in I^{(n)}(\lambda), \ \
\rho^{A}_{H}(\lambda_{1})=\rho^{A\:(1)}_{H_{0}}(\lambda_{2})=0.
\end{array}
\eeq
From {\it H(n-1)} {\it ii)}  and Lemma \ref{2.3} {\it ii)} we get that
\[
\rho^{A}_{H}(\lambda_{1})=0, \ \ \lambda_{1}\in ]-\infty,
\inf\sigma(H)+ (n-1)m[\: \Rightarrow \:\lambda_{1}\in \sigma_{\rm
pp}(H)+\d\G(\kappa_{a}(\omega)).
\]
From Lemma \ref{2.4} {\it ii)} we know that
\[
\rho^{A\:(1)}_{H_{0}}(\lambda_{2})=0\:\Rightarrow \:\lambda_{2}\in
\d\G^{(1)}(\kappa_{a}(\omega)).
\]
Using  (\ref{e2.22bis}) we get that
\beq\label{e2.22}
\rho^{\ext\:(1)}(\lambda)=0, \ \ \lambda\in
]-\infty, \inf\sigma(H)+ nm[  \:\Rightarrow\: \lambda\in \sigma_{\rm
pp}(H)+ \d\G^{(1)}(\kappa_{a}(\omega)).
\eeq

The operators $\G(q^{R})\chi(H)$ and hence $\G(q^{R})\chi(H)[H, \i
A]_{0}\chi(H)$ are compact on $\cH$. Choosing hence $R$ large
enough in (\ref{e2.20}) we obtain using (\ref{e2.21bis}) and the
fact that $I(j^{R})I^{*}(j^{R})=\one$ that \beq
\tilde{\rho}^{A}_{H}(\lambda)\geq  \rho^{\ext\:(1)}(\lambda), \ \
\lambda\in  ]-\infty, \inf\sigma(H)+ nm[. \label{e2.21} \eeq By
Lemma \ref{2.3} {\it i)} this implies first that $\rho^{A}_{H}\geq
0$ on $ ]-\infty, \inf\sigma(H)+ nm[$, i.e. {\it H(n)}  {\it i)}
holds.  Using then (\ref{e2.22}) we obtain that
\[
\tilde{\rho}^{A}_{H}(\lambda)=0, \ \ \lambda\in
]-\infty, \inf\sigma(H)+ nm[  \:\Rightarrow\: \lambda\in \sigma_{\rm
pp}(H)+ \d\G^{(1)}(\kappa_{a}(\omega)),
\]
which proves {\it H(n)} {\it ii)}.  Since {\it H(n)} holds for any $n$
we obtain statement {\it i)} of the theorem. The fact that  ${\rm
dim}\one_{[\lambda_{1}, \lambda_{2}]}(H)<\infty$ if $[\lambda_{1},
\lambda_{2}]\cap \tau=\emptyset$ follows from the abstract results recalled in
Subsect. \ref{sec1.3}. We saw in (\ref{e2.19bis}) that
$\kappa_{a}(\omega)$ is a closed countable set.  Using also 
Remark \ref{tirli}, this  implies by
induction on $n$ that $\tau\cap ]-\infty, \inf\sigma(H)+ nm[$ is a
closed countable set for any $n$.  Finally statement {\it iii)}
follows from Lemma \ref{2.3}. This completes the proof of the theorem. \qed

\subsection{Improved Mourre estimate}\label{sec4.5}
Thm. \ref{mourre} can be rephrased as:
\[
\tau_{A}(H)\subset \sigma_{\rm pp}(H)+ \d\G^{(1)}(\kappa_{a}(\omega)),
\]
which is sufficient for our
purposes. Nevertheless a little attention shows that one should expect a
better result, namely:
\[
\tau_{A}(H)\subset \sigma_{\rm pp}(H)+ \d\G^{(1)}(\tau_{a}(\omega)),
\]
i.e. eigenvalues of $\omega$ away from $\tau_{a}(\omega)$ should
not contribute to the set of thresholds of $H$. In this subsection
we prove this result if there exists a comparison operator
$\omega_{\infty}$ such that hypothesis {\it (C)} holds.

We fix a function $q\in \cinf(\rr)$ such that
\beq
0\leq q\leq 1, \ \ q\equiv 0\hbox{ near }0, \ \ q\equiv 1\hbox{ near
}1.
\label{esimp.6}
\eeq
\begin{lemma}\label{simp1}
Assume {\it (H1)}, {\it (G1)}, {\it (G3)}, {\it (M1)} for $\omega$ and
$\omega_{\infty}$ and {\it (C)}. Set $H_{0}= \d\G(\omega)$,
$H_{\infty}=\d\G(\omega_{\infty})$.
Let $q$ as in (\ref{esimp.6}) and
$\chi\in \coinf(\rr)$. Then:
\beq
(\chi^{2}(H_{0})-\chi^{2}(H_{\infty}))\G(q^{R})\in o(R^{0}),
\label{esimp.5}
\eeq
\beq\label{esimp.4}
\begin{array}{rl}
&\chi(H_{0})[H_{0}, \i A]_{0}\chi(H_{0})\G(q^{R})\\[3mm]
=&\chi(H_{\infty})[H_{\infty}, \i A]_{0}\chi(H_{\infty})\G(q^{R}) +
o(R^{0}),\\[3mm]
\end{array}
\eeq
Assume additionally {\it (G5)}. Then
\beq\label{esimp.3}
\tilde{\rho}^{a}_{\omega}=
\tilde{\rho}^{a}_{\omega_{\infty}}.
\eeq
\end{lemma}
\proof
We will first prove the following estimates:
\beq
\label{esimp.1}
 [\chi(H_{\epsilon}), \G(q^{R})], \ \
\left(\chi(H_{0})-\chi(H_{\infty})\right)\G(q^{R})\in o(R^{0}),
\eeq
\beq\label{esimp.2}
\begin{array}{rl}
&(H_{\epsilon_{1}}+\i)^{-1}[H_{0}-H_{\infty}, \i
A]_{0}\G(q^{R})(H_{\epsilon_{2}}+\i)^{-1}\in o(R^{0})  \\[3mm]
&(H_{\epsilon_{1}}+\i)^{-1}[[H_{\infty}, \i
A]_{0}, \G(q^{R})](H_{\epsilon_{2}}+\i)^{-1}\in o(R^{0}),
\end{array}
\eeq
for $\epsilon, \epsilon_{1}, \epsilon_{2}\in \{0, \infty\}$. If we use
the identities
\[
[\d\G(b_{i}), \G(q^{R})]= \d\G(q^{R}, [b_{i}, q^{R}]), \ \
\d\G(b_{1}-b_{2})\G(q^{R})= \d\G(q^{R}, (b_{1}-b_{2})q^{R}),
\]
for $b_{1}= \omega$, $b_{2}= \omega_{\infty}$, Lemma
\ref{compare}, Lemma \ref{gauge} \textit{(i)} and  the bounds in
Prop. \ref{stup1}, it is easy to see that uniformly in $z\in
\cc\backslash \rr\cap \{|z|\leq R\}$:
\[
\begin{array}{rl}
&[(z-H_{\epsilon})^{-1}, \G(q^{R})]\in
O(R^{-1})|{\rm Im} z|^{-2}, \\[3mm]
&(z-H_{\epsilon_{1}})^{-1}(H_{0}-H_{\infty})\G(q^{R})(z-H_{\epsilon_{2}})^{-1}\in
o(R^{0})|{\rm Im} z|^{-2}.
\end{array}
\]
Using the functional calculus formula (\ref{HS}) this implies
(\ref{esimp.1}). The proof of (\ref{esimp.2}) is similar using
Lemma \ref{compare} and Lemma \ref{gauge} \textit{(v)}. The proof
of (\ref{esimp.4}) is now easy: we move the operator $\G(q^{R})$
to the left, changing $H_{0}$ into $H_{\infty}$ along the way, and
then move $\G(q^{R})$ back to the right. All errors terms are
$o(R^{0})$, by (\ref{esimp.1}), (\ref{esimp.2}). (\ref{esimp.5})
follows from (\ref{esimp.1}).
 If we restrict (\ref{esimp.5}), (\ref{esimp.4})  to
the one-particle sector we obtain that
\[
\begin{array}{rl}
&(\chi^{2}(\omega)-\chi^{2}(\omega_{\infty}))q^{R}\in o(R^{0}), \\[3mm]
&\chi(\omega)[\omega, \i a]_{0}\chi(\omega)q^{R}=
\chi(\omega_{\infty})[\omega_{\infty}, \i a]_{0}\chi(\omega_{\infty})q^{R}+
o(R^{0}).
\end{array}
\]
Using {\it (G5)} and the fact that $(1-q)\in \coinf(\rr)$ we see that
$\chi(H_{\epsilon})(1-q)^{R}$ is compact for $\epsilon=0, \infty$.
Writing $1= (1-q)^{R}+ q^{R}$, we easily obtain (\ref{esimp.3}). \qed

\medskip

\begin{theoreme}\label{mourrebis}
Let $H$ be an abstract QFT Hamiltonian satisfying the hypotheses of
Thm. \ref{mourre}.  Let
$\omega_{\infty}$ be a comparison Hamiltonian on $\ch$  such
that {\it (C1)} holds.  Then the
conclusions of Thm. \ref{mourre} hold for
\[
\tau:=\sigma_{\rm pp}(H)+ \d\G^{(1)}(\tau_{a}(\omega)).
\]
\end{theoreme}

\proof We use the notation in the proof of Thm. \ref{mourre}. We
pick a function $q_{1}$ satisfying (\ref{esimp.6}) such that
$q_{1}j_{\infty}=j_{\infty}$, so that
\[
I^{*}(j^{R})= \one\otimes\G(q_{1}^{R})I^{*}(j^{R}).
\]
Therefore in (\ref{e2.20}) we can insert $\one\otimes\G(q_{1}^{R})$ to
the left of $I^{*}(j^{R})$. If we set
\[
H_{\infty}^{\ext}:= H\otimes\one + \one\otimes H_{\infty},
\]
then using the obvious extension of Lemma \ref{simp1} to $H^{\ext}$
and $H_{\infty}^{\ext}$, we obtain instead of (\ref{e2.20}):
\beq
\begin{array}{rl}
&\chi(H)[H, \i A]_{0}\chi(H)\\[3mm]
=& \G(q^{R})\chi(H)[H, \i A]_{0}\chi(H)\\[3mm]
&+ I(j^{R})\chi(H_{\infty}^{\ext})[H_{\infty}^{\ext}, \i
A^{\ext}]_{0}\chi(H_{\infty}^{\rm ext})\one_{[1,
+\infty[}(N_{\infty})I^{*}(j^{R}) + o(R^{0}).
\end{array}
\label{e2.201}
\eeq
Therefore in the later steps of the proof we can replace $\omega$ by
$\omega_{\infty}$. By assumption $\kappa_{a}(\omega_{\infty})=
\tau_{a}(\omega_{\infty})$ and by Lemma \ref{simp1}
$\tau_{a}(\omega_{\infty})= \tau_{a}(\omega)$. This completes the
proof of the theorem. \qed

\section{Scattering theory for abstract QFT Hamiltonians}\label{sec5}\init
In this section we consider the scattering theory for our abstract 
QFT Hamiltonians. This theory is formulated in terms of {\em
asymptotic Weyl operators}, (see Thm. \ref{4.1bis}) which form 
regular CCR representations over  $\ch_{\rm c}(\omega)$. Using
the fact that the theory is massive, it is rather easy to show that
this representation is of Fock type (see Thm. \ref{4.6}).  The basic
question of scattering theory, namely the asymptotic completeness of
wave operators, amounts then to prove that the space of vacua for the
two asymptotic CCR representations coincide with the space of bound
states for $H$. This will be shown in Thm. \ref{asympt-comp}, using
the propagation estimates of Sect. \ref{sec6}.

In all this section we only consider objects with superscript $+$,
corresponding to $t\to +\infty$. The corresponding
objects with superscript $-$ corresponding to $t\to -\infty$ have the
same properties.

\subsection{Asymptotic fields}\label{sec5.1}
For $h\in \ch$ we set $h_{t}:=\e^{-\i t \omega}h$. Recall that
$\ch_{\rm c}(\omega)\subset \ch$ is the continuous spectral
subspace for $\omega$ and that by hypothesis {\it (S)}  there
exists a  subspace $\ch_{0}$ dense in $\ch_{\rm
c}(\omega)$  such that for all $h\in \ch_{0}$   there
exists $\epsilon>0$  such that
\[
\|\one_{[0, \epsilon]}(\frac{\x}{|t|})\e^{-\i t \omega}h\|\in
O(t^{-\mu}), \qquad \mu>1.
\]

\begin{theoreme}
Let $H$ be an abstract QFT Hamiltonian such that hypotheses {\it (Is)}
for $s>1$ and  {\it (S)} hold. Then:
{\it i)}  For all $h\in \ch_{\rm c}(\omega)$ the strong limits
\beq
W^{+}(h):= \slim_{t\fld +\infty}\e^{\i tH}W (h_{t})\e^{-\i tH}
\label{eas.3bis}
\eeq
exist. They are called the {\em asymptotic Weyl operators}.
The asymptotic Weyl operators can be also defined using the norm limit:
\beq
W^{+}(h)(H+b)^{-n}= \lim_{t\fld +\infty}\e^{\i tH}W (h_{t})
(H+b)^{-n}\e^{-\i tH},
\label{eas.3ter}
\eeq
for $n$ large enough.

{\it ii)} The map
\beq\ch_{\rm c}(\omega)\ni h\mapsto W^+(h)\label{kwer.1}\eeq
is strongly continuous
and for $n$ large enough, the map
\beq
\ch\ni h_{\rm c}(\omega)\mapsto W^+(h)(H+b)^{-n}\label{kwer.2}
\eeq
is norm continuous.
\newline {\it iii)}
The operators $W^+(h)$ satisfy the Weyl commutation relations:
\[
W^{+}(h)W^{+}(g)= \e^{-\i\12{\rm Im}(h|g)}W^{+}(h+g).
\]
{\it iv)}  The Hamiltonian preserves the asymptotic Weyl operators:
\beq
\e^{\i tH}W^+(h)\e^{-\i tH}= W^+(h_{-t}).
\label{eas.3er}
\eeq
\label{4.1bis}
\end{theoreme}
\proof The proof is almost identical to the proof of \cite[Thm.
10.1]{DG}, therefore we will only sketch it.
We have:
\[
W(h_{t})= \e^{-\i tH_{0}}W(h)\e^{\i tH_{0}},
\]
which implies that, as a quadratic form on $\cD(H_{0})$, one has
\beq
\p_{t}W (h_{t})= -[H_{0}, \i W (h_{t})].
\label{eas.2}\eeq
Using (\ref{eas.2}) and the fact that for $n$ large enough $\cD(H^{n})\subset \cD(H_{0})\cap
\cD(V)$, we have, as
quadratic forms on $\cD(H^{n})$:
\[
\p_{t}\e^{\i tH}W(h_{t})\e^{-\i tH}= \e^{\i tH}
 [V, \i W(h_{t})]\e^{-\i tH}.
\]
Integrating this relation we have  as a quadratic form
identity on $\cD(H^{n})$
\beq
\e^{\i tH}W (h_{t})\e^{-\i tH}- W (h)=
\int_{0}^{t}\e^{\i t'H}[V, \i W(h_{t'})]\e^{-\i t'H}\d t'.
\label{eas.01}
\eeq
We claim that for $h\in \ch_{0}$ (see hypothesis {\it (S)}), and  $p\geq {\rm deg}w/2$:
\beq\label{e5.1}
\|[V, W(h_{t})](N+1)^{-p}\|\in L^{1}(\d t).
\eeq
In fact writing $w$ as $\sum_{p+q\leq {\rm deg }(w)}w_{p, q}$, where
$w_{p,q}$ is of order $(p,q)$ and using Prop. \ref{wick-recap-bis}, we obtain that
\[
[\Wick(w_{p,q}), W(h_{t})]= W(h_{t})\Wick(w_{p,q}(t)),
\]
 where $w_{p,q}(t)$ is the sum of the symbols in the r.h.s. of
(\ref{sec.wick.e9})  for $(s,r)\neq (p,q)$. Using {\it (Is)} and {\it
(S)} we obtain writing $\one=\one_{[0, \epsilon]}(\frac{\x}{\t})+
\one_{]\epsilon, +\infty[}(\frac{\x}{\t})$ that
\[
\|w_{p,q}(t)\|_{B^{2}(\ch)}\in L^{1}(\d t),
\]
which proves (\ref{e5.1}) using Lemma \ref{1.1}.

Using then the higher order estimates, we obtain that the identity (\ref{eas.01}) makes sense as an identity
between bounded operators from $\cD(H^{n})$ to $\cH$ for $n$ large
enough.
It  also proves that the
norm limit
(\ref{eas.3ter}) exists for $h\in \ch_0$.
The rest of the proof is
identical to \cite[Thm. 10.1]{DG}. It relies on the bound
\[
\begin{array}{rl}
&\|\left(\e^{\i tH}W(h_{t})\e^{-\i tH}-\e^{\i tH}W(g_{t})\e^{-\i
tH}\right)(H+b)^{-n}\|\\[3mm]
\leq&
\|\left(W(h)-W(g)\right)(N+1)^{-1}\|\|(N+1)(H+b)^{-n}\|\\[3mm]
\leq &C\|h-g\|(\|h\|^{2}+ \|g\|^{2}+ 1).
\end{array}
\] \qed

\medskip

\bet
{\it i)}  For any $h\in\ch_{\rm c}(\omega)$:
\[
\phi^+(h):=-\i\frac{\d}{\d s}W^+(sh)_{|s=0}
\]defines a self-adjoint operator,
called the {\em asymptotic field},  such that
\[W^+(h)=\e^{\i \phi^+(h)}.\]
{\it ii)} The operators $\phi^+(h)$ satisfy in the sense of quadratic forms on
$\cD(\phi^+(h_1))\cap\cD(\phi^+(h_2))$ the canonical commutation relations
\beq[\phi^+(h_2),\phi^+(h_1)]=\i {\rm Im}(h_2|h_1).\label{pop}\eeq
{\it iii)}
\[
\e^{\i tH}\phi^+(h)\e^{-\i tH}=\phi^+(h_{-t}).
\]
{\it iv)}  For  $p\in \nn$, there exists $n\in \nn$ such that for
$h_{i}\in \ch_{\rm c}(\omega),1\leq i\leq p$, 
$\cD(H^{n})\subset \cD(\Pi_{1}^{p}\phi^{+}(h_{i}))$,
\[
\mathop{\Pi}\limits_{i=1}^{p}\phi^{+}(h_{i})(H+\i)^{-n}=\slim_{t\fld
+\infty}\e^{\i tH}\mathop{\Pi}\limits_{i=1}^{p}\phi(h_{i,t})
\e^{-\i tH}(H+\i)^{-n},
\]
and the map
\[
\ch_{\rm c}(\omega)^{p}\ni (h_{1},\dots, h_{p})\mapsto
\mathop{\Pi}\limits_{i=1}^{p}\phi^{+}(h_{i})(H+\i)^{-n}\in B(\cH)
\]
is norm continuous. \label{4.2bis} \eet \proof The proof is very
similar to \cite[Thm. 10.2]{DG} so we will only sketch it.
Properties {\it i)} and {\it ii)} are  standard consequences of
the fact that the asymptotic Weyl operators define a regular CCR
representation  (see e.g. \cite[Sect. 2]{DG}).
 Property {\it iii)} follows from Thm. \ref{4.1bis} {\it
iv)}. It remains to prove {\it iv)}.  For fixed $p$ we pick $n\in \nn$
such that $N^{p/2}(H+b)^{-n}$ is bounded. It follows that
\beq\label{unifi}
\sup_{t\in \rr}
\|\e^{\i tH}\Pi_{1}^{p}\phi(h_{i,t})(H+b)^{-n}\e^{-\i tH}\|<\infty.
\eeq
Let us first establish the
existence of the strong limit
\beq
\slim_{t\fld +\infty}\e^{\i tH}\Pi_{1}^{p}\phi(h_{i,
t})(H+b)^{-n}\e^{-\i tH}= :R(h_{1},\ldots, h_{p}),\:
\hbox{ for }h_{i}\in \ch.
\label{e4.11}
\eeq
If $m$ is large enough such that $H= H_{0}+ V$ on $\cD(H^{m})$, then
as quadratic form on $\cD(H^{m})$ we have:
\[
\DD \Pi_{1}^{p}\phi(h_{i, t})(H+b)^{-n}= [V, \i
\Pi_{1}^{p}\phi(h_{i, t})](H+b)^{-n},
\]
where the Heisenberg derivative $\DD$ is defined in Subsect.
\ref{sec1.3bis}. Next:
\[
[V, \i \Pi_{1}^{p}\phi(h_{i,
t})](H+b)^{-n}= \sum_{j=1}^{p}\Pi_{1}^{j-1}\phi(h_{i,t})[V,
i\phi(h_{j,t})]\Pi_{j+1}^{p}\phi(h_{i,t})(H+b)^{-n},
\]
 as an operator identity on $\cD(H^{m})$.
The term
$[V, \i \phi(h_{t})]$ is by Prop. \ref{wick-recap-bis}
a sum of Wick monomials  with kernels of the form $w_{p,q}|h_{t})$ or
$(h_{t}|w_{p,q}$.

Arguing as in the proof of Thm. \ref{4.1bis}  we see from hypotheses
{\it (S)} and {\it (Is)} for $s>1$ that for $h\in \ch_{0}$
\beq
\|[V, \i \phi(h_{t})](H+b)^{-n}\|\in L^{1}(\d t).
\label{e4.11bis}
\eeq
This proves the
existence of the limit (\ref{e4.11}) for $u\in \cD(H^{m}), h_{i}\in
\ch_{0}$.
 The fact that the map
\beq\label{fifi}
\ch^{p}\ni (h_{1}, \dots, h_{p})\mapsto
\Pi_{j=1}^{p}\phi(h_{j})(H+b)^{-n}\in B(\cH)
\eeq
is norm continuous implies the existence of the limit for $u\in
\cD(H^{m})$ and $h_{i}\in \ch_{\rm c}(\omega)$.
The estimate (\ref{unifi})
shows the existence of (\ref{e4.11}) for all $u\in \cH$.

We prove now {\it iv)}. We recall that
\beq
\sup_{|s|\leq 1, \|h\|\leq
C}\Big\|\Big(\frac{W(sh)-\one}{s}\Big)(N+1)^{-1}\Big\|<\infty,
\label{e4.13}
\eeq
and
\beq
\lim_{s\fld 0}\sup_{\|h\|\leq C}\Big\|\Big(\frac{W(sh)-\one}{s}-
\i\phi(h)\Big)(N+1)^{-1}\Big\|=0.
\label{e4.14}
\eeq
We fix $P\in \nn$ and $M$ large enough so that $N^{P+1}(H+b)^{-M}$ is
bounded and prove {\it iv)} by induction on $1\leq p\leq P$.

 We have to
show that $\cD(H^{M})\subset \cD(\Pi_{1}^{p}\phi^{+}(h_{i}))$ and that
$R(h_{1}, \ldots, h_{p})= \Pi_{1}^{p}\phi^{+}(h_{i})(H+b)^{-M}$.
This amounts to show that
\[
R(h_{1}, \ldots, h_{p})= \slim_{s\fld
0}(\i s)^{-1}(W^{+}(sh_{1})-\one)\Pi_{2}^{p}\phi^{+}(h_{i})(H+b)^{-M}.
\]
Note that  by the induction assumption
$\cD(H^{M})\subset \cD(\Pi_{2}^{p}\phi^{+}(h_{i}))$ and
\beq
\Pi_{2}^{p}\phi^{+}(h_{i})(H+b)^{-M}= \slim_{t\fld
+\infty}\e^{\i tH}\Pi_{2}^{p}\phi(h_{i,t})\e^{-\i tH}(H+b)^{-M}.
\label{e4.12}
\eeq
Using (\ref{e4.12}) and the fact that $\e^{\i tH}W(h_{1,t})\e^{-\i tH}$ is
uniformly bounded in $t$, we have:
\[
\begin{array}{l}
(is)^{-1}(W^{+}(sh_{1})-\one)\Pi_{2}^{p}\phi^{+}(h_{i})(H+b)^{-M}\\[3mm]
= \slim\limits_{t\fld +\infty}\e^{\i tH}(\i s)^{-1}(W(sh_{1,t})-\one)
\Pi_{2}^{p}\phi(h_{i,t})\e^{-\i tH}(H+b)^{-M}.
\end{array}
\]
So to prove {\it iv)}, it suffices to check
 that
\beq
\slim_{s\fld 0}\slim_{t\fld \infty}\e^{\i tH}R(s,t)\e^{-\i tH}=0,
\label{e4.15}
\eeq
for
\[
R(s,t)=\Big(\frac{W(sh_{1,t})-\one}{s}- \i\phi(h_{1,t})\Big)
\Pi_{2}^{p}\phi(h_{i,t})(H+b)^{-M}.
\]
Using (\ref{e4.13}) and the higher order estimates, we see that $R(s,
t)$ is uniformly bounded for $|s|\leq 1, t\in \rr$, and using then
(\ref{e4.14}) we see that $\lim_{s\fld 0}\sup_{t\in \rr}\|R(s,
t)u\|=0$, for $u\in \cD(H^{M})$. This shows (\ref{e4.15}). The norm
continuity result in {\it iv)} follows from  the norm continuity of
the map (\ref{fifi}). \qed

\medskip

Finally the following theorem follows from Thm. \ref{4.2bis} as in
\cite[Subsect. 10.1]{DG}.

\bet {\it i)}
For any $h\in\ch_{\rm c}(\omega)$,
the {\em asymptotic creation and annihilation operators} defined on
 $\cD(a^{+\sharp}(h)):=\cD(\phi^+(h))\cap \cD(\phi^+(\i h))$
 by
\[
\begin{array}{l}
a^{+*}(h):=
\frac{1}{\sqrt{2}}\left(\phi^{+}(h)-\i\phi^{+}(\i h)\right), \\[3mm]
a^{+}(h):=\frac{1}{\sqrt{2}}\left(\phi^{+}(h)+\i \phi^{+}(\i h)\right).
\end{array}
\]
are closed.
\newline
 {\it ii)}
The operators $a^{+\sharp}$
 satisfy  in the sense of quadratic forms on
$\cD(a^{+\#}(h_1))\cap\cD(a^{+\#}(h_2))$
the canonical
commutation relations
\[\begin{array}{l}
[a^{+}(h_1),a^{+*}(h_2)]= (h_1|h_2)\one,\\[3mm]
[a^+(h_2),a^+(h_1)]=
[a^{+*}(h_2),a^{+*}(h_1)]= 0.\end{array}\]
{\it iii)}
\beq
\e^{\i tH}a^{+\sharp}(h)\e^{-\i tH}=a^{+\sharp}(h_{-t}).
\label{eas.5ter}
\eeq
\newline
{\it iv)} For $p\in \nn$, there exists $n\in \nn$ such that for
$h_{i}\in \ch_{\rm c}(\omega), 1\leq i\leq p$,  $\cD((H+\i)^{n})\subset
\cD(\Pi_{1}^{p}a^{+\sharp}(h_{i}))$ and
\[
\Pi_{1}^{p}a^{+\sharp}(h_{i})(H+b)^{-n}
=\slim_{t\to\infty}\e^{\i tH}\Pi_{1}^{p}a^{\sharp}(h_{i,t})
(H+b)^{-n}\e^{-\i tH}.
\]
\label{4.3}
\eet
\subsection{Asymptotic spaces and wave operators}\label{sec5.2}
In this subsection we recall the construction of asymptotic vacuum
spaces and wave  operators taken from \cite[Subsect. 10.2]{DG} and adapted
to our setup.

We define the  {\em asymptotic vacuum space}:
\[
\cK^{+}:= \{u\in \cH\ |\ a^{+}(h)u=0, \:  h\in \ch_{\rm c}(\omega)\}.
\]
The {\em asymptotic space} is defined as
\[\cH^+:=\cK^+\otimes\G(\ch_{\rm c}(\omega)).
\]
The proof of the following proposition is completely analogous to
\cite[Prop. 10.4]{DG}.
\begin{proposition} {\it i)}
$\cK^{+}$ is a closed $H-$invariant space.
\newline
{\it ii)}  $\cK^{+}$ is
included in the domain  of $\Pi_{1}^{p}a^{+\sharp}(h_{i})$ for
$h_{i}\in \ch_{\rm c}(\omega)$.
\newline
{\it iii)}
\[
\cH_\pp(H)\subset\cK^+.
\]
\label{as.4}
\end{proposition}
The {\em asymptotic Hamiltonian} is defined by
\[
H^+:=K^+\otimes\one+\one\otimes\d\Gamma(\omega),\hbox{ for }K^+:
=H\Big|_{\K^+}.
\]
We also define
\beq
\begin{array}{l}
\Omega^{+}:\cH^+\to\cH ,\\[3mm]
\Omega^+ \psi \otimes a^{*}(h_{1})\cdots
a^{*}(h_{p})\Omega:= a^{+*}(h_{1})\cdots a^{+*}(h_{p})\psi,
\ \ \ \ h_1,\dots,h_p\in\ch_{\rm c}(\omega),\ \ \  \psi\in\cK^+.
\end{array}
\label{eas.10a}
\eeq
The map $\Omega^{+}$ is called the {\em wave operator}.
The following theorem is analogous to \cite[Thm. 10.5]{DG}
\begin{theoreme}
\label{as.4bis}
 $\Omega^{+}$ is a unitary map from $\cH^{+}$ to
$\cH$ such that:
\[
\begin{array}{l}
a^{+\sharp}(h)\Omega^{+}= \Omega^{+}\one\otimes a^{\sharp}(h),\ \ \: h\in
\ch_{\rm c}(\omega),\\[3mm]
H\Omega^{+} = \Omega^{+}H^+.
\end{array}
\]
\label{4.6}
\end{theoreme}
\proof
By general properties of regular CCR representations,
(see \cite[Prop. 4.2]{DG}) the
operator $\Omega^+$ is well defined and isometric.
To prove that it is unitary, it suffices to
 show that the CCR representation
$\ch_{\rm c}(\omega) \ni h\mapsto W^{+}(h)$ admits a densely
defined number operator (see e.g. \cite[Subsect. 4.2]{DG}).

Let $n^{+}$ be the quadratic form associated to the CCR representation
$W^{+}$. Let us show that $\cD(n^{+})$ is
dense in $\cH$. We fix $n\in \nn$ such that
\[
a^{+}(h)(H+b)^{-n}=\slim_{t\to +\infty}\e^{\i tH}a(h_{t})\e^{-\i
tH}(H+b)^{-n}, \ \ h\in \ch_{\rm c}(\omega).
\]
For each finite dimensional space ${\mathfrak f}\subset
\ch_{\rm c}(\omega)$ set:
\[
n_{\cf}^{+}(u)=
\sum_{i=1}^{{\rm dim}\cf}\|a^{+}(h_{i})u\|^{2},
\]
for $\{h_{i}\}$ an orthonormal base of ${\mathfrak f}$. We have for
$u\in \cD(H^{n})$:
\[
\begin{array}{rl}

n_{\cf}^{+}(u)&= \lim\limits_{t\fld
+\infty}\sum\limits_{i=1}^{{\rm
dim}\cf}\|a(h_{i,t})\e^{-\i tH}u\|^{2} \\[3mm]&
 = \lim\limits_{t\fld +\infty}(\e^{-\i tH}u| \d\G(P_{\cf,t})\e^{-\i tH}u),
\end{array}
\]
if $P_{\cf,t}$ is the orthogonal projection  on $\e^{-\i t\omega}\cf$.
But $\d\Gamma(P_{\cf,t})\leq N$, so
\[
n_{\cf}^{+}(u)\leq \sup_{t}\|N^{\12}\e^{-\i tH}u\|^{2}\leq
C\|(H+b)^{p}u\|^{2},
\]
for some $p$, by the higher order estimates.
Therefore
\[
\cD(H^{p})\subset \cD(n^{+}),
\] which
for $p$ large enough, which implies that $\cD(n^{+})$  is densely defined.
\qed

\subsection{Extended wave operator}\label{sec5.3}
In Subsect. \ref{sec1.1} we introduced the scattering Hilbert space
$\cH^{\rm scatt}\subset \cH^{\ext}$. Clearly $\cH^{\rm scatt}$ is
preserved by $H^{\ext}$.
We see that $\cH^{+}$ is a subspace of
$\cH^{\rm scatt}$ and
\[
H^{+}=H^{\rm ext}_{\big|\cH^{+}}.
\]
We define the {\em extended wave operator} $\Omega^{\ext, +}:
\cD(\Omega^{\ext, +})\to \cH$ by:
\[
\cD(\Omega^{\ext, +})= D(H^{\infty})\otimes\G_{\rm fin}(\ch_{\rm
c}(\omega)),
\]
and
\[
\Omega^{\ext, +}\psi\otimes a^{*}(h_{1})\cdots a^{*}(h_{p})\Omega:
=a^{*+}(h_{1})\cdots a^{*+}(h_{p})\psi, \ \ \psi\in D(H^{\infty}), \ \
h_{i}\in \ch_{\rm c}(\omega).
\]
Note that $\Omega^{\ext, +}: \cH^{\rm scatt}\to \cH$ is unbounded and:
\[
\Omega^{+}=\Omega^{\ext, +}_{\big|\cH^{+}}.
\]
Considering $\Omega^{+}$ as a partial isometry equal to $0$ on
$\cH^{\rm scatt}\ominus \cH^{+}$, we can rewrite this identity as:
\beq
 \Omega^+=\Omega^{\ext,+}\one_{\cH^+},
\label{rew}\eeq
where $\one_{\cH^+}$ denotes the projection onto $\cH^+$ inside
the space $\cH^{\rm scatt}$.

Moreover using Thm. \ref{4.3} {\it iv)}, we obtain as in \cite[Thm.
10.7]{DG} the following
alternative expression for $\Omega^{\ext, +}$.
\begin{theoreme}
{\it i)} Let $u\in\cD(\Omega^{\ext,+})$. Then the limit
\[
\lim_{t\fld +\infty}\e^{\i tH}I\e^{-\i tH^{\ext}}u
\]
exists and equals $\Omega^{\ext,+}u$.
\newline{\it ii)} Let
$\chi\in C_0^\infty(\rr)$. Then
$\Ran\chi(H^\ext)\subset\cD(\Omega^{\ext, +})$,
$I\chi(H^\ext)$ and $\Omega^{\ext,+}\chi(H^\ext)$ are bounded
operators and
\beq
\slim_{t\fld +\infty}\e^{\i tH}I \e^{-\i tH^\ext}\chi(H^\ext)
=\Omega^{\ext,+}\chi(H^\ext).
\label{limiths}
\eeq
\label{huebner-spohn}
\end{theoreme}

\section{Propagation estimates}\label{sec6}\init
In this section we consider an abstract QFT Hamiltonian $H$ and
fix a weight operator $\x$.  We will prove various propagation
estimates for $H$. The proof of the phase-space estimates will be
 more involved than in \cite{DG}, \cite{DG1}. In fact the
operator playing the role of the acceleration $[\omega, \i
[\omega, \i \x]]$ vanishes in the situation considered in these
papers.

\subsection{Maximal velocity estimates}\label{sec6.1}
The following proposition shows that bosons cannot propagate in the
region $\x> v_{\rm max}t$ where
\[
v_{\rm max}:=\|[\omega, \i \x]\|.
\]
\begin{proposition}
Assume hypotheses {\it (G1)}, {\it (Is)} for $s>1$. Let $\chi\in \coinf(\rr)$.
Then for $R'>R>v_{\rm max}$, one has:
\[
\int_{1}^{\infty}\Big\|\dega{\one_{[R,
R']}(\rt)}^{\12}\chi(H)
\e^{-\i t H}u\Big\|^{2}\frac{\d t}{t}\leq C\|u\|^{2}.
\]
\label{6.1}
\end{proposition}
\proof The proof is almost identical to \cite[Prop. 11.2]{DG} so
we will only sketch it. We fix $G\in \coinf(]v_{\rm max},
+\infty[)$ with $G\geq \one_{[R, R']}$ and set
$F(s)=\int_{s}^{+\infty}G^{2}(t)\d t$. We use the propagation
observable $\Phi(t)= \chi(H)\d\G(F(\frac{\x}{t}))\chi(H)$. We use
that
\[
\begin{array}{rl}
\dd_{0} F(\frac{\x}{t})&= t^{-1}G(\frac{\x}{t})([\omega, \i
\x]-\frac{\x}{t})G(\frac{\x}{t})+ O(t^{-2})\\[3mm]
&\leq -\frac{C_{0}}{t}G^{2}(\frac{\x}{t}) + O(t^{-2})
\end{array}
\]
by Lemma \ref{gauge}.
The term $\chi(H)[V, \i \d\G(F(\frac{\x}{t}))]\chi(H)$ is $O(t^{-s})$
in norm by hypothesis {\it (Is)}, Lemma \ref{1.1} and the higher order
estimates. \qed
\subsection{Phase space propagation estimates}\label{sec6.2}
Set
\[
v:=[\omega, \i \x],
\]
and recall
from hypothesis {\it (G2)} that
\[
[\omega, \i v]= \gamma^{2}+ r_{-1-\epsilon},
\]
where $\gamma\in S^{-\12}_{\epsilon, (1)}$, $r_{-1-\epsilon}\in
S_{(0)}^{-1-\epsilon}$ for some $\epsilon>0$.

We will show that for free bosons the instantaneous velocity $v$
and the average velocity $\xt$ converge to each other when $t\to \pm
\infty$.
\begin{proposition}
Assume {\it (G1)}, {\it (G2)} and {\it (Is)} for $s>1$ and let $\chi\in \coinf(\rr)$ and $0<c_{0}<c_{1}$.
Then
\[
i)\ \ \int_{1}^{+\infty}\|\d\G\left((\xt -v)\one_{[c_{0}, c_{1}]}(\xt)(\xt
-v)\right)^{\12}\chi(H)\e^{-\i
tH}u\|^{2}\d t\leq C\|u\|^{2},
\]
\[
ii) \ \ \int_{1}^{+\infty}\|\d\G\left(\gamma\one_{[c_{0}, c_{1}]}(\xt)\gamma\right)^{\12}\chi(H)\e^{-\i
tH}u\|^{2}\frac{\d t}{t}\leq C\|u\|^{2}.
\]
\label{6.2}
\end{proposition}
\proof We follow the proof of \cite[Prop. 11.3]{DG}, \cite[Prop.
6.2]{DG1} with some modifications due to our abstract setting.

It clearly suffices to prove Prop. \ref{6.2} for $c_{1}>v_{\max}+1$,
which we will assume in what follows. We fix a function $F\in
\cinf(\rr)$, with $F, F'\geq 0$, $F(s)=0$ for $s\leq c_{0}/2$, $F(s),
F'(s)\geq d_{1}>0$ for $s\in [c_{0}, c_{1}]$. We set
\[
R_{0}(s)=\int_{0}^{s}F^{2}(t)\d t,
\]
so that $R_{0}(s)=0$ for $s\leq c_{0}/2$, $R_{0}'(s), R_{0}"(s)\geq
d_{2}>0$ for $s\in [c_{0}, c_{1}]$. Finally we fix another function
$G\in \cinf(\rr)$ with  $G(s)=1$ for $s\leq c_{1}+1$, $G(s)=0$ for
$s\geq c_{1}+ 2$, and set:
\[
R(s):= G(s)R_{0}(s).
\]
The function $R$ belongs to $\coinf(\rr)$ and  satisfies:
\beq\label{chi}
R(s)=0\hbox{ in }[0, c_{0}/2], \ \ R'(s)\geq d_{3}\one_{[c_{0},
c_{1}]}(s)+ \chi_{1}(s),
\ \ R"(s)\geq d_{3}\one_{[c_{0},
c_{1}]}(s)+ \chi_{2}(s),
\eeq
for $\chi_{1}, \chi_{2}\in \coinf(]v_{\rm max}, +\infty[)$ and
$d_{3}>0$. We set
\[
b(t):=R(\xt)-\12\left(R'(\xt)(\xt -v)+ \hc \right),
\]
which satisfies $b(t)\in O(1)$ and use the propagation observable
\[
\Phi(t)= \chi(H)\d\G(b(t))\chi(H).
\]
Using Lemma \ref{gauge} we obtain that:
\beq\label{e6.1}
\p_{t}b(t)=\frac{1}{t}\left( R"(\xt)\frac{\x^{2}}{t^{2}}
-\12 \xt R"(\xt) v -\12 vR"(\xt)\xt\right)+ O(t^{-2}),
\eeq
and
\beq\label{e6.2}
\begin{array}{rl}
[\omega, \i b(t)]=&\frac{1}{t}\left(v R"(\xt)v -\12 \xt R"(\xt)v-\12
vR"(\xt)\xt\right)\\[3mm]
&+ \12 \left(R'(\xt)[\omega, \i v] +\hc\right)+ O(t^{-2}).
\end{array}
\eeq
 Adding (\ref{e6.1}) and (\ref{e6.2}) we obtain:
\[
\dd_{0}b(t)=\frac{1}{t}(\xt -v)R"(\xt)(\xt -v)+ \12
\left(R'(\xt)[\omega, \i v]+ \hc\right)+ O(t^{-2}).
\]
By hypothesis {\it (G2)}, we have:
\[
[\omega, \ i v]= \gamma^{2}+ r_{-1-\epsilon},
\]
for $\gamma\in S^{-\12}_{\epsilon, (1)}$, $r_{-1-\epsilon}\in
S^{-1-\epsilon}_{(0)}$.  Since $0\not\in \supp R'$, we know by Lemma
\ref{gototime} that
\[
R'(\xt)r_{-1-\epsilon}\in O(t^{-1-\epsilon}).
\]
Using that $\gamma\in S^{-\12}_{\epsilon, (1)}$, we get by Lemma
\ref{gauge} {\it vii)} that:
\[
\12 \left(R'(\xt)\gamma^{2}+\hc\right)=  \gamma R'(\xt)\gamma+
O(t^{-3/2+ \epsilon}).
\]
Finally this gives:
\[
\dd_{0}b(t)=\frac{1}{t}(\xt -v)R"(\xt)(\xt -v)+ \gamma
R'(\xt)\gamma+ O(t^{-1-\epsilon_{1}}),
\]
for some $\epsilon_{1}>0$.

 We note that $R'$ and $R"$ are positive,
except for the error terms due to $\chi_{1}, \chi_{2}$ in (\ref{chi}).
To handle these terms we pick $\chi_{3}\in \coinf(]v_{\rm max},
+\infty[)$ such that $\chi_{3}\chi_{i}= \chi_{i}$, $i=1,2$.
Then $[\xt -v, \chi_{3}(\xt)]\in
O(t^{-1})$ and $[\gamma, \chi_{3}(\xt)]\in O(t^{-3/2+ \epsilon})$ by
Lemma \ref{gauge} {\it i)} and {\it vii)}. This yields:
\[
\begin{array}{rl}
\pm \frac{1}{t}(\xt -v)\chi_{2}(\xt)(\xt -v)
=&\pm \frac{1}{t}\chi_{3}(\xt)(\xt -v)\chi_{2}(\xt)(\xt-v)\chi_{3}(\xt)+O(t^{-2})\\[3mm]
\leq& \frac{C}{t}\chi_{3}^{2}(\xt)+ O(t^{-2}),\\[3mm]
\end{array}
\]
\[
\begin{array}{rl}
\pm \gamma \chi_{1}(\xt)\gamma
 =& \pm \chi_{3}(\xt) \gamma
\chi_{1}(\xt) \gamma \chi_{3}(\xt)+ O(t^{-3/2+ \epsilon})\\[3mm]
\leq & \frac{C}{t}\chi_{3}^{2}(\xt)+ O(t^{-3/2+ \epsilon}),
\end{array}
\]
using that $\gamma\in S^{-\12}_{(0)}$ and Lemma \ref{gototime}.
Using again (\ref{chi}), we finally get:
\beq
\begin{array}{rl}
\dd_{0}b(t)\geq& \frac{C_{1}}{t}(\xt -v)\one_{[c_{0}, c_{1}]}(\xt)(\xt
-v)+ C_{1} \gamma\one_{[c_{0}, c_{1}]}(\xt)\gamma \\[3mm]
&-\frac{C_{2}}{t} \chi_{3}^{2}(\xt)+ O(t^{-1-\epsilon_{1}}),
\end{array}
\label{chichi}
\eeq
for some $C_{1}, \epsilon_{1}>0$.

To handle the commutator $[V, \i \d\G(b(t))]$ we note that using Lemma
\ref{gauge} {\it iv)} and the fact that $0\not\in \supp R$, we have
\[
b(t)= \one_{[\epsilon, +\infty[}(\xt)b(t)\one_{[\epsilon, +\infty[}(\xt)+ O(t^{-2})
\]
for some $\epsilon>0$. Using also hypothesis {\it (Is)} for $s>1$,
this implies that if $V=\Wick(w)$ then $\triple\d\G(b(t))w\triple\in L^{1}(\d
t)$. Using the higher order estimates this implies that
\[
\|\chi(H)[V,
\i \d\G(b(t))\chi(H)]\|\in L^{1}(\d t).
\] The rest of the proof is as
in \cite [Prop. 11.3]{DG}. \qed

\subsection{Improved phase space propagation estimates}\label{sec6.3}
In this subsection we will prove improved propagation estimates. We
will use the  following lemma which is an analog of \cite[Lemma 6.4]{DG1} in our
abstract setting.  Its proof will be given in the Appendix.
\begin{lemma}\label{painful}
Assume {\it (H1)}, {\it (G1)}, {\it (G2)} and set $v=[\omega,
\i\x]$ which is a bounded operator on $\ch$. Let $c=(\xt -v)^{2}+
t^{-\delta}$, $\delta>0$ and set $\epsilon_{0}=\inf(\delta,
1-\delta/2)$.  If $J\in \coinf(\rr)$ then:
\[
i)\ \ J(\xt)c^{\12}\in O(1),
\]
\[
ii)\ \ [c^{\12}, J(\xt)]\in O(t^{-1+ \delta/2}).
\]
If $J\in \coinf(\rr\backslash \{0\})$ then for $\delta$ small enough:
\[
iii)\ \ \begin{array}{rl}& J(\xt)\dd_{0}c^{\12}J(\xt)\\[3mm]
=& -\frac{1}{t}J(\xt)c^{\12}J(\xt)+\gamma J(\xt) M(t)J(\xt) \gamma +
O(t^{-1-\epsilon_{1}}),
\end{array}
\]
where $\epsilon_{1}>0$ and $M(t)\in O(1)$.\\
If $J,J_1 \in \coinf(\rr)$ and $J_1\equiv 1$ on $\supp J$, then:
\[
iv)\ \ |J(\xt)(\xt-v)+ \hc|\leq
CJ_{1}(\xt)c^{\12}J_{1}(\xt)+O(t^{-\epsilon_{0}}).
\]
If $J, J_{1}, J_{2}\in \coinf(\rr)$ with $J_{2}\equiv 1$ on $\supp
J$ and $\supp J_1$, then:
\[
v)\ \ \pm (J(\xt)(\xt -v)c^{\12}J_{1}(\xt)+\hc)\leq C(\xt
-v)J_{2}^{2}(\xt)(\xt -v)+ O(t^{-\epsilon_{0}}).
\]
\end{lemma}

\begin{proposition}\label{6.3}
Assume {\it (G1)}, {\it (G2)}, {\it (Is)} for $s>1$.
Let $J\in \coinf(]c_{0}, c_{1}[)$ for $0<c_{0}<c_{1}$ and
$\chi\in\coinf(\rr)$. Then:
\[
\int_{1}^{+\infty}\|\d\G(\left|J(\xt)(\xt -v)+
\hc\right|)^{\12}\chi(H)\e^{-\i tH}u\|^{2}\frac{\d t}{t}\leq C\|u\|^{2}.
\]
\end{proposition}
\proof We fix
$J_{1}\in\coinf(]c_{0}, c_{1}[)$ with $J_1\equiv 1$ on $\supp J$ and
set
\[
b(t)= J_{1}(\xt)c^{\12}J_{1}(\xt), \hbox{ for }c=(\xt -v)^{2}+ t^{-\delta},
\]
and  $\delta>0$ will be chosen small enough later.  We
will use the propagation observable
\[
\Phi(t)= \chi(H)\d\G(b(t))\chi(H).
\]
Note that by Lemma \ref{painful} {\it i)} and the higher order
estimates $b(t), \:\Phi(t)\in O(1)$. We first note that
\[
\chi(H)[V, \i
\d\G(b(t))]\chi(H)\in O(t^{-s}),
\]
using hypothesis {\it (Is)} and Lemma \ref{painful} {\it i)}.
Next
\[
\DD_{0}\d\G(b(t))= \d\G(\dd_{0}b(t)),
\]
\[
\dd_{0}b(t)= \left(\dd_{0}J_{1}(\xt)\right)c^{\12}J_{1}(\xt)+ \hc +
J_{1}(\xt)(\dd_{0}c^{\12})J_{1}(\xt).
\]
By Lemma \ref{painful} {\it iii)} we know that choosing $\delta$ small
enough:
\[
\begin{array}{rl}
&J_{1}(\xt)(\dd_{0}c^{\12})J_{1}\xt)\\[3mm]
=&-J_{1}(\xt)\frac{c^{\12}}{t}J_{1}(\xt)
+ \gamma J_{1}(\xt) M(t)J_{1}(\xt)\gamma + O(t^{-1-\epsilon_{1}}),
\end{array}
\]
for some $\epsilon_{1}>0$ and $M(t)\in O(1)$. By Lemma \ref{painful} {\it iv)} we get
then that
\[
\begin{array}{rl}
&-J_{1}(\xt)(\dd_{0}c^{\12})J_{1}(\xt)\\[3mm]
\geq& \frac{C}{t}\left|J(\xt)(\xt -v)+
\hc\right|-C \gamma J_{1}^{2}(\xt)\gamma -Ct^{-1 -\epsilon_{1}}
\end{array}
\]
for some $\epsilon_{1}>0$. Next by Lemma \ref{gauge}:
\[
\dd_{0}J_{1}(\xt)= -\frac{1}{2t}J_{1}'(\xt)(\xt -v)
+ O(t^{-2}),
\]
which by Lemma \ref{painful} {\it v)} gives for  $J_{2}\in
\coinf(]c_{0}, c_{1}[)$ and $J_{2}\equiv 1$ on $\supp J_{1}$:
\[
\left(\dd_{0}J_{1}(\xt)\right)c^{\12}J_{1}(\xt)+ \hc \geq
-\frac{C}{t}(\xt -v)J^2_{2}(\xt)(\xt -v) + O(t^{-1-\epsilon_{1}})
\]
for some $\epsilon_{1}>0$. Collecting the various estimates, we
obtain finally
\[
-\DD \Phi(t)\geq  \frac{C}{t} \chi(H)\d\G(\left|J(\xt)(\xt -v)+
\hc\right|)\chi(H)-C R_{1}(t)-CR_{2}(t)+ O(t^{-1-\epsilon_{1}}),
\]
where
\[
R_{1}(t)= \chi(H)\d\G(\gamma J_{1}^{2}(\xt)\gamma)\chi(H), \ \
R_{2}(t)= \frac{1}{t}\chi(H)\d\G((\xt -v)J^2_{2}(\xt)(\xt
-v))\chi(H)
\]
are integrable along the evolution by Prop. \ref{6.2}.
 We can then complete the proof as in
\cite[Prop. 6.3]{DG1}. \qed
\subsection{Minimal velocity estimate}\label{sec6.4}
In this subsection we prove the minimal velocity estimate. It says
that for states with energy away from thresholds and eigenvalues of
$H$, at least one boson should escape to infinity.
We recall that as in Subsect. \ref{sec4.4}, $A=\d\G(a)$.

\begin{lemma}\label{6.4}
Let $H$ be an abstract QFT Hamiltonian. Assume {\it (G4)}.
Let $k\in \nn$, $m=1,2$ and $\chi\in \coinf(\rr)$. Then there exists $C$ such that for any $\epsilon>0$ and $q\in
\coinf([-2\epsilon, 2\epsilon])$ with $0\leq q\leq 1$ one has:
\[
\|N^{k}\frac{A^{m}}{t^{m}}\G(q^{t})\chi(H)\|\leq C \epsilon^{m}.
\]
where $q^{t}= q(\xt)$.
\end{lemma}
\proof
Applying Prop. \ref{stup1} {\it ii)} we get
\beq\label{e6.02}
(\d\G(a))^{2m}\leq N^{2m-1}\d\G(a^{2m}).
\eeq
Next
\beq\label{e6.03}
\G(q^{t})\d\G(a^{2m})\G(q^{t})= \d\G((q^{t})^{2},
q^{t}a^{2m}q^{t})\leq \d\G(q^{t}a^{2m}q^{t}),
\eeq
by Prop. \ref{stup1} {\it iv)}. We write using {\it (G4)}:
\[
q^{t}a^{2m}q^{t}= G^{t}\x^{-m}a^{2m}\x^{-m}G^{t}\leq
Ct^{2m}(G^{t})^{2},\ \ m=1,2,
\]
for $G^{t}= G(\xt)$ and $G(s)= s^{m}q(s)$. Using that $|G(s)|\leq
C\epsilon^{m}$ we obtain that
\beq
q^{t}a^{2m}q^{t}\leq C\epsilon^{2m}t^{2m}, \ \ m=1,2.
\label{e6.01}
\eeq
From (\ref{e6.01}) and (\ref{e6.02}), (\ref{e6.03}) we obtain
\beq\label{clay}
\G(q^{t})N^{2k}\d\G(a)^{2m}\G(q^{t})\leq C \epsilon^{2m}t^{2m}N^{2k+2m}.
\eeq
This implies the Lemma using the higher order estimates. \qed

\medskip

\begin{proposition}\label{6.5}
Let $H$ be an abstract QFT Hamiltonian. Assume hypotheses {\it (Gi)},
for $1\leq i\leq 5$, {\it (M1)}, {\it (M2)}, {\it
(Is)} for $s>1$.
Let $\chi\in \coinf(\rr)$ be supported in
$\rr\backslash(\tau\cup \sigma_{\rm pp}(H))$. Then there exists
$\epsilon>0$ such that:
\[
\int_{1}^{+\infty}\left\|\Gamma\left(\one_{[0,\epsilon]}\left(\frac{|x|}{t}
\right)\right)
\chi(H)\e^{-\i t H}u\right\|^{2}\frac{\d t}{t}
\leq C\|u\|^{2}.
\]
\end{proposition}

\proof Let us first prove the proposition for $\chi$ supported near an
energy level $\lambda\in \rr\backslash \tau\cup \sigma_{\rm pp}(H)$.
By Thm. \ref{mourre}, we can find $\chi\in \coinf(\rr)$ equal to $1$
near $\lambda$ such that for some $c_{0}>0$:
\beq
\chi(H)[H, \i A]_{0}\chi(H)\geq c_{0}\chi^{2}(H).
\label{e6.9}
\eeq
Let $\epsilon>0$ be a parameter which will be fixed later. Let $q\in
\coinf(|s|\leq 2\epsilon)$, $0\leq q\leq 1$, $q=1$ near $\{|s|\leq
\epsilon\}$ and let $q^{t}= q(\xt)$.

We use the propagation observable
\[
\Phi(t):= \chi(H)\G(q^{t})\frac{A}{t}\G(q^{t})\chi(H).
\]
We fix  cutoff functions $\tilde{q}\in \coinf(\rr)$, $\tilde{\chi}\in
\coinf(\rr)$ such that
\[
\supp \tilde{q}\subset[-4\epsilon, 4\epsilon],\ 0\leq \tilde{q}\leq 1, \
\tilde{q}q=q,\ \tilde{\chi}\chi=\chi.
\]
By Lemma \ref{6.4} for $m=1$ the observable $\Phi(t)$ is uniformly bounded. We have:
\beq
\begin{array}{rl}
\DD\Phi(t)&= \chi(H)\d\Gamma(q^{t},
\dd_{0}q^{t})\frac{A}{t}\Gamma(q^{t})\chi(H)+ \hc\\[3mm]
&+ \chi(H)[V, \i\Gamma(q^{t})] \frac{A}{t}\Gamma(q^{t})\chi(H)+
\hc\\[3mm]
&+t^{-1}\chi(H)\Gamma(q^{t})[H,
\i A]\Gamma(q^{t})\chi(H)\\[3mm]
&-t^{-1}\chi(H)\Gamma(q^{t})
\frac{A}{t}\Gamma(q^{t})\chi(H)\\[3mm]
&=: R_{1}(t)+ R_{2}(t)+ R_{3}(t)+R_4(t).
\end{array}
\label{e6.3}
\eeq
We have used the fact, shown in the proof of Lemma \ref{commut-1}, that
$\G(q^{t})$ preserves $\cD(H_{0})$ and $\cD(N^{n})$
to expand the commutator $[H, \i \Phi(t)]$ in (\ref{e6.3}).

Let us first estimate $R_{2}(t)$. By Prop.  \ref{wick-recap} and
hypothesis {\it (Is)}
\[
[V, \i \G(q^{t})]\in (N+1)^{n} O_N(t^{-s}), \: s>1,
\]
for some $n$. Therefore by the higher order estimates and Lemma
\ref{6.4} for $m=1$:
\beq
R_{2}(t)\in O(t^{-s}), \: s>1.
\label{e6.4}
\eeq
We estimate now $R_{1}(t)$. By Lemma \ref{gauge} {\it i)}:
\[
\dd_{0}q^{t}= -\frac{1}{2t}\left((\xt -v)q'(\xt) +\hc\right) + r^t=:\frac{1}{t} g^{t}+ r^t,
\]
where $r^t\in O(t^{-2})$. By the higher order estimates
$\|\chi(H)d\G(q^{t}, r^{t})\|\in O(t^{-2})$, which using Lemma
\ref{6.4}  for $m=1$ yields
\[
\|\chi(H)d\G(q^{t}, r^{t})\frac{A}{t}\G(q^{t})\chi(H)\|\in O(t^{-2}).
\]
Then we set
\[
B_{1}:=\chi(H)\d\Gamma(q^{t}, g^{t})(N+1)^{-\12}, \ \ \ B_{2}^{*}:=
(N+1)^{\12} \frac{A}{t}\Gamma(q^{t})\chi(H),
\]
and use the inequality
\beq
\begin{array}{rl}
\chi(H)\d\G(q^{t}, g^{t})\frac{A}{t}\G(q^{t})\chi(H)+\hc&
=B_{1}B_{2}^{*}+ B_{2}B_{1}^{*}\\[3mm] \geq
 -B_{1}B_{1}^{*}
-B_{2}B_{2}^{*}.
\end{array}
\label{Hs129}
\eeq
We can write:
\beq
\begin{array}{rl}
-B_{2}B_{2}^{*}
&=-\chi(H)\tilde{\chi}(H)\Gamma(q^{t})\Gamma(\tilde{q}^{t})\frac{A^{2}}{t^{2}}(N+1)\Gamma(\tilde{q}^{t})\Gamma(q^{t})\tilde{\chi}(H)\chi(H)\\[3mm]
&=\chi(H)\Gamma(q^{t})\tilde\chi(H)
\Gamma(\tilde q^t)\frac{A^{2}}{t^{2}}(N+1)
\Gamma(\tilde q^t)\tilde\chi(H)\Gamma(q^{t})\chi(H) +O(t^{-1})\\[3mm]
&\geq-\epsilon^2C_{1}\chi(H)\Gamma^2(q^t)\chi(H)+O(t^{-1}).
\end{array}
\label{e6.7}
\eeq
In  the first step we use that $[\tilde{\chi}(H), \G(q^{t})]\in
O(t^{-1})$ by
 Lemma \ref{commut-1} and that
$\frac{A^{2}}{t^{2}}(N+1)\Gamma(q^{t})\chi(H)\in O(1)$ by Lemma
\ref{6.4} for $m=2$. In the second step
we use the following estimate analogous to (\ref{clay}):
\[
\tilde\chi(H)
\Gamma(\tilde q^t)\frac{A^{2}}{t^{2}}(N+1)
\Gamma(\tilde q^t)\tilde\chi(H)\leq C_{1}\epsilon^2.
\]

Next we use  Prop. \ref{stup1} {\it iv)} to obtain:
\[
\begin{array}{rl}
B_{1}^{*}B_{1}=&\chi(H)\d\G(q^{t}, g^{t})^{2}(N+1)^{-1}\chi(H)\\[3mm]
\leq& \chi(H)\d\G((g^{t})^{2})\chi(H).
\end{array}
\]
By  Prop. \ref{6.2}, we obtain
\beq
\int_{1}^{+\infty}\|B_{1}\e^{-\i tH}u\|^{2}\frac{\d t}{t}\leq C\|u\|^{2}.
\label{e6.6}
\eeq
To handle $R_{3}(t)$, we  write using Lemma \ref{commut-1}:
\beq
\begin{array}{rl}
R_3(t)&
=t^{-1}\Gamma(q^{t})\chi(H)[H,\i A]\chi(H)\Gamma(q^{t}) +
O(t^{-2})\\[3mm]&
\geq C_{0}t^{-1}\Gamma(q^{t})\chi^{2}(H)\Gamma(q^{t}) -Ct^{-2}\\[3mm]&
\geq C_{0}t^{-1}\chi(H)\Gamma^{2}(q^{t})\chi(H)-Ct^{-2}.
\end{array}
\label{e6.9bis}
\eeq
It remains to estimate $R_{4}(t)$. We write using Lemma \ref{6.4}:
\beq
\begin{array}{rl}
R_{4}(t)&= -t^{-1}\chi(H)\G(q^{t})\frac{A}{t}\G(q^{t})\chi(H)\\[3mm]
&= -t^{-1}\chi(H)\Gamma(q^t)\tilde\chi(H)\G(\tilde q^{t})\frac{A}{t}
\G(\tilde q^{t})\tilde\chi(H)\Gamma(q^t)\chi(H)+ O(t^{-2})\\[3mm]&
\geq -\epsilon C_{2}t^{-1}\chi(H)\Gamma(q^t)^2\chi(H)+O(t^{-2}).
\end{array}
\label{P51D}
\eeq
Collecting (\ref{e6.7}), (\ref{e6.9bis}) and (\ref{P51D}),
we obtain
\beq
\begin{array}{l}
-t^{-1}B_2^*(t)B_2(t)+R_{3}(t)+R_4(t)\\[3mm]
\geq (-\epsilon^2 C_1+C_0-\epsilon C_2)
 t^{-1}\chi(H)\Gamma(q^{t})^2\chi(H)+ O(t^{-2}).
\end{array}
\label{Fw190}
\eeq
We pick now $\epsilon$ small enough
so that $\tilde C_0=-\epsilon^2 C_1+C_0-\epsilon C_2>0$. Using
(\ref{e6.4}),  (\ref{e6.6}) and (\ref{Fw190}) we conclude that
\[
\DD\Phi(t)\geq \frac{\tilde C_{0}}{t}\chi(H)\Gamma^{2}(q^{t})\chi(H) -R(t)
-Ct^{-s},\: s>1.
\]
where $R(t)$ is integrable along the evolution. We finish the proof as
in \cite[Prop. 11.5]{DG}.
\qed

\section{Asymptotic Completeness}\label{sec7}\init
In this section we prove the asymptotic completeness of wave
operators. The first step is the {\em geometric asymptotic
completeness}, identifying  the asymptotic vacua with the subspace of
states living at large times $t$ in $\x\leq \epsilon t$ for
arbitrarily small $\epsilon>0$. In the second step, using the minimal
velocity estimate, one shows that these states have to be bound states
of $H$.
\subsection{Existence of asymptotic localizations}\label{sec7.1}
\bet\label{7.1}
Let $H$ be an abstract QFT Hamiltonian. Assume hypotheses {\it (G1)},
{\it (G2)}, {\it (Is)} for $s> 1$.
Let $q\in C_0^\infty(\rr)$, $0\leq q\leq 1$, $q=1$ on a neighborhood
of zero. Set $q^t=q(\xt)$. Then there exists
\beq
\slim_{t\to\infty}\e^{\i tH}\Gamma(q^t)\e^{-\i tH}
=:\Gamma^+(q).\label{asym}
\eeq
We have
\beq
\Gamma^+(q\tilde q)=\Gamma^+(q)\Gamma^+(\tilde q),
\label{com1}
\eeq
\beq
0\leq\Gamma^+(q)\leq\Gamma^+(\tilde q)\leq \one,\hbox{ if } 0\leq q\leq
\tilde q\leq 1,
\label{com1bis}
\eeq
\beq[H,\Gamma^+(q)]=0.\label{com2}\eeq
\eet
 The proof is completely similar to the proof of \cite[Thm.
12.1]{DG}, using Prop. \ref{6.3}.
An analogous result  is true for the free Hamiltonian $H_{0}$.
\begin{proposition}\label{7.2}
 Assume hypotheses {\it (H1)}, {\it (G1)},
{\it (G2)}.
Let $q\in C^\infty(\rr)$, $0\leq q\leq 1$, $q\equiv 1$ near $\infty$.
 Then there exists
\beq
\slim_{t\to\infty}\e^{\i tH_{0}}\Gamma(q^t)\e^{-\i tH_{0}}
=:\Gamma_{\rm free}^+(q).\label{asym-bis}
\eeq
Moreover if additionally $q\equiv 0$ near $0$ then:
\beq
\Gamma_{\rm free}^+(q)= \Gamma_{\rm free}^+(q)\G(\one_{\rm
c}(\omega)),
\label{symsym}
\eeq
where $\one_{\rm c}(\omega)$ is the projection on the continuous
spectral subspace of $\omega$.
\end{proposition}
\proof  By density it suffices to the existence of the limit
(\ref{asym-bis}) on $\G_{\rm fin}(\ch)$.

Using the  identity  (see e.g. \cite[Lemma 3.4]{DG}):
\[
\frac{\d }{\d t}\G(r_{t})= \d\G(r_{t}, r'_{t}),
\]
we obtain for $a, b\in B(\ch)$:
\[
\G(a)-\G(b)=\int_{0}^{1}\d\G(ta+ (1-t)b, a-b)\d t.
\]
It follows then from Prop.\ref{stup1} that
\[
B(\ch)\ni a\mapsto \G(a)(N+1)^{-1} \in B(\G(\ch))
\]
is norm continuous.  This implies that it suffices to prove the
existence of the limit for $q\in \cinf(\rr)$ $0\leq q\leq 1$
and $q\equiv 1$ near $\infty$, $q\equiv Cst$ near $0$. In particular
$q'\in \coinf(\rr\backslash\{0\})$.
We can then repeat the proof of \cite[Thm.
12.1]{DG}, noting that the only place where $q\equiv 1$ near $0$ is
needed is to control the commutator $[V, \i \G(q^{t})]$ which is
absent in our case. This proves (\ref{asym-bis}).
Restricting (\ref{asym-bis}) to the one-particle sector we obtain the
existence of
\beq\label{asymsym}
q^{+}:=\slim_{t\to +\infty}\e^{\i t \omega}q^{t}\e^{-\i t \omega}.
\eeq
By Lemma \ref{gauge} {\it i)}, we see that $[\chi(\omega), q^{+}]=0$
for each $\chi\in \coinf(\rr)$ hence $q^{+}$ commutes with $\omega$.

If $q\equiv 0$ near $0$ then clearly
\[
\one_{\rm pp}(\omega)q^{+}=q^{+}\one_{\rm pp}(\omega)=0, \hbox{ and
hence }q^{+}= q^{+}\one_{\rm c}(\omega)= \one_{\rm c}(\omega)q^{+}.
\]
We note now that
\[
\G^{+}_{\rm free}(q)= \Gamma(q^{+}),
\]
which implies (\ref{symsym}).  \qed
\subsection{The projection $P_{0}^{+}$.}\label{sec7.2}
\bet
Let $H$ be an abstract QFT Hamiltonian. Assume hypotheses {\it (G1)},
{\it (G2)}, {\it (Is)} for $s> 1$.
Let $\{q_{n}\}\in C_0^\infty(\rr)$ be a decreasing
sequence of functions such that $0\leq q_{n}\leq 1$, $q_{n}=\equiv 1$
on a neighborhood
of $0$ and $\cap_{n=1}^\infty\supp q_n=\{0\}$.
Then
\beq
P_0^+:=\slim_{n\to\infty}\Gamma^+(q_{n})\hbox{ exists}.
\label{pro1}
\eeq
$P_0^+$ is an orthogonal projection independent on the choice of the
sequence $\{q_{n}\}$.  Moreover:
\[
[H,P_0^+]=0.
\]
Moreover if {\it (S)} holds:
\beq
\Ran P_0^+\subset\cK^+.
\label{incl}
\eeq
\label{P61}
\eet
The range of  $P_{0}^{+}$ can be interpreted as the space of
states asymptotically containing no bosons away from the origin.

\proof The proof is analogous to \cite[Thm. 12.3]{DG}. We will
only detail (\ref{incl}). Let $n\in \nn$ such that
$\cD(H^{n})\subset \cD(a^{+ *}(h))$ for all $h\in \ch_{\rm
c}(\omega)$. We will show that for $u\in {\rm Ran}P_{0}^{+}$:
\[
(H+b)^{-n}a^{+}(h)u=0, \ \ h\in \ch_{\rm c}(\omega).
\]
Since $h\mapsto (H+b)^{-n}a^{+}(h)$ is norm continuous by Thm.
\ref{4.2bis}, we can assume that $h\in \ch_{0}$. By {\it (S)} and
the fact that $u\in {\rm Ran}P_{0}^{+}$ we can  choose $q\in
\coinf(\rr)$ with $0\leq q\leq 1$ such that:
\[
u=\lim_{t\to +\infty}\e^{\i tH}\G(q^{t})\e^{-\i tH} u, \ \
q^{t}h_{t}\in o(1).
\]
Then:
\[
\begin{array}{rl}
(H+b)^{-n}a^{+}(h)u=&\lim_{t\to +\infty}\e^{\i
tH}(H+b)^{-n}a(h_{t})\G(q^{t})\e^{-\i tH}u\\[3mm]
=&\lim_{t\to +\infty}\e^{\i tH}(H+b)^{-n}\G(q^{t})a(q^{t}h_{t})\e^{-\i
tH}u\\[3mm]
=&0,
\end{array}
\]
using that $(N+1)^{-1}a(q^{t}h_{t})\in o(1)$ and the higher order
estimates. \qed

\subsection{Geometric inverse wave operators}\label{sec7.3}
Let $j_0\in C_0^\infty(\rr)$,
$j_\infty\in C^\infty(\rr)$, $0\leq j_0,
j_\infty$, $j_0^2+j_\infty^2\leq 1$, $j_0=1$ near $0$
(and hence $j_\infty=0$
near $0$). Set $j:=(j_0,j_\infty)$, $j^t=(j_0^t,j_\infty^t)$.

As in Subsect. \ref{sec1.1}, we introduce
 the operator $I(j^t):
\cH^{\ext}\to\cH$.
\bet
Assume {\it (G1)}, {\it (G2)},  {\it (Is)} for $s>1$. Then:

{\it i)} The following limits exist:
\beq
\slim_{t\fld +\infty}\e^{\i tH^\ext}I^*(j^t) \e^{-\i tH},
\label{gg1}\eeq
\beq\slim_{t\fld +\infty}\e^{\i tH} I(j^t)\e^{-\i tH^\ext}.
\label{gg2}\eeq
If we denote (\ref{gg1}) by $\invWave^+(j)$, then (\ref{gg2}) equals
$\invWave^+(j)^*$ and $\|\invWave^+(j)\|\leq 1$.
\newline {\it ii)} For any bounded Borel function $F$ one has
\[
\begin{array}{l}
\invWave^+(j)F(H)= F(H^\ext)\invWave^+(j).
\end{array}
\]
 {\it iii)} Let $q_0,q_\infty\in C^\infty(\rr)$,
 $\nabla q_0,\nabla q_\infty\in C_0^\infty(\rr)$,
$0\leq q_0,q_\infty\leq 1$, $q_0\equiv 1$ near $0$ and
$q_{\infty}\equiv 1$ near $\infty$. Set
$\tilde j:=(\tilde j_0,\tilde j_\infty):=(q_0j_0,q_\infty j_\infty)$. Then
\[
\Gamma^+(q_0)\otimes\Gamma^{+}_{\rm free}(q_{\infty})\invWave^+(j)=
\invWave^+(\tilde j).
\]
{\it iv)}
Assume additionally that   $j_0+j_\infty= 1$. Then ${\rm
Ran}\invWave^+(j)\subset \cH^{\rm scatt}$ and if $\chi\in C_0^\infty(\rr)$:
\[\Omega^{\ext,+}\chi(H^\ext)\invWave^+(j)=\chi(H).
\]
\label{supermain}
\eet
Note that statement  {\it iv)} of Thm. \ref{supermain} makes sense
since ${\rm Ran}\invWave^+(j)\subset \cH^{\rm scatt}$ and
$\chi(H^{\ext})$ preserves $\cH^{\rm scatt}$.

\proof
Statements {\it i)}, {\it ii)}, {\it iii)} are proved exactly as in
\cite[Thm. 12.4]{DG}, we detail only {\it iv)}.

We pick $q_{\infty}\in \cinf(\rr)$ with $q_{\infty}\equiv 1$ near
$\infty$, $q_{\infty}\equiv 0$ near $0$ and
$q_{\infty}j_{\infty}=j_{\infty}$. Applying {\it iii)} for
$q_{0}\equiv 1$, we obtain by {\it iii)} that $\one\otimes
\G^{+}_{\rm free}(q_{\infty})\invWave^+(j)=\invWave^+(j)$.
Applying then  (\ref{symsym}) we get that $\one\otimes\G(\one_{\rm
c}(\omega))\invWave^+(j)=\invWave^+(j)$ i.e. ${\rm
Ran}\invWave^+(j)\subset \cH^{\rm scatt}$.  The rest of the proof
of {\it iv)} is as in \cite[Thm. 12.4]{DG}. \qed

\subsection{Geometric asymptotic completeness}
\label{sec7.4}
In this subsection we will show that
\[
\Ran P_0^+=\cK^+.
\]
We call this property {\em geometric asymptotic completeness}. It
will be convenient to work in the scattering  space $\cH^{\rm
scatt}$ and to treat $\Omega^+$ as a partial isometry
$\Omega^+:\cH^{\rm scatt}\to\cH$, as explained in Subsect.
\ref{sec5.3}. \bet Assume {\it (G1)}, {\it (G2)},  {\it (S)}, {\it
(Is)} for $s>1$. Let $j_n=(j_{0,n},j_{\infty,n})$ satisfy the
conditions of Subsect. \ref{sec7.3}. Additionally, assume that
$j_{0,n}+j_{\infty,n}=1$ and that for any $\epsilon>0$, there
exists $m$ such that, for $n>m$, $\supp
j_{0,n}\subset[-\epsilon,\epsilon]$.
 Then
\[
\Omega^{+*}=\wlim_{n\to\infty}\invWave^+(j_n).
\]
Besides
 \[
\cK^+=\Ran P_0^+.
\]
\label{P51A}
\eet
\proof
The proof is analogous to \cite[Thm. 12.5]{DG}. Since it is in
important step, we will give some details.
If  $q\in C_0^\infty(\rr)$ is such that $q=1$ in a neighborhood of $0$,
$0\leq q\leq 1$ then for sufficiently big $n$ we have $qj_{0,n}=j_{0,n}$.
 Therefore, for sufficiently big $n$
by Thm. \ref{supermain} {\it iii)}
\[
(\Gamma^+(q)\otimes\one) W^+(j_n)-W^+(j_n)=0.
\]
Hence
\beq
\wlim_{n\to\infty}\Big(P_0^+\otimes\one W^+(j_n)-W^+(j_n)\Big)=0.
\label{pzero}
\eeq
Let  $\chi\in \coinf(\rr)$. We have
\[
\begin{array}{rll}
\Omega^{+*}\chi(H)&=
\Omega^{+*}\Omega^{\ext,+}\chi(H^\ext)W^+(j_n)&\ (1)\\[3mm]
&=\wlim_{n\to\infty}\Omega^{+*}\Omega^{\ext,+}\chi(H^{\ext})
\invWave^+( j_n)&\ (2)\\[3mm]
&=\wlim_{n\to\infty}\Omega^{+*}\Omega^{\ext,+}\chi(H^{\ext})
P_0^+\otimes\one\invWave^+( j_n)&\ (3)\\[3mm]
&=\wlim_{n\to\infty}P_0^+\otimes\one\chi(H^{\ext})
\invWave^+( j_n)&\ (4)\\[3mm]
&=\wlim_{n\to\infty}P_0^+\otimes\one
\invWave^+( j_n)\chi(H)&\ (5)\\[3mm]
&=\wlim_{n\to\infty}
\invWave^+( j_n)\chi(H)&\ (6).
\end{array}
\]
We use Thm. \ref{supermain} in step (1), (\ref{pzero}) in step (3),
${\rm Ran}P_{0}^{+}\subset \cK^{+}$ in step (4), Thm. \ref{supermain}
{\it ii)} in step (5) and (\ref{pzero}) again in step (6).
Clearly this implies that:
\[
\Omega^{+*}=\wlim_{n\to\infty}
\invWave^+( j_n).
\]
Therefore by (\ref{pzero})
\[
\Ran\Omega^{+*}\subset\Ran P_0^+\otimes\Gamma(\ch)\subset
\cK^+\otimes\Gamma(\ch).
\]
But by construction
\[
\Ran\Omega^{+*}=
\cK^+\otimes\Gamma(\ch).\]
Hence $\cK^+\otimes\Gamma(\ch)=\Ran P_0^+\otimes\Gamma(\ch)$,
and therefore $\cK^+=\Ran P_0^+$. \qed

\subsection{Asymptotic completeness}
\label{sec7.5}
In this subsection, we will prove asymptotic completeness.
\bet\label{asympt-comp}
Assume hypotheses {\it (Hi)}, $1\leq i\leq 3$, {\it (Gi)}, $1\leq
i\leq 5$, {\it (Mi)} $i=1,2$, {\it (Is)} for $s>1$ and {\it (S)}.
Then:
\[
\cK^{+}= \cH_{\rm pp}(H).
\]
\eet
 \proof
By Proposition \ref{as.4} and geometric asymptotic completeness
we already know that
\[
\cH_\pp(H)\subset\cK^+=\Ran P_0^+.
\]
It remains to prove that $P_{0}^{+}\leq \one_{\rm pp}(H)$.  Let
$\chi\in \coinf(\rr\backslash(\tau\cup \sigma_{\rm pp}(H)))$.
We deduce from Prop. \ref{6.5} in Subsect. \ref{sec6.4}
that there exists
$\epsilon>0$ such that for $q\in \coinf([-\epsilon, \epsilon])$
with $q(x)=1$ for $|x|<\epsilon/2$ we
have
\[
\int_{1}^{+\infty}\|\G(q^{t})\chi(H)\e^{-\i tH}u\|^{2}\frac{\d
t}{t}\leq c\|u\|^{2}.
\]
Since $\|\G(q^{t})\chi(H)\e^{-\i tH}u\|\to \|\G^{+}(q)\chi(H)u\|$,
we have $\Gamma^{+}(q)\chi(H)=0$. This implies that
\[
P_{0}^{+}\leq \one_{\tau\cup\sigma_{\rm pp}}(H).
\]
Since $\tau$
is a closed countable set and $\sigma_{\rm pp}(H)$
can accumulate only
at $\tau$, we see that $\one_{\rm pp}(H)= \one_{\tau\cup\sigma_{\rm pp}}(H)$. This completes the proof of the theorem. \qed

\appendix
\section{Appendix}\label{appendo}\init
\subsection{Proof of Lemma \ref{gauge}}
To prove {\it i)} we restrict the quadratic form
$[F(\frac{\x}{R}), \omega]$ to $\cS$. Using (\ref{HS}), we get
\beq\label{sarko}
\begin{array}{rl}
[F(\frac{\x}{R}), \omega]=&\frac{\i}{2\pi R}\int_{\cc}\partial_{\,
\overline z}\tilde{F}(z)
(z-\frac{\x}{R})^{-1}[\x, \omega](z-\frac{\x}{R})^{-1}\d z\wedge \d\,\overline z,\\[3mm]
=&\frac{\i}{2\pi R}\int_{\cc}\partial_{\,
\overline z}\tilde{F}(z)
(z-\frac{\x}{R})^{-2}[\x, \omega]\d z\wedge \d\,\overline z\\[3mm]
&+\frac{\i}{2\pi R^{2}}\int_{\cc}\partial_{\,
\overline z}\tilde{F}(z)
(z-\frac{\x}{R})^{-2}{\rm ad}^{2}_{\x}\omega(z-\frac{\x}{R})^{-1}\d z\wedge \d\,\overline z
\end{array}
\eeq where the right hand sides are   operators on $\cS$.  Since
$\ad^{2}_{\x}\omega\in S^{0}_{(0)}$, we see that the last term
belongs to $R^{-2}S^{0}_{(0)}$. Using  the bound
$\frac{\x}{R}(z-\frac{\x}{R})^{-1}=O(|{\rm Im}z|^{-1})$ for $z\in
\supp\tilde{F}$, we see that the last term belongs also to
$R^{-1}S^{-1}_{(0)}$. This proves {\it i)} for $k=0$.

Replacing $\omega$ by $[\omega, \x]$ and using that
$\ad^{2}_{\x}[\omega, \x]\in S^{(0)}_{(0)}$ we get also {\it i)} for $k=1$.

{\it ii)} follows from {\it i)} for $k=0$ since $\cS$ is a core for $\omega$.  {\it iii)}  and {\it iv)} are
proved similarly.  {\it v)} is proved as {\it i)},  replacing
$\omega$ by $[\omega, \i a]_{0}$ and using only the first line of
(\ref{sarko}). To prove {\it vi)} we restrict again the quadratic form
$[F(\frac{\x}{R}), \omega^{2}]$ to $\cS$ and get:
\beq
\begin{array}{rl}
&[F(\frac{\x}{R}), \omega^{2}]\\[3mm]
=&\frac{\i}{2\pi R}\int_{\cc}\partial_{\,
\overline z}\tilde{F}(z)
(z-\frac{\x}{R})^{-1}[\x, \omega^{2}](z-\frac{\x}{R})^{-1}\d z\wedge \d\,\overline z,\\[3mm]
=&\frac{\i}{2\pi R}\int_{\cc}\partial_{\,
\overline z}\tilde{F}(z)
(z-\frac{\x}{R})^{-1}\left(2[\x, \omega]\omega+ [\omega, [\x,
\omega]]\right)(z-\frac{\x}{R})^{-1}\d z\wedge \d\,\overline z,
\end{array}
\label{sarko-bis}
\eeq
where the right hand sides are   operators on $\cS$.
Note that $[\omega, [\x,
\omega]]$ is bounded by {\it (G2)}. We use next that $\omega(z-\xr)^{-1}\omega^{-1}\in O(|{\rm
Im}z)|^{-2}$ uniformly in $R\geq 1$ to obtain {\it vi)}.

To prove {\it vii)}, we pick another function $F_{1}\in \coinf(\rr\backslash \{0\})$ such
that $F_{1}F=F$ and note that
\[
[F(\frac{\x}{R}), b]= F(\frac{\x}{R})[F_{1}(\frac{\x}{R}), b]+
[F(\frac{\x}{R}), b]F_{1}(\frac{\x}{R}).
\]
Applying again (\ref{HS}), we get
\[
[F(\frac{\x}{R}), b]=\frac{\i}{2\pi R}\int_{\cc}\partial_{\,
\overline z}\tilde{F}(z)
(z-\frac{\x}{R})^{-1}[\x, b](z-\frac{\x}{R})^{-1}\d z\wedge \d\,
\overline z,
\]
and the analogous formula for $[F_{1}(\frac{\x}{R}), b]$.
We  use then that $[\x, b]\in S^{-\mu+ \delta}_{(0)}$ and Lemma
\ref{gototime},  moving  powers of $\x$ through the resolvents either
to the left or to the right to obtain {\it vii)}. \qed

\subsection{Proof of Lemma \ref{compare}.}

We use the identity:
\[
\omega^{-\12}= c_{0}\int_{0}^{+\infty}s^{-\12}(\omega+ s)^{-1}\d s,
\]
to get:
\[
\omega^{\12}[F(\xr), \omega^{-\12}]=
c_{0}\int_{0}^{+\infty}s^{-\12}\omega^{\12}(\omega+ s)^{-1}[F(\xr),
\omega](\omega+ s)^{-1}\d s\in O(R^{-1}),
\]
since $\omega\geq m>0$. Hence
\[
\begin{array}{rl}
&\omega^{-\12}(\omega-\omega_{\infty})F(\xr)\omega^{-\12}\\[3mm]
=&\omega^{-\12}(\omega- \omega_{\infty})\omega^{-\12}F(\xr)+
\omega^{-\12}(\omega-
\omega_{\infty})\omega^{-\12}\omega^{\12}[F(\xr), \omega^{-\12}]
\\[3mm]
= &\omega^{-\12}(\omega-
\omega_{\infty})\omega^{-\12}\x^{\epsilon}\x^{-\epsilon}F(\xr)+
O(R^{-1})\\[3mm]
=& O(R^{-\epsilon})+ O(R^{-1}).
\end{array}
\]
The second statement of the lemma is obvious. \qed
\subsection{Proof of Lemma \ref{painful}.}

 Since by {\it (G1)} $[v, \x]$ extends from $\cS$ as a bounded
operator on $\ch$ and $\cS$ is a core for $\x$, we get that $v$
preserves $\cD(\x)$. Since  $\xt-v$ is selfadjoint on $\cD(\x)$ we get
\[
\cD(c)=\cD((\xt-v)^{2})=\{u\in \cD(\x)|\ \  (\xt-v)u\in \cD(\x)\}=
\cD(\x^{2}),
\]
so $c$ is selfadjoint on $\cD(\x^{2})$. Since  $v\in
S^{0}_{(0)}$ we get by Lemma \ref{gototime} that   $J(\xt)cJ(\xt)\in O(1)$
which proves {\it i)}.

Let us now prove {\it ii)}.  We first consider the commutator $[c,
J(\xt)]$ for $J\in \coinf(\rr)$. We have
\[
\begin{array}{rl}
[c, J(\xt)]= &(\xt -v)[v, J(\xt)] + [v, J(\xt)](\xt -v)\\[3mm]
=& t^{-1}(\xt -v)J'(\xt)[v, \x]+ t^{-1}J'(\xt)[v, \x](\xt -v)\\[3mm]
+ &(\xt-v)M(t)+ M(t)(\xt -v),
\end{array}
\]
where $M(t)\in t^{-2}S^{0}_{(0)}\cap t^{-1}S^{-1}_{(0)}$ by Lemma
\ref{gauge} {\it i)}.  This implies that the last two terms in the
r.h.s. are $O(t^{-2})$. Using then that  $[v, J'(\xt)]\in O(t^{-1})$
and $[[v, \x], \xt]\in O(t^{-1})$  since $v\in S^{0}_{(3)}$, we see
that
\[
\begin{array}{l}
(\xt -v)J'(\xt)[v, \x]= J'(\xt)M_{1}(t)+ O(t^{-1}), \\[3mm]
  J'(\xt)[v,\x](\xt -v)= J'(\xt)M_{2}(t)+ O(t^{-1}),
\end{array}
\]
where $M_{i}(t)\in O(1)$.  This shows that:
\beq
[c, J(\xt)]= \frac{1}{t}J'(\xt) O(1)+ O(t^{-2}).
\label{e6.3b}
\eeq
We will use the following identities valid for $\lambda>0$:
\beq
\lambda^{-\12}= c_{0}\int_{0}^{+\infty}s^{-\12}(\lambda+ s)^{-1}\d s,
\ \ \lambda^{\12}= c_{0}\int_{0}^{+\infty}s^{-\12}\lambda(\lambda+
s)^{-1}\d s,
\label{e6.5}
\eeq
and
\beq
\lambda^{-\frac{3}{2}}= 2c_{0}\int_{0}^{+\infty}s^{-\12}(\lambda+
s)^{-2}\d s,
\label{e6.6b}
\eeq
which follows by differentiating the first identity of (\ref{e6.5})
w.r.t. $\lambda$.  A related obvious bound is:
\beq
\int_{0}^{+\infty}s^{-\12} (t^{-\delta}+ s)^{-n}\d s=
O(t^{(n-\12)\delta}), \ \ n\geq 1.
\label{obvious}
\eeq

From (\ref{e6.5}) we obtain that
\beq\label{e6.7b}
c^{\12}= c_{0}\int_{0}^{+\infty}s^{-\12}c(c+ s)^{-1}\d s, \hbox{ as a
strong integral on }\cD(c).
\eeq
Therefore
\[
[c^{\12}, J(\xt)]= c_{0}\int_{0}^{+\infty}s^{-\12}\left([c, J(\xt)](c+
s)^{-1}-c(c+ s)^{-1}[c,
J(\xt)](c+s)^{-1}\right)\d s
\]
We use the bounds
\beq\label{urk}
\|c(c+s)^{-1}\|\leq 1, \ \ \|(c+s)^{-1}\|\leq
(t^{-\delta}+ s)^{-1},
\eeq
 and (\ref{e6.3b}) to obtain
\[
\|[c^{\12}, J(\xt)]\|\leq Ct^{-1}\int_{0}^{+\infty}s^{-\12}(t^{-\delta}+
s)^{-1}\d s=O(t^{-1+ \delta/2}),
\]
by (\ref{e6.5}), which proves {\it ii)}.

To prove {\it iii)} we first compute
\beq\label{e6.7bis}
\dd_{0}c= -\frac{2}{t}(\xt-v)^{2} -\left([\omega, \i v](\xt -v)+
\hc\right) -\delta t^{-\delta-1}.
\eeq
We first rewrite the second term in the r.h.s. in a convenient way:

by {\it (G2)}, we have
\[
[\omega, \i v]= \gamma^{2}+ r_{-1-\epsilon}, \ \ \gamma\in
S^{-\12}_{\epsilon, (1)}, \ \ r_{-1-\epsilon}\in
S^{-1-\epsilon}_{(0)}.
\]
Since $v\in S^{0}_{(0)}$, we get first that:
\beq
(\xt -v)r_{-1-\epsilon}\in O(t^{-1})S^{-\epsilon}_{(0)}+
S^{-1-\epsilon}_{(0)}.
\label{airk}
\eeq
We claim also that
\beq
[\gamma, \xt -v]\in O(t^{-1})S^{-\12+ \epsilon}_{(0)} + S^{-3/2 +
2\epsilon}_{(0)}.
\label{iark}
\eeq
Clearly $[\gamma, \x]\in S^{-\12+ \epsilon}_{(0)}$. To handle
$[\gamma, v]$ we use the Lie identity and write:
\beq\label{ork}
\i\:[\gamma, v]= -[\gamma, [\omega, \x]]= [\omega, [\x, \gamma]]+
[\x, [\omega, \gamma]]  \in S^{-3/2 +2\epsilon}_{(0)},
\eeq
which proves (\ref{iark}). By Lemma \ref{gototime} {\it i)}, we get that
\[
\gamma[\gamma, \xt -v], \ \ [\gamma, \xt -v]\gamma\in t^{-1}S^{-1+
\epsilon}_{(0)} + S^{-2+ 2\epsilon}_{(0)},
\]
and hence using that $0<\epsilon<\12$:
\[
[\omega, \i v](\xt -v)+ \hc.= 2\gamma(\xt -v)\gamma+ R_{2}(t),
\]
where $R_{2}(t)\in  O(t^{-1})S^{-\epsilon_{1}}_{(0)}+
S^{-1-\epsilon_{1}}_{(0)}$, for some $\epsilon_{1}>0$.
We set now:
\[
R_{0}(t)= -\frac{2c}{t},\ \ R_{1}(t)= -(\delta-2)t^{-\delta-1}, \ \
R_{3}(t)=-2\gamma(\xt -v)\gamma,
\]
and rewrite (\ref{e6.7bis}) as
\[
\dd_{0}c= \sum_{i=0}^{3}R_{i}(t).
\]
Using (\ref{e6.7b}), we obtain as a strong integral on $\cD(c)$:
\[
\begin{array}{rl}
\dd_{0}c^{\12}=& c_{0}\int_{0}^{+\infty}s^{-\12}\left(
\dd_{0}c(c+ s)^{-1}-c(c+s)^{-1}\dd_{0}c(c+s)^{-1}\right)\d s\\[3mm]
=&\mathop{\sum}\limits_{i=0}^{3}c_{0}\int_{0}^{+\infty}s^{-\12}\left(
R_{i}(t)(c+ s)^{-1}-c(c+s)^{-1}R_{i}(t)(c+s)^{-1}\right)\d s\\[3mm]
=:& \mathop{\sum}\limits_{i=0}^{3}I_{i}(t).
\end{array}
\]
Using (\ref{e6.5}) we obtain
\[
I_{0}(t)= -\frac{1}{t}c^{\12}, \ \ I_{1}(t)=
Ct^{-\delta-1}c^{-\12}=O(t^{-\delta/2-1}).
\]
It remains to handle the terms $J(\xt)I_{i}(t)J(\xt)$ for $i=2,3$. We
write them as:
\[
\begin{array}{rl}
J(\xt)I_{i}(t)J(\xt)=&c_{0}\int_{0}^{+\infty}s^{-\12}J(\xt)
R_{i}(t)(c+ s)^{-1}J(\xt)\d s\\[3mm]
&-c_{0}\int_{0}^{+\infty}s^{-\12}J(\xt)c(c+s)^{-1}R_{i}(t)(c+s)^{-1}J(\xt)\d s.
\end{array}
\]
 We will need to use the fact that $O\not\in \supp J$.
To do this we claim that if
$J, J_{1}\in \coinf(\rr)$  with $J_{1}\equiv 1$ near $\supp J$ then:
\beq
J(\xt)(c+ s)^{-1}(1-J_{1})(\xt)\in O(t^{-2}(t^{-\delta}+s)^{-2})+
O(t^{-2}(t^{-\delta}+s)^{-3}),
\label{oinkoink}
\eeq
\beq
J(\xt)c(c+ s)^{-1}(1-J_{1})(\xt)\in O(t^{-2}(t^{-\delta}+s)^{-1})+
O(t^{-2}(t^{-\delta}+s)^{-2}).
\label{oinkoinkoink}
\eeq

We pick  $T_{1} \in \coinf(\rr)$,   $T_{1}\equiv 1$ on $\supp J_{1}'$, $T_{1}\equiv 0$ on $\supp
J$. We write using (\ref{e6.3b}):
\[
\begin{array}{rl}
&J(\xt)(c+s)^{-1}(1-J_{1})(\xt)\\[3mm]
= &J(\xt)(c+s)^{-1}[c,J_{1}(\xt)](c+s)^{-1}\\[3mm]
=& J(\xt)(c+s)^{-1}T_{1}(\xt)O(t^{-1})(c+s)^{-1}+
J(\xt)(c+s)^{-1}O(t^{-2})(c+s)^{-1}\\[3mm]
=& J(\xt)(c+s)^{-1}[T_{1}(\xt, c)](c+s)^{-1}O(t^{-1})(c+s)^{-1}+
J(\xt)(c+s)^{-1}O(t^{-2})(c+s)^{-1}\\[3mm]
=& J(\xt)(c+s)^{-1}O(t^{-1})(c+s)^{-1}O(t^{-1})(c+s)^{-1}+
J(\xt)(c+s)^{-1}O(t^{-2})(c+s)^{-1}.
\end{array}
\]
We obtain (\ref{oinkoink}) using the
bound $\|(c+s)^{-1}\|\leq (t^{-\delta}+s)^{-1}$. (\ref{oinkoinkoink})
follows from (\ref{oinkoink}) using that $c(c+s)^{-1}= \one
-s(c+s)^{-1}$.

We hence fix a cutoff $J_{1}\in \coinf(\rr\backslash \{0\})$ such that
$J_{1}\equiv 1$ on $\supp J$ and set
\[
\tilde{R}_{i}(t)= J_{1}(\xt)R_{i}(t)J_{1}(\xt),
\]
and denote by $\tilde{I}_{i}(t)$ the analogs of
$I_{i}(t)$ for $R_{i}(t)$ replaced by $\tilde{R}_{i}(t)$.

We claim that:
\beq
\label{irk}
J(\xt)\left(I_{i}(t)-\tilde{I}_{i}(t)\right)J(\xt)\in O(t^{-2 +
5\delta/2}), \ \ i=2,3.
\eeq
To prove (\ref{irk}), we note that $\tilde{I}_{i}(t)$ is obtained from
$I_{i}(t)$ by inserting $J_{1}(\xt)$ to the left and right of
$R_{i}(t)$ under the integral sign. The error terms under the integral
sign coming from this insertion are estimated using (\ref{oinkoink}),
(\ref{oinkoinkoink}) and the fact that $R_{i}(t)\in O(1)$ for $i=2,
3$, since $\gamma\in S^{-\12}_{(0)}$.  The integrals of these error
terms are estimated using (\ref{obvious}), which by a painful but
straightforward computation gives (\ref{irk}).

By Lemma \ref{gototime} {\it ii)}, we know that $\tilde{R}_{2}(t)\in
O(t^{-1-\epsilon_{1}})$ for some $\epsilon_{1}>0$ small enough, hence
using the bounds (\ref{urk}) and (\ref{obvious}), we obtain that for $\delta>0$ small
enough
\[
\tilde{I}_{2}(t) \hbox{ and hence }J(\xt)I_{2}(t)J(\xt)\in O(t^{-1-\epsilon_{2}}), \ \ \epsilon_{2}>0.
\]
To treat $\tilde{I}_{3}(t)$, we use that
\[
\tilde{R}_{3}(t)= \gamma^{*}_{t}(\xt -v)\gamma_{t}, \hbox{ for
}\gamma_{t}= \gamma J_{1}(\xt).
\]

We claim that
\beq
[\gamma_{t}, c]\in O(t^{-3/2+ \epsilon}).
\label{iti}
\eeq
Let us prove this claim. We write:
\[
[\gamma_{t}, c]= (\xt -v)[\gamma_{t}, \xt -v]+ [\gamma_{t}, \xt
-v](\xt -v),
\]
and
\[
[\gamma_{t}, \x]= [\gamma, \x]J_{1}(\xt), \ \ [\gamma_{t}, v]=
[\gamma, v]J_{1}(\xt)+ \gamma[J_{1}(\xt), v].
\]
Now
\[
(\xt -v)[\gamma, \x]J_{1}(\xt), \ \ [\gamma, \x]J_{1}(\xt)(\xt -v)\in
O(t^{-\12+ \epsilon}).
\]
This follows from the fact that  $[\gamma, \x]\in S^{-\12+
\epsilon}_{(0)}$, $0\not\in \supp J_{1}$ and Lemma \ref{gototime} {\it
ii)}.  Similarly we saw in
(\ref{ork}) that $[\gamma, v]\in S^{-3/2+ 2\epsilon}_{(0)}$, which
implies that:
\[
(\xt -v)[\gamma, v]J_{1}(\xt), \ \ [\gamma, v]J_{1}(\xt)(\xt -v)\in
O(t^{-3/2+ 2\epsilon}).
\]
Finally using Lemma \ref{gauge} {\it i)} we write:
\[
[J_{1}(\xt ,v)]= \frac{1}{t}J_{1}'(\xt)[\x, v]+ M(t), \ M(t)\in
O(t^{-2})S^{0}_{(0)}\cap O(t^{-1})S^{-1}_{(0)}.
\]
Since
$\gamma\in S^{-\12}_{(0)}$ and $[\x, v]\in S^{0}_{(0)}$, we get that
\[
(\xt -v)\gamma J_{1}'(\xt)[\x, v], \ \ \gamma J_{1}'(\xt)[\x,
v](\xt -v)\in O(t^{-\12}),
\]
and since $M(t)\in O(t^{-2})S^{0}_{(0)}\cap O(t^{-1})S^{-1}_{(0)}$:
\[
(\xt -v)\gamma M(t), \ \ \gamma M(t)(\xt -v)\in O(t^{-2}).
\]
Collecting the various estimates we obtain (\ref{iti}).

From the estimate of $[\gamma_{t},c]$ we obtain:
\beq
[\gamma_{t}, (c+ s)^{-1}]\in O(t^{-3/2 + \epsilon}(t^{-\delta}+
s)^{-2}),
\label{urkurk}
\eeq
\beq
[\gamma_{t}, c(c+ s)^{-1}]\in O(t^{-3/2 + \epsilon}(t^{-\delta}+s)^{-1}).
\label{urkbis}
\eeq
We now write:
\[
\begin{array}{rl}
\tilde{I}_{3}(t)=&c_{0}\int_{0}^{+\infty}s^{-\12}
\gamma^{*}_{t}(\xt -v)\gamma_{t}(c+ s)^{-1}\d s\\[3mm]
&-c_{0}\int_{0}^{+\infty}s^{-\12}c(c+s)^{-1}\gamma^{*}_{t}(\xt -v)\gamma_{t}(c+s)^{-1}\d s
\end{array}
\]
We first move $\gamma_{t}$ to the right in the two integrals using
(\ref{urkurk}) and the fact that
\[
\gamma_{t}^{*}(\xt -v)= J_{1}(\xt)\gamma(\xt-v)\in O(1),
\]
since $\gamma\in S^{-\12}_{(0)}$. We obtain errors terms of size
$O(t^{-3/2 + \epsilon+ 5\delta/2})$ using (\ref{obvious}). We then
move $\gamma_{t}^{*}$ to the left in the second integral using
(\ref{urkbis}) and the fact that
\[
\|(\xt -v)(c+ s)^{-1}\|\leq \|c^{\12}(c+ s)^{-1}\|\leq t^{\delta/2}.
\]
We obtain error terms of size $O(t^{-3/2+ \epsilon+ \delta})$ using
again (\ref{obvious}).
Hence for $\delta>0$ small enough, we get:
\[
\begin{array}{rl}
\tilde{I}_{3}(t)=& c_{0}\int_{0}^{+\infty}\gamma^{*}_{t}s^{-\12}
(\xt -v)(c+ s)^{-1}\gamma_{t}\d s\\[3mm]
&-c_{0}\int_{0}^{+\infty}\gamma^{*}_{t}s^{-\12}c(c+s)^{-1}(\xt -v)(c+s)^{-1}\gamma_{t}\d s\\[3mm]
&+ O(t^{-1-\epsilon_{1}})\end{array}
\]
for some $\epsilon_{1}>0$. The integrals can be computed exactly since
$\xt -v$ commutes with $c$ and are equal to $C_{1}(\xt -v)c^{-\12}$
for some constant $C_{1}$ and hence $O(1)$. This yields:
\[
\begin{array}{rl}
J(\xt)\tilde{I}_{3}(t)J(\xt)=&
J(\xt)\gamma^{*}_{t}M(t)\gamma_{t}J(\xt)+ O(t^{-1-\epsilon_{1}})\\[3mm]
=& J(\xt)\gamma M(t)\gamma J(\xt)+ O(t^{-1-\epsilon_{1}}),
\end{array}
\]
for $M(t)\in O(1)$. Using also (\ref{irk}), the same equality holds
for $J(\xt)I_{3}(t)J(\xt)$. Finally we use that
$\gamma J(\xt)\in O(t^{-\12})$, by Lemma \ref{gototime} {\it ii)} and
$[\gamma, J(\xt)]\in O(t^{-3/2+ \epsilon})$, to get:
\[
J(\xt)\gamma M(t)\gamma J(\xt)= \gamma J(\xt)M(t)J(\xt)\gamma +
O(t^{-2+ \epsilon}).
\]
Hence
\[
J(\xt)I_{3}(t)J(\xt)= \gamma J(\xt)M(t)J(\xt)\gamma +
O(t^{-1-\epsilon_{1}}),
\]
which completes the proof of {\it iii)}.

Let us now prove {\it iv)}. Set
\[
B_{0}= J(\xt)(\xt -v)+ \hc, \ \ B_{1}=
J_{1}(\xt)c^{\12}J_{1}(\xt).
\]
By Lemma \ref{gauge} we have:
\[
\begin{array}{rl}
B_{0}^{2}=& 4(\xt -v) J^{2}(\xt)(\xt -v)+ O(t^{-1})\\[3mm]
\leq & C(\xt -v)J_{1}^{4}(\xt)(\xt -v)+ O(t^{-1})\\[3mm]
=&CJ_{1}^{2}(\xt)(\xt -v)^{2}J_{1}(\xt)+ O(t^{-1})\\[3mm]
=&C J_{1}^{2}(\xt)c J_{1}^{2}(\xt) + O(t^{-\delta})\\[3mm]
=&C J_{1}(\xt)c^{\12}J_{1}^{2}(\xt)c^{\12}J_{1}(\xt)+O(t^{-\epsilon_{0}})\\[3mm]
=& CB_{1}^{2}+ O(t^{-\epsilon_{0}}),
\end{array}
\]
where we used {\it ii)} in the last step. Applying then Heinz theorem
we obtain that
\[
|B_{0}|\leq C(B_{1}^{2}+ t^{-\epsilon_{0}})^{\12}\leq CB_{1}+
Ct^{-\epsilon_{0}/2},
\]
which proves {\it iv)}.

To prove {\it v)} we set
\[
B_{2}= J(\xt)(\xt -v)c^{\12}J_{1}(\xt)+ \hc.
\]
Using {\it ii)} and Lemma \ref{gauge}, we get:
\[
\begin{array}{rl}
\pm B_{2}= &\pm\left((\xt -v)J J_1(\xt)c^{\12} +\hc\right)+ O(t^{-1+ \delta/2})\\[3mm]
=&\pm\left(c^{\12}(\xt -v)c^{-\12}J J_1(\xt)c^{\12}+ \hc\right) +
O(t^{-1+ \delta/2})\\[3mm]
\leq& Cc + O(t^{-1+ \delta/2})\\[3mm]
\leq &C(\xt -v)^{2}+ O(t^{-\epsilon_{0}}),
\end{array}
\]
since $(\xt -v)c^{-\12}$ is bounded with norm $O(1)$. Since
$B_{2}= J_{2}(\xt)B_{2}J_{2}(\xt)$ we get
\[
\pm B_{2}\leq C J_{2}(\xt)(\xt-v)^{2}J_{2}(\xt) +
O(t^{-\epsilon_{0}})= C(\xt -v)J_{2}^{2}(\xt)(\xt -v)+
O(t^{-\epsilon_{0}}),
\]
by Lemma \ref{gauge}.
\qed

\end{document}